\documentclass[apj,numberedappendix,twocolumn,chicago,onecolappendix]{emulateapj}
\usepackage{hyperref}
\usepackage{graphicx}
\usepackage{color}
\usepackage{bm}
\usepackage{amsmath,amssymb}
\usepackage{ulem}
\usepackage{slashed}
\usepackage{enumerate} 
\usepackage{lmodern}
\usepackage[T1]{fontenc}

\usepackage{subfigure}

\newcommand\lsim{\mathrel{\rlap{\lower4pt\hbox{\hskip1pt$\sim$}}
        \raise1pt\hbox{$<$}}}
\newcommand\gsim{\mathrel{\rlap{\lower4pt\hbox{\hskip1pt$\sim$}}
        \raise1pt\hbox{$>$}}}

\newcommand{\D}{\mathrm{d}}

\newcommand{\Qmat}{\mathbf{Q}}

\def\ffrac#1#2{{\textstyle\frac{#1}{#2}}}
\def\eq{{\rm eq}}

\newcommand{\LA}{\left\langle}
\newcommand{\RA}{\right\rangle}
\newcommand{\In}{\rm in}
\newcommand{\Out}{\rm out}
\newcommand{\parsec}{\,{\rm pc}}

\begin{document}

\title{Isotropic--Nematic Phase Transitions in Gravitational Systems}
\author{Zacharias Roupas}
\affiliation{Institute of Physics, E\"otv\"os University, P\'azm\'any P.~s.~1/A, Budapest, 1117, Hungary}

\author{Bence Kocsis}
\affiliation{Institute of Physics, E\"otv\"os University, P\'azm\'any P.~s.~1/A, Budapest, 1117, Hungary}

\author{Scott Tremaine}
\affiliation{Institute for Advanced Study, Princeton, NJ 08540, USA}

\begin{abstract}
We examine dense self-gravitating stellar systems dominated by a central potential, such as 
nuclear star clusters hosting a central supermassive black hole. 
Different dynamical properties of these systems evolve on vastly different timescales. In particular, the orbital-plane orientations are typically driven 
into internal thermodynamic equilibrium by vector resonant relaxation before the orbital eccentricities  or semimajor axes relax. 
We show that the statistical mechanics of such systems exhibit a striking resemblance to liquid crystals, with analogous ordered-nematic and 
disordered-isotropic phases. The ordered phase consists of bodies orbiting in a disk in both directions, with the disk thickness depending on temperature, while the disordered phase corresponds to a nearly isotropic distribution of the orbit normals. 
We show that below a critical value of the total angular momentum, the system undergoes a first-order phase transition between the ordered and disordered phases. At the critical point the phase transition becomes second-order while for higher angular momenta there is a smooth crossover. 
We also find metastable equilibria containing two identical disks with mutual inclinations between $90^{\circ}$ and $180^\circ$. 
\end{abstract}

\maketitle

\section{Introduction}\label{s:Intro}

The thermodynamics and statistical mechanics of a self-gravitating gaseous or stellar system is an intriguing subject with a long history \citep{Tolman_1934,LyndenBell:1966bi,Lynden-Bell_1968,Shu_1978,tremaine+henon_1986,Padmanabhan_1990} and many recent developments \citep{Gurzadian_1986,Nakamura_2000,Votyakov_2002a,deVega_Sanchez_2002a,deVega_Sanchez_2002b,Arad_2005,Chavanis_2006,Axenides:2012bf,Touma+Tremaine_2014,
BarOr+Alexander_2014,Roupas:2014sda,Tremaine_2015}.
Astrophysical applications include star clusters, galaxies, dark matter halos, and galaxy clusters. 
Statistical mechanics offers the hope of describing the \textit{macroscopic} equilibrium structure of astrophysical systems without tracking the \textit{microscopic}, particle-by-particle, evolution. The traditional approach, in which thermodynamic equilibrium is found by maximizing the entropy, does not work straightforwardly in self-gravitating systems for several reasons. 
First, there is no global entropy maximum for an isolated self-gravitating gaseous or stellar system \citep{Antonov_1962}, although long-lived metastable thermal equilibrium states (i.e., local entropy maxima) can exist if the system is confined to a finite volume \citep{Lynden-Bell_1968,Chavanis_2005}. 
In addition, depending on the number of objects and the size of the system, the stochastic process by which the system evolves toward statistical equilibrium is often too slow for the system to approach the equilibrium state during its lifetime and therefore relaxation remains incomplete
\citep{Binney+Tremaine_2008}. 

In the traditional thermodynamic approach, employed in most of the above references, the fundamental microscopic entities are stars, molecules, or elementary particles, considered to be point masses, which we label ``bodies''. 
However, in contrast to systems with only short-range forces, the phase-space trajectories of bodies in self-gravitating systems are in many cases endowed with a special structure 
where the location of each body is restricted over long intervals to a bounded region of phase space, usually of lower dimensionality \citep{Sridhar+Touma_1999}. This occurs if the dynamical system admits isolating integrals of motion \citep{Kandrup_1998,Binney+Tremaine_2008} and allows the introduction of the familiar concept of the \textit{orbit}. 
\cite{Rauch+Tremaine_1996} showed that certain degrees of freedom that describe the orbits may relax much faster than others, in a process they called \textit{resonant relaxation} (see also \citealt{Chavanis_2012,Fouvry_2015}). For example, bodies bound to a dominant central point mass execute eccentric orbits described by an angular-momentum vector and an eccentricity vector, which relax much faster than the semimajor axis. The phase-space distribution may then attain 
partial thermodynamic equilibrium, in which the rapidly evolving degrees of freedom are driven to internal thermodynamic equilibrium while others are frozen-in at their initial values. 

Another example is a system dominated by a spherically symmetric potential, in which the directions of the angular-momentum vectors relax much faster than either the magnitudes of the angular-momentum vectors or the energies of the orbits (equivalently, much faster than either the eccentricities or semimajor axes).  This process is called \textit{vector resonant relaxation} (VRR) \citep{Rauch+Tremaine_1996,Hopman+Alexander_2006,Gurkan+Hopman_2007,Eilon_2009,Kocsis+Tremaine_2011,Kocsis_2015}, and is the subject of this work.

Gravitational systems that relax through VRR are conceptually similar to liquid crystals, as we will describe in more detail later in this paper. In normal liquids the microscopic degrees of freedom are the positions and momenta of the molecules, while their size may be ignored. However, in certain liquids, the molecules' size and shape gives rise to new types of macroscopic structures. The liquid may acquire properties similar to a crystal: a macroscopic alignment of the molecules' orientation with discrete rotational symmetries. For example in the \textit{nematic phase} of a liquid crystal, axisymmetric molecules align in parallel or antiparallel configurations \citep{Singh_2002}. Similarly, in gravitational systems subject to VRR, the  basic microscopic entities, the orbits, behave as solid bodies. Their orientational degrees of freedom relax into an internal thermodynamic equilibrium. The correspondence with liquid crystals is due to the similarity between the Coulomb-type  electro- and magnetostatic interaction acting between axisymmetric molecules and the Newton\-ian gravitational interaction acting between orbits.
We shall show that there is a gravitational phase transition between disk+halo and isotropic phases\footnote{Note that this is different from the gravitational phase transitions commonly discussed in the literature, which usually refer to a transition between collapsed (core-halo structure) and diffuse phases in spherical systems \citep{Aronson_1972,Stahl_1994,Chavanis_2006}.}, which strongly resembles the nematic-isotropic phase transition of liquid crystals.

Furthermore, we identify one more connection with condensed-matter physics, the existence of negative-temperature equilibria for which entropy decreases with increasing energy \citep{Kocsis+Tremaine_2011}. Generally, negative-temperature equilibria may arise if the energy has an upper bound. In nuclear spin systems, the relaxation time of spin-spin interactions is typically much smaller than that of spin-lattice interactions \citep{Ramsey_1956}, leading to a thermodynamic equilibrium for the former while the latter remain frozen at their initial values \citep{LandauLifshitz}.
These systems admit negative-temperature configurations, in which most spins are anti-aligned with respect to the magnetic field.  In VRR the role of spin is played by the orbital angular momentum and that of the magnetic field by the thermodynamic variable conjugate to the total angular momentum, the ``rotation'' (Eq.\ \ref{eq:omega}).

For a specific example of VRR, let us consider nuclear star clusters (NSCs), which consist of stars orbiting a supermassive black hole (SMBH).  NSCs are found at the centers of galaxies and include the densest stellar systems in the Universe \citep{Norris_2014}. 
The number density of various stellar types in NSCs increases steeply inwards with radius as $r^{-1.5}$ to $r^{-2.9}$ within the radius of influence of the SMBH, which is $\sim 1\parsec$ for the NSC at the center of the Milky Way 
\citep{Schodel_2009,Schodel_2014,Bartko_2009,Yelda_2014,Chatzopoulos_2015}.
In the Milky Way's NSC, the low-mass old stars are spherically distributed. There is also a prominent population of young stars at radii between about 0.03 and $0.5\parsec$.\footnote{Based on their colors and spectra these are O-type main-sequence stars, which have an estimated age of 4-6 Myr.} The distribution of orbital planes of the young stars exhibits several structures with debated physical origin, including a disk in the inner region $0.03$--$0.12\parsec$, a clumpy, disordered, roughly spherical structure in the intermediate region $0.12$--$0.2\parsec$ and a warped, twisted disk in the outer region $0.2$--$0.5\parsec$ \citep{Bartko_2009,Genzel_2010}. A fraction of stars in the intermediate region reside in a disk that rotates clockwise as seen from the Sun; a smaller fraction appear to reside in a disk that rotates counterclockwise, with orbit normals separated by $\sim 100^\circ$ from the normal to the clockwise disk; and the rest constitute a roughly spherical distribution \citep{Levin_2003,Bartko_2009}. \cite{Yelda_2014} estimate that $25\%$ of young stars are in the clockwise disk and $75\%$ are spherically distributed. The nearly identical ages of the young stars in both disks suggest a common origin, but why is the distribution of orbital planes so complex? Is it possible that this complicated structure can exist in statistical equilibrium, or is this a transient feature? 

A further motivation to understand the relaxation processes in dense stellar populations comes from the emerging field of gravitational-wave astronomy. The recent LIGO discovery of gravitational waves from the merger of two black holes has opened a new window on the Universe \citep{2016PhRvL.116f1102A}. Most of the massive stars formed in the past history of the Galaxy would by now have turned into neutron stars and stellar-mass black holes. The number density of stellar-mass black holes is expected to be up to a billion times higher in NSCs than in the Galactic field so NSCs may dominate the rate of black hole-black hole mergers detectable by LIGO \citep{O'Leary_2009}. An understanding of the expected distribution of black holes in galactic nuclei may influence gravitational-wave search strategies and eventually help to interpret detections. 
 
The main stellar-dynamical processes in NSCs may be ordered as follows. The spherical potential due to the central SMBH of mass $M_\bullet$ ($4\times10^6M_\odot$ for the Milky Way) supports eccentric Keplerian orbits with a characteristic orbital time $t_{\rm orb}\sim (G M_\bullet/r^3)^{-1/2}$, between $10^2$ and $10^3$~yr for distances (more precisely, semimajor axes) $r$ between $\sim 0.03$ and 0.5 pc. The gravitational field of the spherical distribution of old, low-mass stars causes in-plane apsidal precession for each eccentric orbit with a characteristic precession time $t_{\rm in-plane} \sim t_{\rm orb} M_\bullet/(N m)$ or $10^4$--$10^5$ yr in this region, where $N$ is the enclosed number of stars and $m$ is their typical mass. The stellar orbits conserve their angular-momentum vector and orbital energy over timescales of this order. However on longer timescales the time-varying higher multipole moments of the gravitational field of the stars drive chaotic mixing in phase space, through three distinct processes that operate on different timescales (see Figure 1 in \citealt{Kocsis+Tremaine_2011}):
\begin{itemize}
 \item $t_{\rm vrr} \sim t_{\rm orb} M_\bullet/(\sqrt{N} m)$ ($\sim 10^6$--$10^7$~yr), the 
diffusion time of angular-momentum vector directions or orbital planes, called the \textit{vector resonant relaxation} time,
 \item $t_{\rm srr} \sim t_{\rm orb} M_\bullet/m$ ($\sim 10^8$--$10^{10}$~yr), the diffusion time of the magnitude of the angular momenta or eccentricity, called the \textit{scalar resonant relaxation} time, 
 \item $t_{\rm 2-body} \sim t_{\rm orb} M_\bullet^2/(Nm^2)$ ($\sim 10^9$--$10^{10}$~yr), the diffusion time of energy or semimajor axis, called the \textit{two-body relaxation} time.
 \end{itemize}
The age of the young stars observed in the Galactic center is 4--6 Myr, long enough that we expect that VRR has strongly affected the distribution of orbital planes, but short enough that the eccentricities and semimajor axes remain frozen at their initial values. 

More generally, in most of the volume of NSCs inside the sphere of influence of the central SMBH $M_\bullet \gg N m$ and $N\gg 1$, and if these inequalities are satisfied we have the timescale hierarchy \citep{Kocsis+Tremaine_2011}:
\begin{equation}\label{eq:hierarchy}
t_{\rm orb}\ll t_{\rm in-plane} \ll t_{\rm vrr} \ll t_{\rm srr} \ll t_{\rm 2-body}\, .
\end{equation}
This hierarchy implies that different dynamical properties of the system relax on different timescales and they may reach internal statistical equilibrium independently from one another (see \citealt{Sridhar+Touma_2016a,Sridhar+Touma_2016b} for a mathematically rigorous derivation of some of these results).

In this paper, we examine the equilibrium distribution of orbital planes after the VRR process is completed ($t\gtrsim t_{\rm vrr}$) but before scalar resonant relaxation begins ($t\lesssim t_{\rm srr}$). VRR is driven by the gravitational interaction averaged over two much faster motions, the orbital motion around the SMBH and the in-plane precession induced by the spherical distribution of old stars \citep{Kocsis_2015}. Since precessing eccentric stellar orbits trace out axisymmetric punctured disks, the gravitational interaction is to be calculated between two such  disks, in which the local surface density is proportional to the residence time of the star at the given location. 
Since $t_{\rm srr}$ and $t_{\rm 2-body}$ are much longer than the timescale of interest, the eccentricity and semimajor axis are nearly conserved during this process and so the surface density, as a function of radius, of the punctured disk is fixed, though its orientation is not\footnote{During VRR, the relative change in angular momentum is ${\Delta |L|}/{|L|} = (t_{\rm vrr}/t_{\rm srr})^{1/2} \sim N^{-1/4}$, and the change in energy is ${\Delta E}/{E} = (t_{\rm vrr}/t_{\rm 2-body})^{1/2} \sim (Nm/M)^{1/2}N^{-1/4}$. The former sets the change of $\sqrt{1-e^2}$ where $e$ is eccentricity and the latter sets the change in semimajor axis since $E=-G M_\bullet/(2a)$.}. This leads to an averaged or effective Hamiltonian that determines the time evolution of the orientations of the orbit normals or angular-momentum vector directions (see Appendix \ref{s:app:RR}). The VRR Hamiltonian is given by $\frac12 N(N-1)$ terms corresponding to pairwise interactions between the punctured disks. Each pairwise interaction is further decomposed into a sum over multipoles. In this decomposition, the coupling constants depend on the conserved quantities of the two disks, mass, semimajor axis, and eccentricity, which are set by their initial distribution. In the terminology of statistical physics mass, semimajor axis, and eccentricity are quenched random variables, so the coupling constants in the Hamiltonian are random matrices as in spin glasses. If the external perturbations of the galaxy on the NSC may be neglected (as we shall assume in this paper), the energy corresponding to the VRR Hamiltonian and the total angular-momentum vector are also  conserved.

Our main goal in this paper is to describe the mean-field theory of VRR for the dominant quadrupolar interaction, arguing that the qualitative features of the theory would be similar if higher order multipoles were included. The statistical equilibrium is described by a distribution function in orbit-normal space that extremizes the Boltzmann entropy. As we already noted the quadrupolar mean-field model of liquid crystals, the Maier-Saupe model \citep{Maier+Saupe_1958,Plischke+Bergersen_2006}, is analogous to the mean-field model of VRR for a one-component cluster in which the bodies have the same masses, semimajor axes, and eccentricities. 

The model we investigate in this paper is also reminiscent of the Hamiltonian mean-field model \citep{Campa_2014}, which describes the dynamics of $N$ identical rigid rotors interacting via a cosine potential. Both are exactly soluble and exhibit phase transitions and other interesting behavior. In both cases the simplicity arises because the interactions of all pairs of bodies are identical, and the dynamical behavior is entirely determined by a few moments of the distribution function. 

The remainder of the paper is organized as follows. In Section~\ref{s:mean_vrr} we formulate the mean-field theory of vector resonant relaxation. In Section \ref{s:one_comp}, we restrict our attention to the special case of a cluster composed of bodies of a single mass, eccentricity and semimajor axis; this restriction simplifies the physics so that we can explore the thermodynamics and statistical mechanics analytically. We find all of the equilibrium solutions in the microcanonical and canonical ensembles; we present the axisymmetric equilibria in Section~\ref{s:ax-sym}, and the non-axisymmetric ones in Section~\ref{s:biax}. In Section \ref{sec:g_canonical} we discuss a different ensemble, which we call the $\omega T N$-ensemble, in which the system is embedded in a bath with which it can exchange not only energy but also angular momentum. Section \ref{sec:multi} contains a brief discussion of how these results can be applied to separable multi-component systems in which the interactions between components are weak. We discuss our conclusions in Section~\ref{s:con}.

\section{Mean-Field Theory of Vector Resonant Relaxation}\label{s:mean_vrr}

We wish to describe the angular distribution of orbits of gravitating bodies bound to a spherically symmetric potential, usually dominated by a central point mass. To this end we adopt the VRR Hamiltonian obtained by \cite{Kocsis_2015} and reproduced in Appendix~\ref{s:app:RR}. This Hamiltonian is derived by introducing a canonical transformation from Cartesian positions and momenta to action-angle variables, called Delaunay variables, and averaging over the rapidly varying angles, the mean anomaly and argument of periapsis (see also \citealt{Sridhar+Touma_2016a,Sridhar+Touma_2016b,Sridhar+Touma_2016c}). To leading order in the ratio of the stellar mass to the central mass, the VRR Hamiltonian is the average potential energy between the annuli or punctured axisymmetric disks that the precessing elliptical trajectories cover over times long compared to the apsidal precession time ($t_{\rm in-plane}$ in Section~\ref{s:Intro}). The surface density of the annulus at any point is inversely proportional to the time the body resides at that position during its orbit. Since the semimajor axis and eccentricity are approximately conserved during VRR, these annuli conserve their intrinsic properties (i.e., surface-density distribution, inner and outer radii), but their orientations can vary due to VRR. Since we have averaged over 2 phase-space coordinates and an additional 2 phase-space coordinates are frozen-in (conserved), the relaxation is restricted to a 2-dimensional space, which can be taken to be determined by the direction of the unit vector parallel to the angular momentum of the orbit (i.e., the normal to the annulus), which we denote by $\bm{n}$. Alternative but less convenient coordinates are the inclination and the longitude of the node of the orbit. In summary, the basic structures of VRR are concentric orbit annuli with distinct masses, inner, and outer radii, which behave as rigid bodies pinned to the dominant central point mass, and precess due to their mutual gravitational torques.

An important property of the annulus, which is conserved during VRR, is its scalar angular momentum
\begin{equation}\label{eq:l}
	l=m \sqrt{G M_\bullet a (1-e^2)}\,;
\end{equation}
here we have denoted the mass of each body by $m$, its semimajor axis by $a$, its eccentricity by $e$, and the mass of the central object (e.g., the SMBH in a NSC) by $M_\bullet$. 

We assume that the system is comprised of $K$ distinct groups of bodies. The $i^{\rm th}$ group contains $N_i$ bodies of identical mass $m_i$, semimajor axis $a_i$, and eccentricity $e_i$. We assume the ordering $a_1 \leq a_2 \leq \cdots \leq a_n$. This model can be thought of as representing an approximate coarse-grained distribution, assuming that the bodies are grouped into bins with similar mass, semimajor axis, and eccentricity. We shall refer to these groups as ``components''. Since every orbit of the same component has the same scalar angular momentum $l_i$, we can visualize our construction in angular-momentum space as follows: the tips of the angular-momentum vectors of different bodies in the same component lie on a thin spherical shell, and since the scalar angular momentum of each body is conserved during VRR, the angular momenta can only move along this spherical shell, so the bodies in a given shell can interact with bodies in other shells but can only relax within their own shell. 

We denote the distribution function or number per unit solid angle within shell $i$ by $f_i(\bm{n})$. We neglect two-body or higher-order correlation functions (the mean-field approximation). 

Let us summarize the simplifying assumptions used in this paper.

\begin{enumerate} 

 \item\label{i:assumption:precession} The long-term evolution is driven by the mutual gravitational torques between concentric, axially symmetric, time-independent structures, the time-averaged orbits, which change in orientation in response to these torques. This assumption is generally valid for masses which execute circular orbits in an arbitrary spherical potential or general orbits in a smooth spherical potential that is not exactly quadratic or Keplerian. 
 \item\label{i:assumption:quadrupole} The multipole expansion of the interaction potential between orbits is truncated at the quadrupole order. The quadrupole is generally the strongest multipole interaction for near-Keplerian orbits, whether radially overlapping or not, but the cumulative effects of higher order multipoles may be significant for radially overlapping orbits \citep{Kocsis_2015}. 
 \item\label{i:assumption:meanfield}  There are $K$ components, each comprised of $N_i\gg 1$ bodies with the same scalar angular momentum $l_i$, and the quadrupole coupling  ($J_{ij}$ in Eq.~\ref{eq:Econt}) between any body in component $i$ and any body in component $j$ is the same.
 \item\label{i:assumption:correlations} Multi-body correlations are negligible, so the evolution can be described by a mean-field model. This assumption is expected to be valid if $N_i$ is sufficiently large. 
 \item\label{i:assumption:Nconservation} The total number of bodies, the mass of each body, and their total vector angular momentum are conserved, except in Section \ref{sec:g_canonical}, where the bodies are allowed to exchange angular momentum with a surrounding reservoir.
 \item\label{i:assumption:conservation} Either the total energy of the system (i.e., the total gravitational potential energy arising from pairwise interactions of orbits) is conserved or the VRR temperature is fixed (see Section~\ref{s:E_In}). These assumptions correspond to the microcanonical and canonical ensembles, respectively. 
 
 \end{enumerate}
 
After presenting the framework for a general multi-component model in this section, we will focus most of the paper on the simple case of a one-component system\footnote{In Section \ref{sec:multi} we will discuss briefly how these results can be extended to multi-component systems in which the interactions between different components are much weaker than the interactions within a component.}, which exhibits a rich phase diagram and provides an analytically tractable starting point for comparisons with future numerical studies of multi-component systems.

\subsection{Basic definitions}

\label{sec:basic}

In the following, Greek indices label coordinates and Latin indices label stellar components. We use bold characters to denote 3-dimensional vectors, and normal (non-bold) symbols for their norm ($X=\Vert\bm{X}\Vert$). If we write out the Cartesian coordinates of some vector or tensor which has a stellar component index, we put the component index in parentheses in superscript to reduce clutter, e.g., $X^{(i)}_{\mu}$ denotes a Cartesian coordinate of the vector $\bm{X}_i$.

The system is characterized by the total number of bodies $N$, the total angular momentum $\bm{L}$, and the total orbit-averaged interaction energy $E$. In spherical coordinates, the number of bodies in the $i^{\rm th}$ component with $\bm{n}$ between $(\theta,\phi)$ and $(\theta+d\theta,\phi+d\phi)$ is $f_i(\bm{n}) d\Omega$, where $d\Omega=\sin\theta\, d\theta d\phi$ is the solid angle, and therefore the number of bodies of the $i$\textsuperscript{th} component is
\begin{equation}\label{eq:Ns}
	N_i = \int f_i(\bm{n}) d\Omega,
\end{equation}
where unless otherwise noted the integral is over the full unit $2$-sphere. The total number of bodies is 
\begin{equation}\label{eq:Ntot}
N = \sum_{i=1}^K N_i.
\end{equation}

The angular momentum of one orbit in the $i^{\rm th}$ component is 
\begin{equation}
	\bm{l} = l_i\bm{n}\quad\mbox{where}\quad l_i= m_i\sqrt{GM_\bullet a_i(1-e_i^2)}, 
\end{equation}
and the total angular-momentum vector of all bodies in the $i^{\rm th}$ component is
\begin{equation}\label{eq:li}
	\bm{L}_i = l_i \int f_i(\bm{n}) \bm{n}\, d\Omega 
\end{equation}
The total angular momentum is
\begin{equation}\label{eq:Ltot}
	\bm{L} = \sum_{i = 1}^K \bm{L}_i.
\end{equation}

The total interaction energy due to the VRR Hamiltonian up to the quadrupole order is given in Appendix \ref{s:app:RR}:
\begin{equation}\label{eq:Econt}
	E = - \ffrac{1}{2} \sum_{i,j=1}^K J_{ij}\iint f_i(\bm{n})f_j(\tilde{\bm{n}}) g(\bm{n},\tilde{\bm{n}})d\Omega d\tilde{\Omega}, 
\end{equation}
with $d\tilde{\Omega} = \sin\tilde{\theta}d\tilde{\theta}d\tilde{\phi}$ and  
\begin{equation}\label{eq:g}
	g(\bm{n},\tilde{\bm{n}}) \equiv \left(\bm{n} \cdot \tilde{\bm{n}} \right)^2 - \ffrac{1}{3}.
\end{equation}
The Hamiltonian (\ref{eq:Econt}) resembles that of the Maier-Saupe model for liquid crystals \citep{Maier+Saupe_1958,Plischke+Bergersen_2006}. We assume that $J_{ij}$ are known constant parameters of the model, which are determined by the quadrupole potential energies of interacting time-averaged orbits. For example, for circular orbits
\begin{equation}\label{eq:Jmean}
  J_{ij} = \frac{3G}{8}\frac{m_i m_j \min(a_i,a_j)^2}{\max(a_i,a_j)^3};
\end{equation}
see Appendix~\ref{s:app:RR} for the more general case of eccentric orbits in a near-Keplerian potential. The array $ J_{ij}$ is symmetric. 

Eq.~(\ref{eq:g}) may be rewritten as 
\begin{equation}
	g(\bm{n},\tilde{\bm{n}})
	= q_{\mu\nu}(\bm{n}){q}_{\mu\nu}(\tilde{\bm{n}})
\end{equation}
where here and in the following we suppress the summation over the Greek indices (Cartesian coordinates), and we have introduced the traceless quadrupole moment tensor of angular momenta
\begin{equation}\label{eq:Qcont0}
q_{\mu\nu}(\bm{n}) \equiv n_\mu n_\nu -\ffrac{1}{3}\delta_{\mu\nu}
\end{equation}
where $\delta_{\mu\nu}$ is the Kronecker-$\delta$. Let us introduce the mean-field sum
\begin{equation}\label{eq:Qcont}
Q_{\mu\nu}^{(i)} = \LA q_{\mu\nu} \RA^{(i)}\equiv \frac{1}{N_i}\int f_i(\bm{n}) q_{\mu\nu}(\bm{n})\, d\Omega
\end{equation}
where we define the ensemble average of $\bm{X}$ over component $i$ as
\begin{equation}\label{eq:average}
	\LA \bm{X}\RA^{(i)} = \frac{1}{N_i}\int f_i(\bm{n})\bm{X} \, d\Omega.
\end{equation}
The orbit-averaged energy of a single body in the $i^{\rm th}$ component with unit normal $\bm{n}$ is, from Eq.~(\ref{eq:Econt}),
\begin{equation}
	\label{eq:epsiloni_Q}
		\varepsilon_i(\bm{n}) = - q_{\mu\nu}(\bm{n}) \sum_{j = 1}^K  {J}_{ij}N_j Q_{\mu\nu}^{(j)} =  -{n}_{\mu} {V^{(i)}_{\mu\nu}} n_{\nu},
\end{equation}
where we defined
\begin{equation}
	\label{eq:V}
		V^{(i)}_{\mu\nu}= \sum_{j = 1}^K  J_{ij} N_jQ_{\mu\nu}^{(j)}
\end{equation}
which is traceless. The total energy is
\begin{equation}\label{eq:Etot}
	E = \ffrac{1}{2} \sum_{i=1}^K N_i \LA \varepsilon_i \RA = -\ffrac{1}{2}\sum_{i,j}^K J_{ij} N_iN_jQ_{\mu\nu}^{(i)}Q_{\mu\nu}^{(j)}\,.
\end{equation}

The one-body energy (\ref{eq:epsiloni_Q}) has the following properties.
\begin{enumerate}
 \item It has inversion symmetry, $\varepsilon_i(\bm{n}_i)=\varepsilon_i(-\bm{n}_i)$. 
 \item \label{i:eigenvalues}
 Since $n_{\mu}n_{\nu}$ has eigenvalues $\{0,0,1\}$, the average of such matrices has eigenvalues between 0 and 1. Thus $Q^{(j)}_{\mu\nu} = \LA n_{\mu}n_{\nu}\RA^{(j)} -\frac{1}{3}\delta_{\mu\nu}$ has eigenvalues between $\frac23$ and $-\frac13$; similarly, the superposition $-V^{(i)}_{\mu\nu}$ has eigenvalues between  $-\frac23 \sum_j J_{ij} N_j$ and $\frac13 \sum_j J_{ij} N_j$. The energy of each body is confined between these bounds with local minima (maxima) along the eigenvector of $-V^{(i)}_{\mu\nu}$ associated with its smallest (largest) eigenvalue and along the opposite directions, and saddle points along the eigenvector associated with the intermediate eigenvalue and the opposite direction.
 \item The equipotential curves with fixed $\varepsilon_i$ are generally ellipses on the unit sphere (i.e., the intersection of the unit sphere with an ellipsoid having the same center), which enclose the local minima and maxima.
\item The eigenvectors of $V_{\mu\nu}^{(i)}$ are fixed points of the time evolution of $\bm{n}$ for any body in component $i$, provided that $V_{\mu\nu}^{(i)}$ is constant. The eigenvectors corresponding to the smallest and largest eigenvalue are stable fixed points, and the intermediate eigenvector is unstable. 
\item The time evolution of $\bm{n}$ is integrable if $V_{\mu\nu}^{(i)}$ is constant in time.  
\end{enumerate}
However the precession of the angular momenta typically changes $Q_{\mu\nu}^{(j)}$ and $V_{\mu\nu}^{(i)}$ in time which usually leads to chaotic evolution. We expect, and shall assume, that this chaotic evolution causes the distribution function to relax toward a state of maximum entropy.

\subsection{Statistical equilibrium}\label{sec:StEq}

We calculate the equilibrium distributions $f_{\eq,i}$ that extremize the Boltzmann entropy  \citep{Jaynes_1965}
\begin{equation}\label{eq:S_boltz}
	S = -k\sum_{i=1}^K\int f_i(\bm{n}) \ln f_i(\bm{n}) \,d\Omega,
\end{equation}
 for fixed energy ${E}$, fixed total angular momentum $\bm{L}$, and fixed number of bodies in each component $N_i$. The use of this formula for the entropy in the mean-field approximation can be justified in the general case of the self-gravitating gas \citep{Padmanabhan_1990, Katz2003} (see also \citealt{Miller_1973,Sormani_Bertin_2013} for critical analysis), and also for VRR equilibria as we show in Section \ref{s:VRRequilibrium}. 
 
 We use the method of Lagrange multipliers. For perturbations $\delta f_i$ about the equilibrium distributions, the extremum must satisfy
\begin{equation}\label{eq:deltaS_L1}
	\delta S/k + \sum_{i=1}^K\alpha_i \delta N_i - \beta\delta {E}  + \gamma_\mu \delta {L}_\mu = 0,
\end{equation}
where $\alpha_i$, $\beta$, and $\gamma_\mu$ are Lagrange multipliers corresponding to the constraints and $\delta$ denotes the first-order variation with respect to $\{f_1(\bm{n}), f_2(\bm{n}), \dots, f_K(\bm{n})\}$. We may identify the Lagrange multiplier $\beta$ with the inverse temperature $\beta = 1/(kT)$, while $\bm{\gamma}$ is the thermodynamic variable conjugate to $\bm{L}$ with respect to entropy. 
We emphasize that in this paper ``temperature'' $T$ refers to the VRR temperature, which is the inverse Lagrange multiplier enforcing conservation of VRR energy in Eq.~(\ref{eq:deltaS_L1}) when maximizing the Boltzmann entropy; thus $\beta^{-1}=kT$. It is also the inverse derivative of the Boltzmann entropy with respect to VRR energy for a series of equilibria.

For the quantity enforcing the conservation of total angular momentum, it will later prove to be convenient to replace $\bm{\gamma}$ by $\bm{\omega}$, which we define as
\begin{equation}\label{eq:omega}
	\bm{\gamma} = \beta \bm{\omega}
\end{equation}
where $\bm{\omega}$ has units of angular velocity, and we refer to it as ``rotation''\footnote{For an ideal gas, $\bm{\omega}$ describes rotation in the sense that the mean velocity of the gas at any position $\bm{r}$ is $\bm{\omega}\times\bm{r}$. However, in VRR $\omega$ is not simply related to the mean angular velocity. } (see Section \ref{sec:g_canonical}, and \citealt{Votyakov_2002a,Votyakov_2002b,Martino_2003}). 
We will use this quantity to illustrate the analogy with paramagnetism and spin systems.

From the definition of $S$, $E$, $N_i$ and $\bm{L}$, we get to first order (for the second-order variation see Appendix \ref{s:app:S2})
\begin{align}	
\delta S/k &= -\sum_{i=1}^K\int (\delta f_i) ( 1 + \ln f_{{\rm eq},i}) \,d\Omega\\
\delta N_i &= \int (\delta f_i) \, d\Omega\,,\\
\label{eq:deltaE} \delta {E} &= \sum_{i=1}^K \int (\delta f_i) \,\varepsilon_i \, d\Omega\,, \\
\delta {\bm{L}} &= \sum_{i=1}^K l_i \int (\delta f_i) \bm{n} \, d\Omega\,,
\end{align}
where $f_{\eq}(\bm{n})$ is the equilibrium distribution function. Eq.~(\ref{eq:deltaS_L1}) becomes
\begin{equation}\label{eq:variation1}
	\sum_{i=1}^K \int \delta f_i ( 1+\ln f_{\eq,i} - \alpha_i +\beta \varepsilon_i - {l}_i\gamma_\mu n_\mu) d\Omega = 0
\end{equation}
Since the variations $\delta f_i$ are independent, the quantities in parentheses must vanish, which implies 
\begin{equation}
	f_{\eq,i} = e^{-1+\alpha_i-\beta \varepsilon_i + {l}_i\gamma_\mu n_\mu}.
\end{equation} 
Using the constraint on $N_i$, Eq.~(\ref{eq:Ns}), we may eliminate $\alpha_i$ to get finally
\begin{equation}\label{eq:fi}
	f_{\eq,i}(\bm{n}) = N_i\frac{e^{-\beta \varepsilon_i(\bm{n})+ {l}_i\gamma_\mu n_\mu}}{\phantom{\Big|}\int e^{-\beta \varepsilon_i(\bm{n}) + {l}_i\gamma_\mu n_\mu} d\Omega}.
\end{equation}
where $\varepsilon_i$ depends on the orbit normal $\bm{n}$ through the tensor $q_{\mu\nu}$ as given by Eqs.~(\ref{eq:Qcont0}), (\ref{eq:Qcont}), and (\ref{eq:epsiloni_Q}). In the following, we drop the subscript ``eq'' for brevity.

Eq.\ (\ref{eq:Qcont}) gives the self-consistency equations
\begin{equation}\label{eq:self_cons}
    Q_{\mu\nu}^{(i)}= \frac{\int (n_{\mu}n_{\nu} - \frac13 \delta_{\mu\nu}) e^{\beta 
    n_{\rho} V^{(i)}_{\rho\sigma} n_{\sigma} + {l}_i\gamma_\sigma n_\sigma} d\Omega}
	{\phantom{\Big|}\int e^{ \beta
    n_{\rho} V^{(i)}_{\rho\sigma} n_{\sigma} + {l}_i\gamma_\sigma n_\sigma}d\Omega} ,	
\end{equation}
where $V_{\mu\nu}^{(i)}$ is defined in Eq.~(\ref{eq:V}). These self-consistency equations are subject to the constraints (\ref{eq:Ltot}) and (\ref{eq:Etot}) for given $\bm{L}$ and $E$, which determine $\beta$ and $\bm{\gamma}$. These equations define the equilibria of the system and may be solved numerically, or in many cases analytically (see Appendix \ref{s:app:partition}).

\subsection{On inequivalence of statistical ensembles}\label{s:E_In}

\label{sec:inequiv}

In statistical physics the term \textit{canonical ensemble} is used to describe a system in equilibrium with a large heat reservoir having a fixed temperature. A \textit{microcanonical ensemble}, on the other hand, describes an isolated system with constant total energy. 
The two ensembles are 
equivalent in the thermodynamic limit $N\rightarrow \infty$ for systems that are governed by so-called ``short-range'' interactions \citep{Campa_2014}. 
The essential property of short-range systems is that they can be divided in an arbitrary manner into subsystems whose mutual interaction energy can be neglected with respect to the total energy of the system. Thus the subsystem energies are additive \citep{Padmanabhan_1990}.

For systems with long-range interactions, such as the self-gravitating gas, the two ensembles may be inequivalent. The inequivalence leads to different stability properties for the two ensembles (see also Appendix \ref{s:app:S2}). A second condition must also be met for inequivalence to appear\footnote{More specifically, inequivalence may appear in regions where the entropy is not a concave function of energy for a series of equilibria.}: 
the existence of a first-order phase transition in the canonical ensemble  \citep{Ellis_2000,Barre_2001,Ellis_2004,Bouchet_2005,Touchette_2005}. The phase transition in the canonical ensemble is replaced by a stable region with negative specific heat in the microcanonical ensemble \citep{Lynden-Bell_1968,Padmanabhan_1990,Campa_2014}.
In the canonical ensemble the phase transition takes place in an out-of-equilibrium process, where the system absorbs or emits the whole latent heat from the heat bath needed for the transition. 
The two phases cannot coexist in equilibrium because the mixed phase has a higher free energy than a single pure phase in the canonical ensemble  due to the interaction energy \citep{Chavanis_2006,Campa_2014}, and there cannot be a phase separation.
In other words, there are negative specific heat equilibria in the phase transition region which are therefore unstable in the canonical ensemble\footnote{\label{foot:canon}A system with negative specific heat cannot be in equilibrium with a heat bath. Any energy loss of the system to a colder heat bath will increase its temperature, making it lose even more energy, while any energy absorption of the system from a hotter heat bath will make the system colder causing a further energy absorption.}.
However, these equilibria are stable in the microcanonical ensemble, and the system can lie in any of those configurations between the two phases.

In the case of inequivalence, the equilibrium state depends on whether the system is under conditions of constant temperature (external environment has a strong influence which resembles a heat bath) in this state, in which case we refer to the canonical ensemble, or under conditions of constant energy (isolated system), in which case we refer to the microcanonical ensemble. 

The self-gravitating gas is generally non-additive and presents inequivalence of ensembles \citep{Padmanabhan_1990,Campa_2014}. On the other hand, in VRR, the spatial and velocity distributions of bodies are not determined by the relaxation process, apart from the orientation of the orbits, leaving open the possibility for specific distributions to be approximately additive with respect to VRR energy. 
However, in more typical cases, VRR in a multi-component system is non-additive. For the one-component systems that are the focus of this paper, VRR is not only non-additive, but also presents a first-order phase transition and therefore inequivalence of ensembles. For this reason the microcanonical and canonical ensembles will be studied separately.

\section{One-component systems}
\label{s:one_comp}

In the next three sections, we restrict ourselves to one-component systems; these are the simplest VRR systems yet they exhibit a remarkably rich phenomenology. 
The quadrupole mean-field VRR Hamiltonian of the one-component system is equivalent to the Maier-Saupe model of liquid crystals, which allows us to follow standard textbooks on the subject \citep{Plischke+Bergersen_2006}\footnote{The main difference is that we also require the conservation of total angular momentum.}. First, we write down the one-component version of the relevant equations of Section~\ref{s:mean_vrr}, in particular the mean-field quadrupole moment, the total energy, the total angular momentum, the distribution function, and the self-consistency equation, and evaluate the entropy and the free energy. We solve the resulting system of equations in the following section.

We adopt a coordinate basis in which the axes are aligned with the eigenvectors of
$Q_{\mu\nu}$. In the next subsection \ref{s:VRRequilibrium} and Appendix \ref{s:app:Qmunu} we show that at equilibrium $\bm{L}$ and $\bm{\gamma}$ are parallel to one of these eigenvectors. Therefore, without loss of generality, we may choose the third coordinate axis to lie along $\bm{L}$. 
In this case we may write 
\begin{align}\label{eq:Qmatrix}
Q_{\mu\nu} &= \LA n_{\mu}n_{\nu}\RA - \ffrac13 \delta_{\mu\nu}
\nonumber\\&
=\left(
\begin{array}{ccc}
	-\frac{1}{2}Q + \frac{1}{2}W & 0 & 0 \\
	0 &	-\frac{1}{2}Q-\frac{1}{2}W & 0 \\
	0 &	0 & Q
\end{array}
\right)
\end{align}
where we define
\begin{align}
\label{eq:Q_def}	Q &= \LA q \RA = \frac{1}{N} \int q f(\bm{n})  d\Omega\,,\\
\label{eq:W_def}	W &= \LA w \RA = \frac{1}{N} \int w f(\bm{n})  d\Omega\,, \\
\label{eq:q_def}	q &= \cos^2\theta -\ffrac{1}{3}\,,\\
\label{eq:w_def}	w &= \sin^2\theta \cos 2\phi\,.
\end{align}
Here $(\theta,\phi)$ are spherical coordinates in orthogonal axes aligned with the eigenvectors of $Q_{\mu\nu}$ and $\bm{n} = (\sin\theta \cos \phi,  \sin\theta \sin \phi, \cos \theta)$. We have introduced the angular-momentum vector direction coordinates $(q,w)$, which are proportional to the $\ell=2$ real spherical harmonics\footnote{Specifically, $q=(16\pi/45)^{1/2}Y_{2,0}(\bm{n})$ and $w=(16\pi/15)^{1/2}\times Y_{2,2}(\bm{n})$, where $Y_{\ell,m}(\bm{n})$ are real spherical harmonics, which form an orthonormal basis on the sphere, $\int Y_{\ell, m}(\bm{n})Y_{\ell',m'}(\bm{n})d\Omega = \delta_{\ell,\ell'}\delta_{m, m'}$.}. These coordinates are defined in the region 
\begin{equation}\label{eq:allowed}
-\ffrac13 \leq  q  \leq \ffrac23 \quad\mbox{and}\quad q  -\ffrac23   \leq \,w  \leq \ffrac23 -  q.
\end{equation}

From Eq.~(\ref{eq:epsiloni_Q}), the mean-field potential energy of a single body is simply 
\begin{equation}\label{eq:varepsilon2}
	\varepsilon(\bm{n}) = -JNQ_{\mu\nu} q_{\mu\nu}(\bm{n}) =  -\ffrac{3}{2}JN\left[ Q q(\bm{n})  + \ffrac{1}{3} W w(\bm{n}) \right]\,.
\end{equation}
All of our results are parameterized by $J$, so its value need not be specified for this analysis. For concreteness $J_{ij}$ is given for general near-Keplerian orbits in Appendix~\ref{s:app:RR}, and for circular orbits Eq.\ (\ref{eq:Jmean}) implies 
\begin{equation}
	J = \frac{3G m^2}{8a}\,.
\end{equation}

At equilibrium, the total energy from Eq.~(\ref{eq:Etot}) is 
\begin{equation}\label{eq:Eav}
	E = -\ffrac{3}{4}JN^2\left( Q^2 + \ffrac{1}{3} W^2 \right)\,. 
\end{equation}

In Appendix~\ref{s:app:Rot} we show that the extrema of the entropy have the property that the Lagrange multiplier $\bm{\gamma}$ and the total angular momentum $\bm{L}$ are parallel. Without loss of generality we may label the direction of $\bm{L}$ as the positive $z$-axis.
Letting
\begin{equation}
	s = \cos\theta\,,
\end{equation} 
we get $\bm{\gamma} \cdot \bm{n} = \gamma s$
and the total scalar angular momentum from Eq.~(\ref{eq:li}) is
\begin{equation}\label{eq:Lav}
	\frac{L}{N l} = \LA s \RA = \frac{1}{N} \int f \cos\theta\, d\Omega\,.
\end{equation} 
Thus,  $\LA s\RA$ is the dimensionless total angular momentum of the system normalized to the configuration in which all angular momenta are aligned. This satisfies $0 \leq L/(Nl) \leq 1$, where $L/(Nl) = 1$ represents a razor-thin disk in physical space with all bodies orbiting counterclockwise as seen from the positive $z$-axis, and $L = 0$ represents configurations with an equal net angular momentum in clockwise and counterclockwise orbits. 

The parameters $Q$ and $W$ are bounded by the inequalities\footnote{These follow from the bounds $0\leq \LA n_\mu^2\RA\leq 1$ (since $n_\mu$ are Cartesian coordinates of a unit vector) and $0\leq\LA \cos\theta\RA^2\leq \LA \cos^2\theta\RA\leq 1$ (due to the Cauchy--Schwarz inequality).} 
\begin{equation}\label{eq:QWbound}
-\frac13 + \left(\frac{L}{Nl}\right)^2 \leq Q \leq \frac23 \,, \quad \ Q -\frac23   \leq W \leq \frac23 - Q\,, 
\end{equation} 
and the total energy is bounded by the inequalities
\begin{equation}\label{eq:Ebound}
\begin{array}{ll}
\displaystyle -\frac{1}{3}  \leq  \frac{E}{JN^2} \leq 0 &\displaystyle\mathrm{~if~} \frac{L}{Nl} \leq \frac{1}{\sqrt{3}}\,,\\
\displaystyle -\frac{1}{3} \leq  \frac{E}{JN^2} \leq -\frac{3}{4} \left[\left(\frac{L}{Nl}\right)^2 - \frac{1}{3} \right]^2 &\displaystyle\mathrm{~if~} \frac{L}{Nl} \geq \frac{1}{\sqrt{3}}\,.
\end{array}
\end{equation}

The mean-field equilibrium distribution function (\ref{eq:fi}) is written as
\begin{align}
	f(\bm{n}) &= N \frac{ e^{ JN \beta Q_{\mu\nu}n_{\mu}n_{\nu} + l \gamma_{\mu} n_{\mu} }}{\phantom{\Big|}\int d\Omega\, e^{ J N\beta Q_{\mu\nu}n_{\mu}n_{\nu} + l \gamma_{\mu} n_{\mu}}} 
    \nonumber\\&=
    N \frac{ e^{\frac{3}{2} J N \beta \left( Q q + \frac{1}{3} W w \right) + l \gamma s }}{\phantom{\Big|}\int d\Omega\, e^{ \frac{3}{2} J N \beta \left( Q q  + \frac{1}{3} W w \right) + l\gamma s }}\,.
\label{eq:f_qw}
\end{align}
In these equations the angular dependence of $f$ is implicit in $q$, $w$, and $s$. We now evaluate $Q$ and $W$ using their definitions in Eqs.\ (\ref{eq:Q_def}) and (\ref{eq:W_def}). In so doing we use the identity $q=s^2-\frac13$ and simplify the integrals over the azimuth angle $\phi$ using modified Bessel functions
\begin{equation}\label{eq:Bessel}
	I_n(z) = \frac{1}{\pi}\int_0^\pi e^{z\cos\phi}\cos (n\phi)d\phi\,.
\end{equation}
The self-consistency equations (\ref{eq:Q_def})--(\ref{eq:W_def}) become 
\begin{align}
\label{eq:q_order}
	Q &= \frac{\int_{-1}^1 (s^2-\frac{1}{3}) I_0 e^{\frac{3}{2} J N \beta Q s^2 + l\gamma s}  ds}{\phantom{\Big|}\int_{-1}^1 I_0 e^{\frac{3}{2} J N \beta \LA q \RA s^2 + l\gamma s} ds } \\
\label{eq:w_order}
	W &= \frac{\int_{-1}^1 (1-s^2) I_1  e^{\frac{3}{2} J N \beta Q s^2 + l\gamma s} ds }{\phantom{\Big|}\int_{-1}^1  I_0 e^{\frac{3}{2} J N \beta Q s^2 + l\gamma s} ds}\,, 
\end{align}
where the arguments of the Bessel functions are $\frac{1}{2}JN \beta W(1-s^2)$.
The moments $Q$ and $W$ are the order parameters of the system. Configurations with $W = 0$ are 
axisymmetric around the $z$-axis while configurations with $W=\pm 3\,Q$ are axisymmetric around the $x$- and $y$-axes, respectively (recall that the positive $z$-axis points along the total angular-momentum vector). Therefore $W$ may be regarded as a measure of the deviation from axisymmetry around  the angular-momentum axis. Configurations with $W=Q= 0$ are isotropic. 

The entropy of equilibrium states is calculated from Eqs.~(\ref{eq:S_boltz}), (\ref{eq:Q_def}), (\ref{eq:W_def}), and (\ref{eq:f_qw}) to be
\begin{equation}\label{eq:S_one}
S_{\rm eq}(E,L)/k = 2\beta E - \gamma L + N \ln Z_0 - N\ln N.
\end{equation}
From now on we drop the constant $N\ln N$ term. In this equation 
\begin{equation}\label{eq:Z}
Z_0 = \int e^{-\beta \varepsilon +l \gamma  s} d\Omega =2\pi\int_{-1}^{1}I_0 e^{\frac{3}{2} J N \beta Q  (s^2-\frac{1}{3})+ l\gamma s} \,ds\,.
\end{equation}
The argument of the Bessel function $I_0$ is given below Eq.~(\ref{eq:w_order}). 
The total energy $E$ and angular momentum $\bm{L}=L\bm{e}_z$ in Eq.~(\ref{eq:S_one}) are given in Eqs.~(\ref{eq:Eav}) and (\ref{eq:Lav}), respectively. 

In what we call the canonical ensemble, the system can exchange energy but not angular momentum with a reservoir. The corresponding thermodynamic potential is the  Helmholtz free energy $F=E-TS$. Using Eq.~(\ref{eq:S_one}) we get
\begin{equation}\label{eq:F_vrr_can_1}
\beta F_{\rm eq}(T,L) = \beta E - S_{\rm eq}/k =  -\beta E + \gamma L - N \ln Z_0 \,.
\end{equation}
In this case, the parameter $\gamma = \gamma (T, L)$ appearing explicitly here and implicitly in $Z_0$ (see Eq.~\ref{eq:Z}) is to be determined by Eqs.~(\ref{eq:Lav}) and (\ref{eq:f_qw}). 
In what we call the $\omega TN$-ensemble (see Eq.~\ref{eq:omega} and Section \ref{sec:g_canonical}), the system can exchange both energy and angular momentum with a reservoir. Then the variables $\beta$ and $\gamma$ conjugate to $E$ and $L$ are conserved during the evolution of the system. The corresponding thermodynamic potential, analogous to the Gibbs free energy, is
\begin{equation}\label{eq:G_vrr_can_1}
G_\eq(T,\omega) = E - T S_{\rm eq} - \omega L = E - N kT \ln Z_0 
\end{equation}

Typically we would like to solve the system of equations~(\ref{eq:q_order})--(\ref{eq:w_order}) and (\ref{eq:Lav}) for $Q$, $W$, and $\gamma$, given values of the angular momentum $L$, and the energy $E$ (for the microcanonical ensemble) or temperature $T$ (for the canonical ensemble). It is helpful to observe that $Q$, $W$, and $\gamma$ only appear in the combinations (see Appendix \ref{s:app:partition}) 
\begin{equation}\label{eq:kappadef}
	\kappa_1 = \ffrac{3}{2}\beta JN Q \; ,\; 
    \kappa_2 = \ffrac{1}{2}\beta JN W \; ,\; c = l\gamma\,.
\end{equation}
Thus the ratio of Eqs.~(\ref{eq:q_order}) and (\ref{eq:w_order}) gives 
\begin{equation}\label{eq:kapparatio}
\frac{\kappa_1}{\kappa_2} =
\frac{\int_{-1}^1 (s^2-\frac{1}{3}) I_0 e^{\kappa_1 s^2 + c s}  ds}{\phantom{\Big|}\int_{-1}^1 (1-s^2) I_1  e^{\kappa_1 s^2 + c s} ds },
\end{equation}
where the argument of the Bessel functions is $\kappa_2(1-s^2)$. Moreover the angular momentum constraint (\ref{eq:Lav}) can be written as
\begin{equation}\label{eq:s_cons}
	\frac{L}{Nl} = 
    \frac{\int_{-1}^1 s I_0 e^{\kappa_1 s^2 + c s}  ds }{\phantom{\Big|}\int_{-1}^1 I_0 e^{\kappa_1 s^2 + c s}  ds}\,.
\end{equation}
This last equation is a monotonic and therefore invertible function of  $c\geq 0$ for given $\kappa_1$ and $\kappa_2$, mapping onto the interval $[0,1)$. Thus for any triple $(\kappa_1, \kappa_2, L)$, we can solve this numerically, and substitute the resulting value of $c(\kappa_1,\kappa_2,L)$ into Eq.~(\ref{eq:kapparatio}) to get a relation between $\kappa_1$ and $\kappa_2$, although this relation may have multiple solutions for $\kappa_2$ at a given value of $\kappa_1$. For each $(\kappa_1,\kappa_2)$ pair we can evaluate the temperature, energy or other quantities using Eqs.~(\ref{eq:q_order}), (\ref{eq:kappadef}), and (\ref{eq:Eav}). Once we have determined the value(s) of $(\kappa_1,\kappa_2)$ corresponding to the desired temperature, the order parameters $Q$ and $W$ 
the entropy, and the free energies 
follow from Eqs.~(\ref{eq:kappadef}), 
(\ref{eq:S_one}), (\ref{eq:F_vrr_can_1}) and (\ref{eq:G_vrr_can_1}.  
In Appendix~\ref{s:app:partition} we simplify the integrals analytically in Eqs.~(\ref{eq:q_order}), (\ref{eq:Z}), and (\ref{eq:s_cons}) for axisymmetric states (see Eqs.~\ref{eq:Zaxi}--\ref{eq:T''}) and derive the asymptotic behavior of macroscopic variables.

\subsection{Positive-temperature equilibria}\label{s:VRRequilibrium}

\begin{figure*}[tbp]
\begin{center}
		\includegraphics[scale = 0.45]{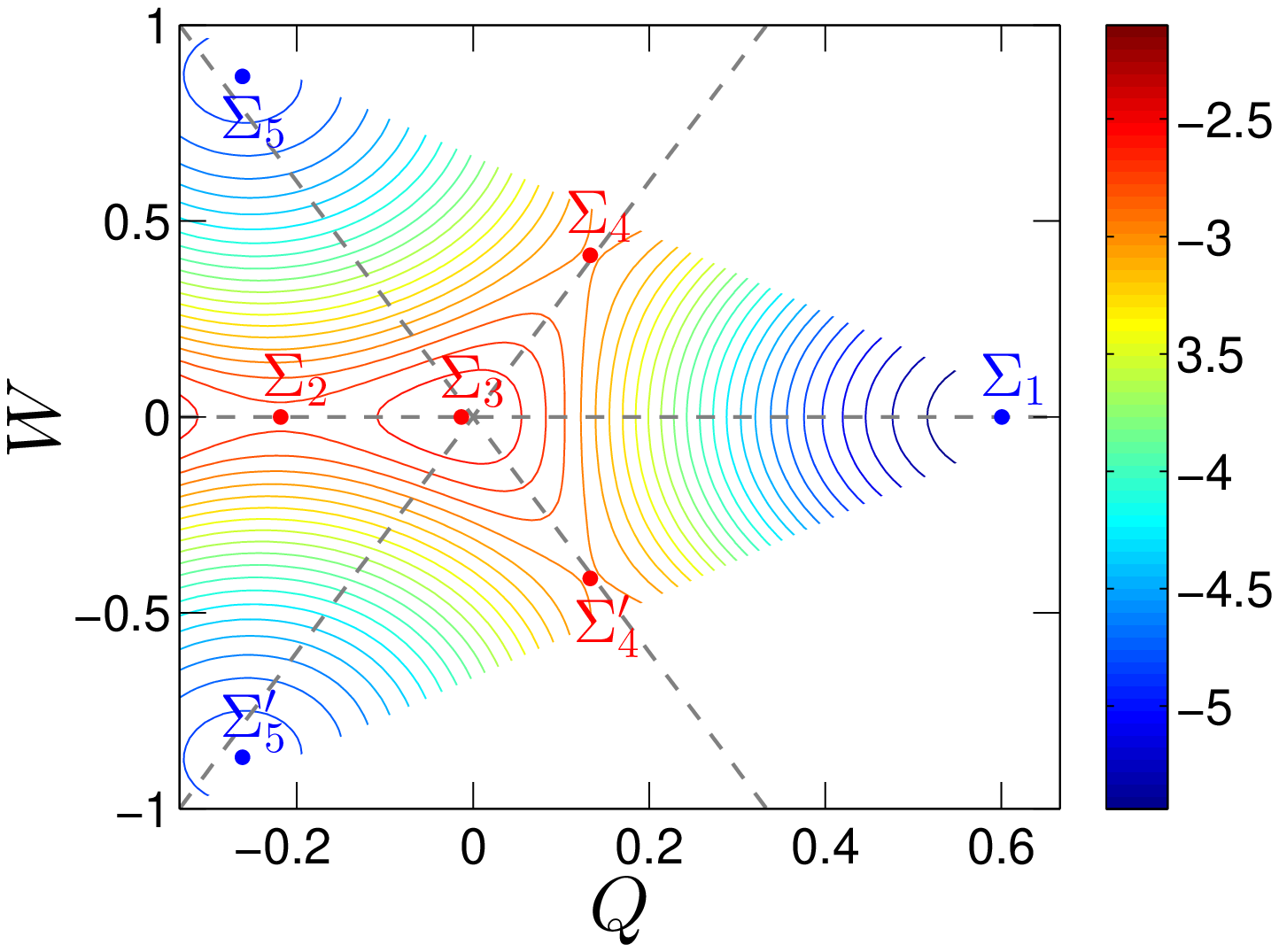} 
		\includegraphics[scale = 0.45]{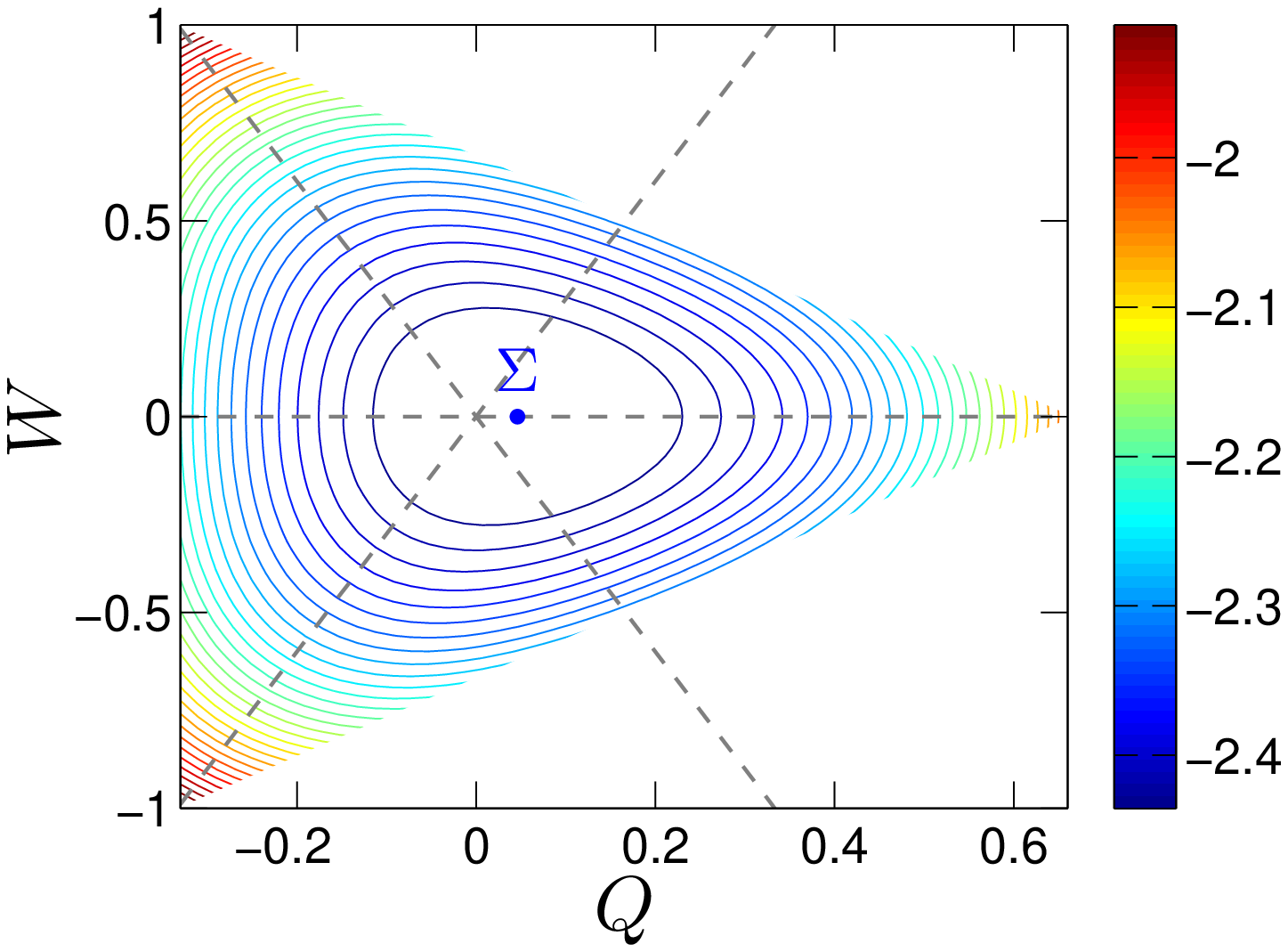} \\ 
        \hspace{-0.7cm}\shortstack{\includegraphics[scale = 0.25]{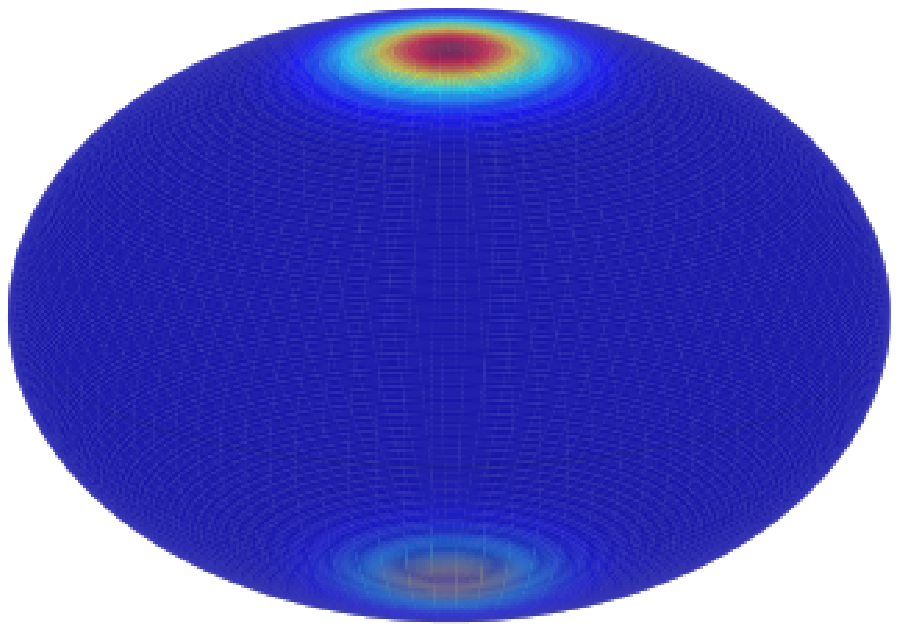}\\[-4ex]\hspace{0.7em}\normalsize{$\Sigma_1$}}
        \hspace{-0.7cm}\shortstack{\includegraphics[scale = 0.25]{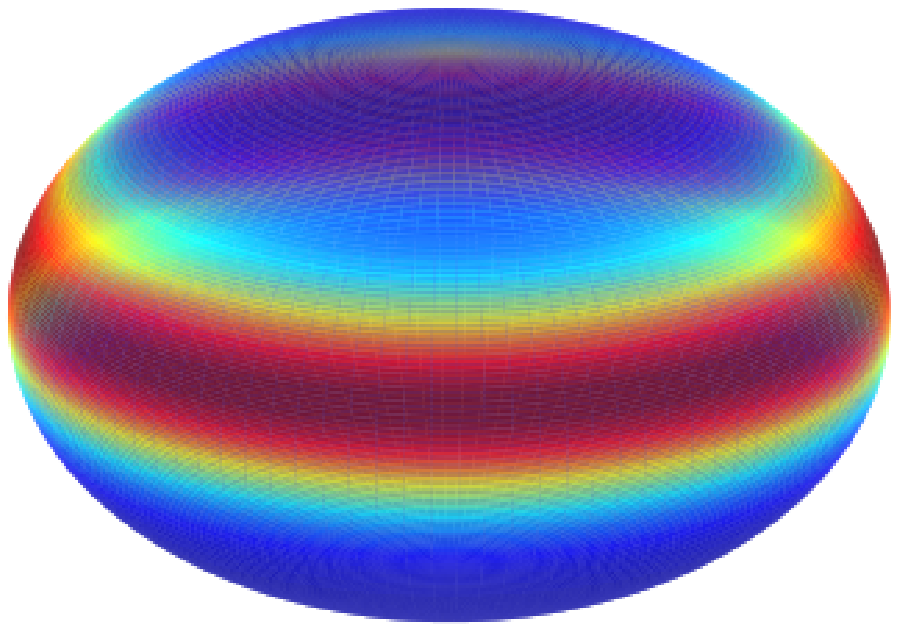}\\[-4ex]\hspace{0.7em}\normalsize{$\Sigma_2$}}        
        \hspace{-0.7cm}\shortstack{\includegraphics[scale = 0.25]{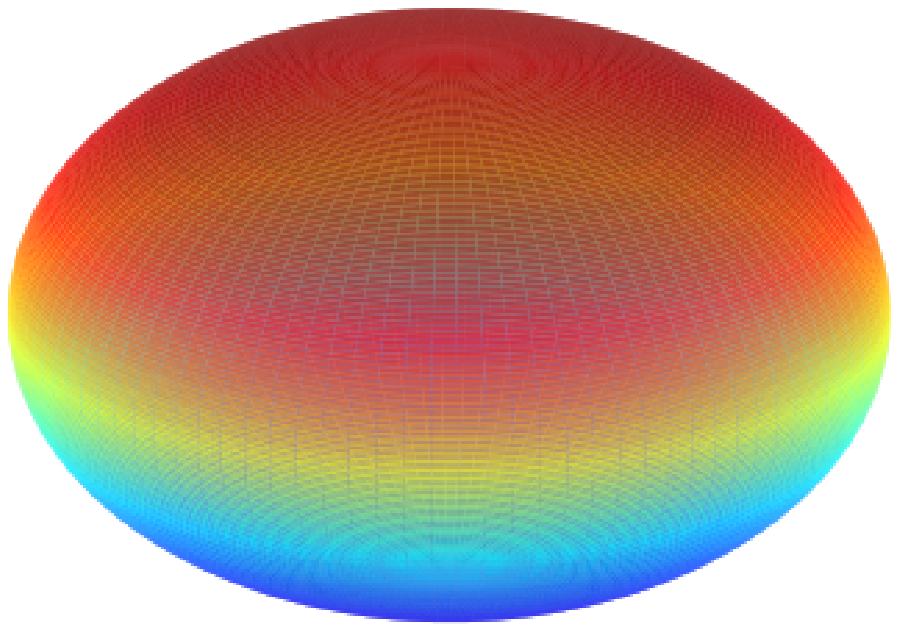}\\[-4ex]\hspace{0.7em}\normalsize{$\Sigma_3$}}
        \hspace{-0.7cm}\shortstack{\includegraphics[scale = 0.25]{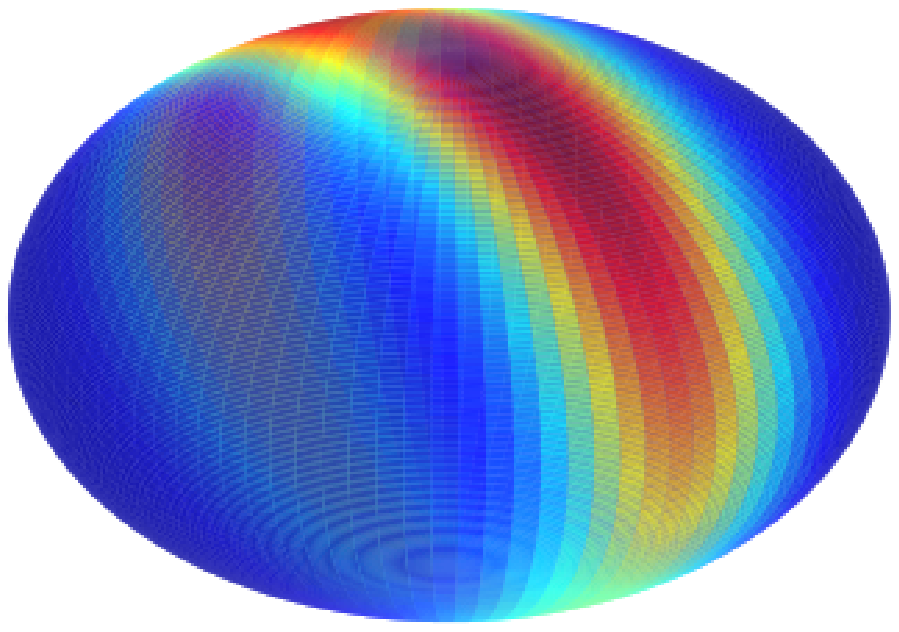}\\[-4ex]\hspace{0.7em}\normalsize{$\Sigma_4$}}
        \hspace{-0.7cm}\shortstack{\includegraphics[scale = 0.25]{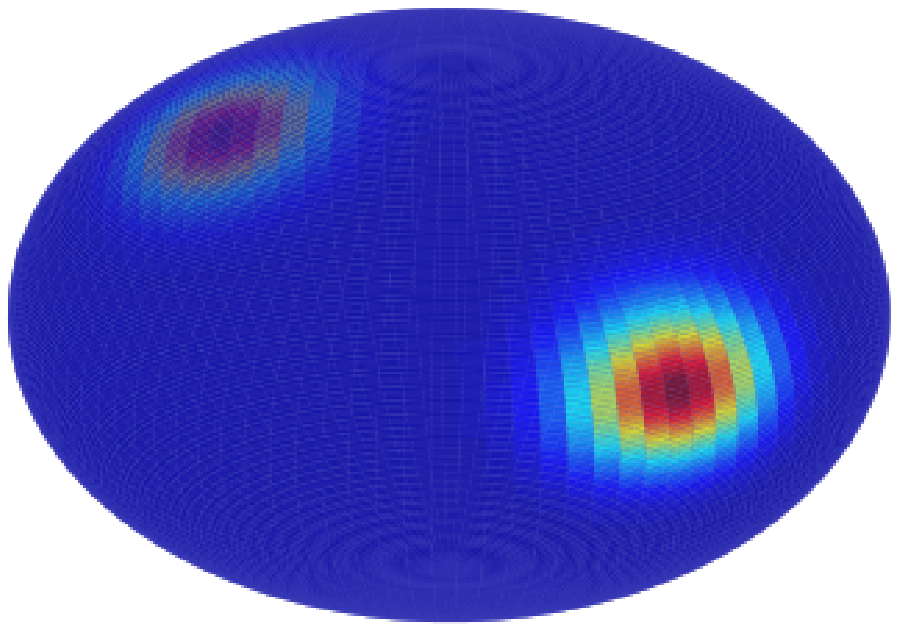}\\[-4ex]\hspace{0.7em}\normalsize{$\Sigma_5$}}       
\caption{ 
    \textit{Top:} 
   Contours of canonical free energy $F/(NkT)$ with respect to the order parameters $Q=\LA\cos^2\theta\RA-1/3$, $W=\LA\sin^2\theta\cos 2\phi\RA$ for total angular momentum $L/(Nl)=0.2$. The temperature is fixed in both panels; the left panel shows a low-temperature system ($kT=0.05JN$) and the right panel shows a high-temperature system ($kT=0.2JN$). The plot assumes a one-component system of bodies on orbits with the same fixed semimajor axis and eccentricity. The equilibria are the extrema of the free energy. The contour colors denote the value of $F/(NkT)$ as shown in the bars to the right of each panel. Black dashed lines denote configurations that are axisymmetric around the $x$, $y$, and $z$ axes. \textit{Top left}: At low temperature, five inequivalent equilibrium states may be identified, denoted by the points $\Sigma_1$, $\Sigma_2$, $\Sigma_3$, $\Sigma_4$ and $\Sigma_5$ (states labeled with primes may be obtained by transformations of the form $\phi\rightarrow\phi+\frac{1}{2}\pi$). In particular, $\Sigma_1$, $\Sigma_2$ and $\Sigma_3$ are axisymmetric around the $z$-axis (the direction of the total angular momentum) with $W = 0$, while $\Sigma_4$ and $\Sigma_5$ are non-axisymmetric states. The equilibrium with the lowest free energy is $\Sigma_1$, which corresponds to a thin disk in physical space having 40\% of its bodies on retrograde orbits. $\Sigma_3$ represents a nearly isotropic ``disordered'' equilibrium, which is unstable at this low temperature because it is a maximum of the free energy. The equilibria $\Sigma_2$  and $\Sigma_4$ are also unstable as they are saddle points. $\Sigma_5$ is a local minimum of the free energy which represents two concentric disks of equal mass with a mutual inclination of $157^{\circ}$ (Eq.~\ref{eq:cosi}). \textit{Top right}: At high temperature there is only one equilibrium, denoted by $\Sigma$, which represents a nearly isotropic, stable, disordered state. \textit{Bottom:} The distribution function of angular-momentum unit vectors $f(\bm{n})$ for the five thermodynamic equilibrium states marked in the top left panel. The density increases on a linear scale from blue to red in all five panels, with a different density range for different panels.
    }
	\label{fig:F_q_w}
\end{center} 
\end{figure*}

In this Section we calculate the positive-temperature VRR mean-field free energy (\ref{eq:F_vrr_can_1}) as the steepest-descent approximation to the partition function and we describe the equilibrium states and their stability properties. The negative-temperature equilibria are discussed in Section \ref{s:neg_T} and Appendix \ref{s:app:S2}.

We work in the $\omega$TN-ensemble, that is we assume that the system is embedded in a heat bath with constant temperature $T$ and rotation $\omega$, with which it can exchange VRR-energy and angular momentum but not bodies. Therefore, in this ensemble $T$, $N$ and $\omega$ are held fixed (see also Section \ref{sec:g_canonical}). The partition function is 
\begin{align}\label{eq:Xi_fg_g_can}
\Xi =& \int d\Omega_1\cdots d\Omega_N \nonumber \\
&\ \times \exp \bigg[\ffrac{1}{2}\beta J\sum_{i,j=1}^{N}\sum_{\mu,\nu=1}^3 q_{\mu\nu}^{(i)} q_{\mu\nu}^{(j)} +  l\sum_{i=1}^{N}\sum_{\mu=1}^3 \gamma_\mu n_\mu^{(i)}\bigg];
\end{align}
strictly the sum should not contain terms with $i=j$ but $\sum_{\mu\nu}q_{\mu\nu}^{(i)}q_{\mu\nu}^{(i)}=\frac{2}{3}$ for any $\Omega_i$ so these only contribute a constant factor to the partition function.
Note that (i) the Latin indices run through different bodies in the single-component case and not through different components as in Section \ref{s:mean_vrr}; (ii) for clarity, in this Section we explicitly write out sums over Cartesian coordinates indicated by Greek indices. 

We will use the Hubbard--Stratonovich method to calculate this integral. We may write the first sum in Eq.~(\ref{eq:Xi_fg_g_can}) as
\begin{equation}\label{eq:Ksum}
	 \sum_{i,j=1}^{N}\sum_{\mu,\nu=1}^3 q_{\mu\nu}^{(i)} q_{\mu\nu}^{(j)} = \sum_{\mu,\nu=1}^3\bigg(  \sum_{i=1}^{N} q_{\mu\nu}^{(i)} \bigg)^2 \equiv N^2 \sum_{\mu,\nu=1}^3K_{\mu\nu}^2.
\end{equation}
Now let $Q_{\mu\nu}$ be an arbitrary $3\times3$ matrix. For $T>0$ and $J>0$ we can write 
\begin{align}
	&\exp\bigg(\ffrac{1}{2}\beta JN^2\!\!\sum_{\mu,\nu=1}^3K_{\mu\nu}^2\bigg) = \left(\frac{\beta J N^2}{2\pi}\right)^\frac{9}{2}\!\!\int_{-\infty}^{+\infty} \!\prod_{\sigma,\lambda=1}^3 dQ_{\sigma\lambda} 
    \nonumber\\
    \label{eq:K_expon}
    &\times \exp\bigg(\!\!-\ffrac{1}{2}\beta J N^2 \sum_{\mu,\nu}Q_{\mu\nu}^2 + \beta JN^2\!\sum_{\mu,\nu}Q_{\mu\nu} K_{\mu\nu} \bigg)
\end{align}
which can be shown by completing the square of $Q_{\mu\nu}$ on the right-hand side. 
Then, Eqs.\ (\ref{eq:Xi_fg_g_can}) and (\ref{eq:K_expon}) become
\begin{align}
\Xi &= \left(\frac{\beta J N^2}{2\pi}\right)^\frac{9}{2} \!\!\int \!\!\prod_{\sigma,\lambda=1}^3 dQ_{\sigma\lambda} 
\exp\bigg(\!\!-\ffrac{1}{2}\beta J N^2 \!\sum_{\mu,\nu}Q_{\mu\nu}^2 \bigg)\nonumber \\
&\quad \times \bigg\{\int d\Omega\, \exp\Big[ \beta JN\!\sum_{\mu,\nu}Q_{\mu\nu}q_{\mu\nu} 
+l\sum_\mu \gamma_\mu n_\mu\Big]\bigg\}^N\nonumber \\
&= \left(\frac{\beta J N^2}{2\pi}\right)^\frac{9}{2} \!\!\int \!\!\prod_{\sigma,\lambda=1}^3 dQ_{\sigma\lambda} \exp\left[- \beta Ng(Q_{\mu\nu},\omega, T)\right]
\label{eq:Xi_fg_g_can_Q}.
\end{align}
Here we have used the fact that the integrals over $d\Omega_1\cdots d\Omega_N$ are separable and the integral is over the unit sphere, $n_{\mu}n_{\mu}=1$. We have also used the definition (\ref{eq:Qcont0}) and in the last line we introduced
\begin{align}
	\beta g =\, &-\ln \int d\Omega \nonumber \\ &\times 
\exp\Big[ \ffrac{1}{2}\beta JN\sum_{\mu\nu} Q_{\mu\nu}(2q_{\mu\nu}-Q_{\mu\nu}) + l\sum_\mu \gamma_\mu n_\mu\Big].
    \label{eq:G_vrr_can_2}
\end{align}
In this formula we identify $\{Q_{\mu\nu}\}$ as the order parameters. 

We now specialize to the thermodynamic limit $N\rightarrow \infty$. 
We note that the integral in Eq.~(\ref{eq:G_vrr_can_2}) is invariant as $N\to\infty$ for fixed $Q_{\mu\nu}$ as long as $\beta$ is rescaled as $\beta\propto N^{-1}$. With this rescaling $\beta g$ is independent of $N$. But since the integrand in (\ref{eq:Xi_fg_g_can_Q}) is $\exp( - N \beta g)$, this can be evaluated asymptotically for large $N$ by the method of steepest descent. We have
\begin{equation}
\ln\Xi=-\beta N \inf_{Q_{\mu\nu}} g(Q_{\mu\nu},\omega,T) + \mbox{O}(\ln N).
\end{equation}

The equilibrium free energy $G_\eq$ is given by $\ln\Xi=-\beta G_\eq$ so by dropping the fractional correction of order $(\ln N)/N$ we have 
\begin{equation}
G_\eq(T,\omega)=N\inf_{Q_{\mu\nu}} g(Q_{\mu\nu},\omega,T).
\end{equation}
Note that the extrema of $g(Q_{\mu\nu},\omega,T)$ occur when $\partial g/\partial Q_{\mu\nu} = 0$, and these conditions give exactly the self-consistency equations (\ref{eq:self_cons}), that is, the same distribution function as specified by use of the Boltzmann entropy (\ref{eq:S_boltz}).

In Appendix \ref{s:app:Qmunu}  we show that a coordinate system aligned with $\bm{\omega}$ can always be rotated in such a way that the
off-diagonal elements of $Q_{\mu\nu}$ are zero at equilibrium and the system is stable with respect to perturbations in the off-diagonal elements in this coordinate system.
Therefore we can finally write
\begin{equation}
	G_{\rm eq}(T,\omega) = N\inf_{Q, W} g(Q,W,\omega,T)\; ,
\end{equation}
where
\begin{align}\label{eq:G_vrr_can_22}
\beta g(Q,W,\omega,T) &=-\ln\int d\Omega\,\exp\big[\ffrac{3}{2}\beta JN(Qq+\ffrac{1}{3}Ww)\nonumber \\
&\quad -\ffrac{3}{4}\beta JN(Q^2+\ffrac{1}{3}W^2)+l\gamma s\big]\,.
\end{align}
which gives exactly Eq.\ (\ref{eq:G_vrr_can_1}).  

More generally, a global minimum of free energy with respect to the order parameters is a \textit{stable} equilibrium, while a local minimum is \textit{metastable}. Saddle points and local/global maxima are \textit{unstable} equilibria. A similar result holds for the free energy of the canonical ensemble $F$, given by
\begin{equation}
	F(Q,W,T,L) = G(Q,W,\omega,T) + \omega L
\end{equation}
where $G(Q,W,\omega,T)=N g(Q,W,\omega,T)$ and now $\omega$ is determined implicitly by the requirement that the total angular momentum is $L$.

The top left panel of Figure \ref{fig:F_q_w} shows the free energy in the canonical ensemble as a function of the order parameters $Q$ and $W$ for $L/(Nl) = 0.2$ at a low temperature, $kT/(JN)=0.05$. 
Points $\Sigma_1$--$\Sigma_5$ are the five distinct equilibrium configurations found at this temperature; the corresponding angular-momentum unit vector distribution functions $f(\bm{n})$ are shown below the panel. The dashed lines represent axisymmetric configurations. In particular, 
\begin{itemize}
\item $\Sigma_1$, $\Sigma_2$ and $\Sigma_3$ are axisymmetric equilibria. $\Sigma_4$ and $\Sigma_5$ are non-axisymmetric (even though they lie close to the dashed lines). 
 \item  $\Sigma_1$ is stable since it is the global minimum of the free energy. We call this the ``ordered phase'' of the system or ``uniaxial'', since it possesses one axis of symmetry (following the terminology of liquid crystals;  \citealt{Gramsbergen_1986,Luckhurst+Sluckin_2015}).
 \item $\Sigma_2$ is a saddle point, which is unstable with respect to perturbations that are not axisymmetric with respect to the $z$-axis.
  \item $\Sigma_3$ is a local maximum and hence an unstable equilibrium at this low temperature; however it is stable at higher temperature or at negative-temperature. It is nearly isotropic and we call it the ``disordered phase''. 
 \item  $\Sigma_4$ is a saddle point and therefore an unstable equilibrium.
 \item $\Sigma_5$ is a local minimum and hence a metastable equilibrium. We call this state ``biaxial'', since at low temperatures the bodies are organized into two distinct disks with different symmetry axes (Section~\ref{s:biax}). The state is long-lived for a significant range of temperatures and total angular momenta. 
\end{itemize}
The top right panel of Figure~\ref{fig:F_q_w} shows the canonical free energy at high temperature for the same mean angular momentum, $L/(Nl)=\LA s\RA = 0.2$. All equilibria merge at the point labeled $\Sigma$ which represents a stable disordered phase. 

The geometry of the equilibrium states may be understood from the bottom panels in Figure~\ref{fig:F_q_w}, or from the eigenvalues of $\LA n_\mu n_\nu\RA$ which from Eqs.~(\ref{eq:Qcont}) and (\ref{eq:Qmatrix})--(\ref{eq:w_def}) are $-\frac12 Q+\frac{1}{2}W+\frac13$, $-\frac12Q-\frac{1}{2}W+\frac13$, and $Q+\frac13$. 
For example, a razor-thin disk with all angular momenta pointing along the positive $z$-axis has distribution function $f(\bm{n})= N\delta( \bm{n} - \bm{e}_z)$, where $\bm{e}_z=(0,0,1)$ is the unit vector along the $z$-axis, and so $\LA n_\mu n_\nu\RA$ has eigenvalues $\{0,0,1\}$. In contrast, an isotropic distribution has $f(\bm{n}) = N/(4\pi)$ and eigenvalues $\{\frac{1}{3},\frac{1}{3},\frac{1}{3}\}$. For the stable equilibrium $\Sigma_1$ in Figure~\ref{fig:F_q_w} the eigenvalues are $\{0.03, 0.03, 0.93\}$. In physical space, this represents a relatively thin, axisymmetric disk containing both prograde and retrograde bodies, oriented perpendicular to the $z$-axis. The metastable equilibrium $\Sigma_5$ has eigenvalues $\{0.9, 0.03, 0.07\}$ which in physical space represents two disks with $157^{\circ}$ mutual inclination (see Eq.~\ref{eq:cosi} below). 

\subsection{Negative-temperature equilibria}\label{s:neg_T}

We remark that the internal energy given by Eq.~(\ref{eq:Eav}) is bounded from above (Eq.~\ref{eq:Ebound}).
It is known that in this case, negative-temperature equilibrium configurations may exist for isolated systems \citep{Ramsey_1956,Pathria+Beale_2012}. Negative absolute temperature states were first introduced by \cite{Onsager_1949} in a statistical approach to turbulent flow aiming to explain long-lasting large vortices. 
This phenomenon may also arise in quantum systems with an upper energy bound; 
such a bound may arise for some degrees of freedom as long as they evolve in isolation from the motional degrees of freedom, as in the case of nuclear spins. 
Indeed, negative-temperature configurations were first produced in the laboratory in nuclear spin systems \citep{Purcell_1951,Ramsey_1951,Ramsey_1956}. The spin-spin relaxation timescale is about six orders of magnitude smaller than that of the spin-lattice interactions in the system of \cite{Purcell_1951}. A negative spin-temperature state was achieved for the spinorial degrees of freedom by cooling the system to very low temperature in the presence of a magnetic field and then rapidly reversing its orientation. The system maintained negative spin temperature for several minutes until relaxation of the motional degrees of freedom restored positive temperature \citep{Ramsey_1956}. 

Recently, a negative-temperature state was achieved in the laboratory for the motional degrees of freedom of ultra-cold quantum boson gases \citep{Braun52}. 
This achievement triggered a  debate \citep{Schneider_2014,Hanggi_2015,Frenkel_2015,Campisi_2015,Cerino_2015,Poulter_2016}, initiated by \cite{Dunkel+Hilbert_2014}, on whether negative absolute thermodynamic temperatures are observable after all or maybe the Boltzmann's entropy formula should be abandoned in favor of Gibbs' entropy, which allows only positive temperatures. However, the arguments in favor of the validity of Boltzmann's formula are more convincing at least for the case of ensemble-equivalence where it was proven recently that Boltzmann's formula is appropriate and negative temperatures do occur \citep{Buonsante_2016}.

Negative-temperature states in nuclear spin systems present a close analogy with the quadrupole VRR system discussed here. The VRR degrees of freedom, the unit normals $\bm{n}$, are similar to spinorial degrees of freedom. 
Orbits averaged over a precession period represent fixed density annuli which are characterized by their normals and relax similarly to spin vectors in quantum mechanics; the difference is that the interaction is a sum of terms proportional to $(\bm{n}_i\cdot \bm{n}_j)^2$ for VRR and $\bm{s}_i\cdot \bm{s}_j$ for spin systems. 
The analogy goes even further since $\bm{\omega}$, defined in Eq.~(\ref{eq:omega}), enters the distribution function through $\bm{\omega}\cdot \langle\bm{n}\rangle$ similar to a paramagnetic term, $\bm{B}\cdot\LA \bm{s} \RA$.

Negative-temperature equilibria also arise for an isolated thin circular disk of bodies undergoing VRR \citep{Kocsis+Tremaine_2011}. That model is recovered in the limit $L/(Nl) \approx 1$ of the model presented here. 
With this perspective, negative-temperature equilibria arise naturally for VRR. 

At negative temperature, we find a single solution to the self-consistency equation (\ref{eq:q_order})--(\ref{eq:w_order}) at fixed angular momentum $\bm{L}$. In Appendix \ref{s:app:S2}, we show that these negative-temperature equilibria are stable in both the canonical and microcanonical ensembles. These states are always axisymmetric (Appendix \ref{s:app:partition:rotating}), and they are similar to the large positive-temperature disordered state shown in the top right panel of Figure~\ref{fig:F_q_w} although negative-temperature states always exhibit a population inversion: the occupation number is an increasing function of one-body VRR-energy (Eq.~\ref{eq:varepsilon2}).

\bigskip
Before closing this section, let us introduce some quantities for further use. We denote by $N_{+}$ the number of orbits with angular-momentum vectors on the $s\geq 0$ hemisphere (i.e. prograde orbits with respect to $\bm{L}$) and by $N_{-}$ the number of orbits with $s \leq 0$ (retrograde orbits with respect to $\bm{L}$)
\begin{align}
	\label{eq:prob_pm}
    N_+ = N  \frac{\int_{0}^1  I_0 e^{\kappa_1 s^2 + c s} ds}{\int_{-1}^1 I_0 e^{\kappa_1 s^2 + c s} ds}\,,\quad 
    N_- = N \frac{\int_{-1}^0  I_0 e^{\kappa_1 s^2 + c s} ds}{\int_{-1}^1 I_0 e^{\kappa_1 s^2 + c s} ds}\,.
\end{align}
We also introduce the expectation value of a quantity $X$ over bodies on the $s\geq 0$ hemisphere as
\begin{equation}\label{eq:X_pm}
\LA X \RA_+ = \frac{1}{N_+} \int_{s\geq 0}  X f\,d\Omega
\end{equation}
and similarly $\LA X \RA_-$ for the $s<0$ hemisphere. In particular, we shall use $L_{\pm}$ to denote the mean angular momentum of the prograde and retrograde orbits with respect to the $z$-axis.

\section{Zero angular momentum}\label{s:s0}

It is instructive to start with the configurations having zero total angular momentum, $L = \gamma = 0$ in Eqs.~(\ref{eq:Lav})--(\ref{eq:f_qw}). In the one-component case, this is completely equivalent to the Maier-Saupe model of liquid crystals. 

\begin{figure}[tbp]
\begin{center}
\includegraphics[scale = 0.45]{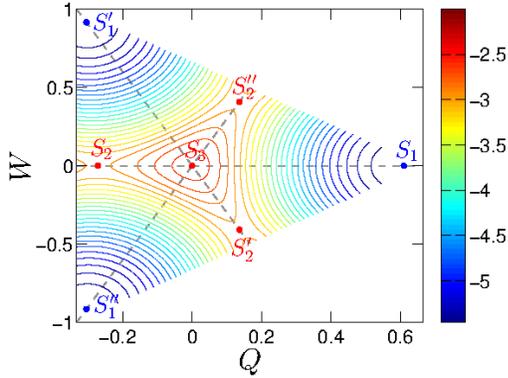} 
	\caption{
    Same as Figure~\ref{fig:F_q_w} with $kT/(JN)=0.05$ but for zero total angular momentum $L = 0$. The three dashed lines denote configurations that are axisymmetric around the $x$, $y$, and $z$ axes. Only three inequivalent equilibria exist, marked by $S_1$, $S_2$, $S_3$. Other equilibria denoted by primes may be obtained by swapping the $x$, $y$, $z$ axes. $S_1$ is a stable equilibrium (up to a possible global rotation of the distribution function) which corresponds to the ordered phase. $S_2$ is a saddle point and hence unstable for perturbations not axisymmetric around the $z$-axis, which drive the system towards $S_1'$ or $S_1''$. $S_3$ is a local maximum of the free energy and hence an unstable equilibrium which corresponds to the disordered isotropic state. See left
 panel of Figure~\ref{fig:f_s_L0} for the distribution functions corresponding to $S_1$ and $S_2$ at a somewhat higher temperature, $kT/(JN)=0.1$.}
\label{fig:F_q_w_L0_T5e-2}
\end{center} 
\end{figure}

Figure \ref{fig:F_q_w_L0_T5e-2} shows the free energy $F/(NkT)$ as a function of the order parameters $Q$ and $W$ at the same temperature $kT/(JN)=0.05$ as in the left panel of Figure~\ref{fig:F_q_w}. We may observe a similar structure as in that Figure with correspondence $\Sigma_1\rightarrow S_1$, $\Sigma_2\rightarrow S_2$, $\Sigma_3\rightarrow S_3$, $\Sigma_4\rightarrow S''_2$, $\Sigma_5\rightarrow S'_1$. The main difference is that all equilibria are exactly axisymmetric ($W = 0$ or $Q = \pm \frac{1}{3} \,W$, see Eq.~\ref{eq:Qmatrix}).  The primed and double primed states may be obtained by a $90^{\circ}$ rotation of the unprimed distribution. 

\begin{figure*}[tbp]
\begin{center}
 		\includegraphics[scale = 0.5]{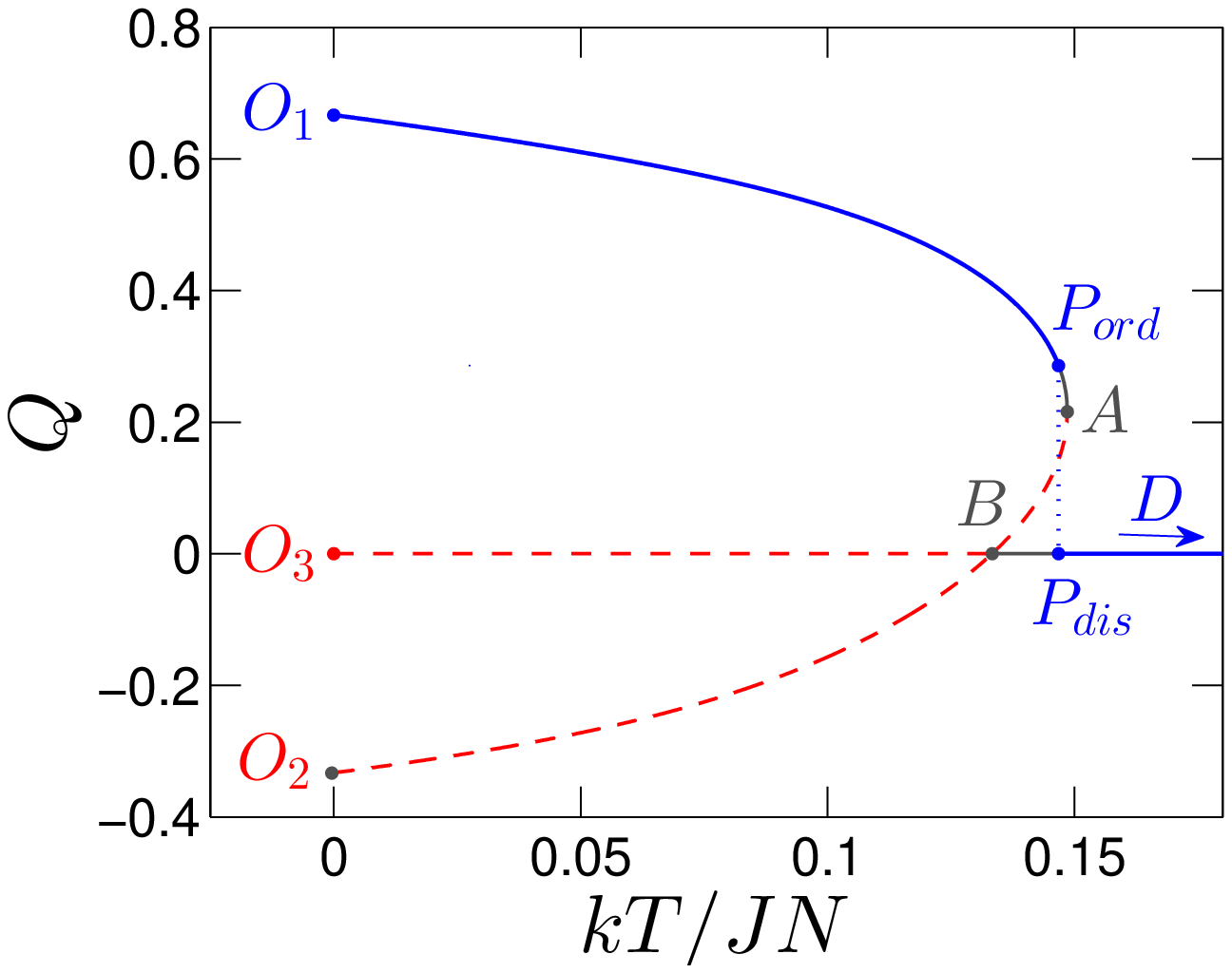} 
 		\includegraphics[scale = 0.5]{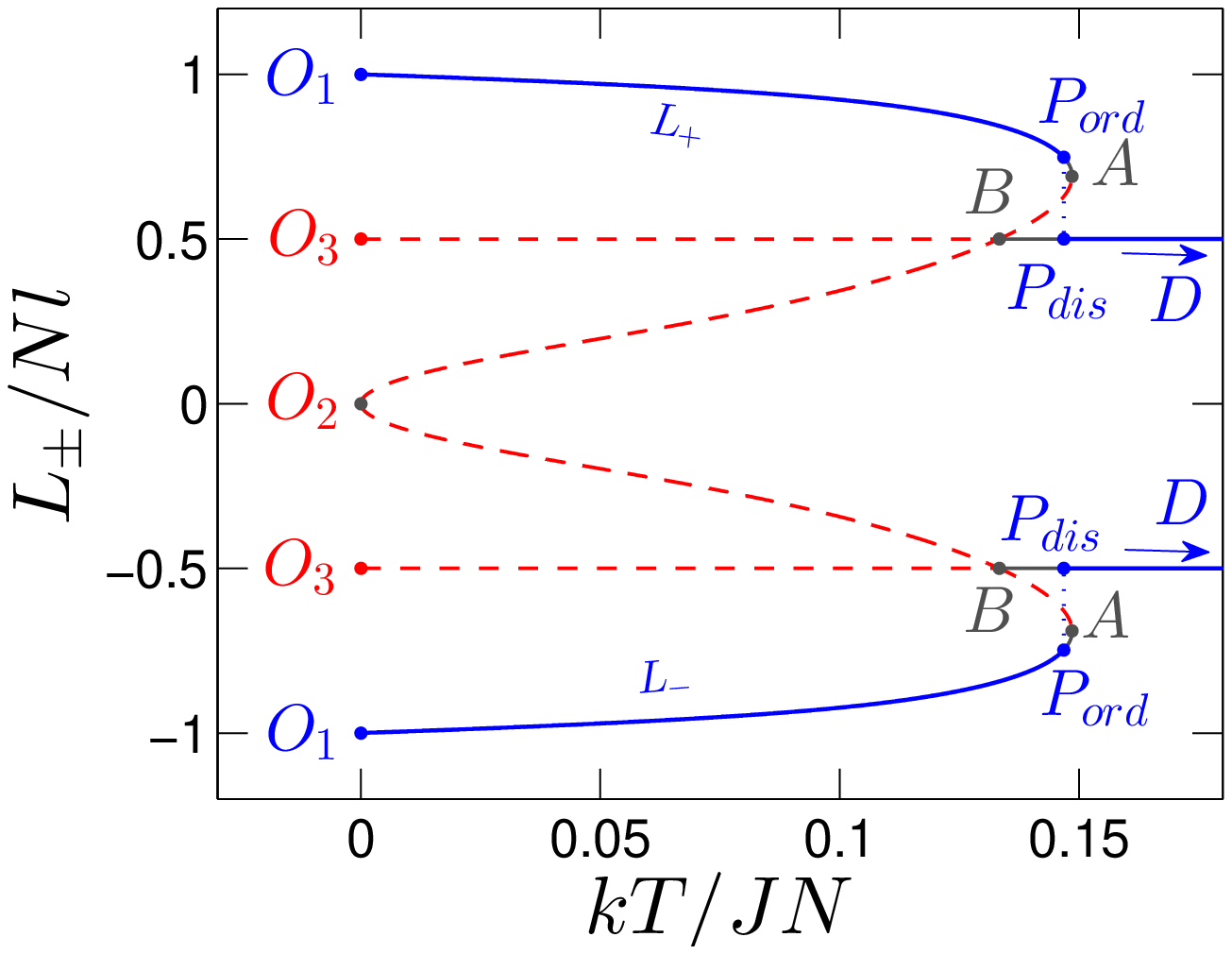} 
	\caption{ Axisymmetric equilibria for zero total angular momentum, $L = 0$, and positive temperature $T$. Blue solid (red dashed) lines denote stable (unstable)  equilibria. \textit{Left: }The order parameter $Q=\LA\cos^2\theta\RA-1/3$ as a function of temperature.
    \textit{Right: }The angular momenta $L_{\pm}/(Nl)=\LA s\RA_\pm$ of clockwise ($-$) and counterclockwise ($+$) orbits with respect to the $z$-axis, as a function of temperature. 
    There are three axisymmetric zero-temperature states, $O_{1,2,3}$. At arbitrarily small positive temperature, the branch starting at $O_1$ is stable, while those near $O_2$ and $O_3$ are unstable.
    The $O_1$ state corresponds to a razor-thin disk comprised of equal numbers of bodies on clockwise and counterclockwise orbits ($Q = 2/3$, $L_{\pm}/(Nl) = \pm 1$). The $O_2$ state has angular-momentum vectors distributed uniformly in the $x$--$y$ or equatorial plane ($Q = -1/3$, $L_{\pm} = 0$); thus the number density in physical space varies with polar angle as $1/\sin\theta$.  The distribution of orbit normals in the $O_3$ state is isotropic, and the number density in physical space is spherically symmetric ($Q = 0$, $L_{\pm}/NL = \pm 1/2$). As the temperature is increased from zero to $kT/(JN)=0.05$, the equilibria $O_1$, $O_2$ evolve along the blue and red lines to the equilibria $S_1$, $S_2$, and $S_3$ shown in Figure~\ref{fig:F_q_w_L0_T5e-2}. In the canonical ensemble, a first-order phase transition occurs at $kT_P/(JN) = 0.146796$ between the ordered or disk+halo phase corresponding to branch $O_1P_{\rm ord}$, and the disordered or isotropic phase corresponding to branch $P_{\rm dis} D$ (where the point $D$ is supposed to lie at infinite temperature). 
    The gray branches $B P_{\rm dis}$ and $P_{\rm ord}A$ are metastable as shown by Figure~\ref{fig:F_T_ABmeta_L0}. In the canonical ensemble the one-component system is either in $P_{\rm ord}$ or $P_{\rm dis}$ and the dotted curve between $P_{\rm ord}$ and $P_{\rm dis}$ does not represent an equilibrium sequence. In the microcanonical ensemble the branches $P_{\rm ord} AB P_{\rm dis}$ are stable and the system passes from one phase to the other at point $B$ with a continuous second-order phase transition (see Figure~\ref{fig:Fbeta_L0} below). The dotted curve represents an equilibrium sequence for a separable multi-component system (see Section~\ref{sec:multi}). 
Here $kT_{B}/(JN)=2/15=0.133333$, $kT_{P}/(JN)=0.146796$, $kT_{A}/(JN)=0.148556$. 
}
    \label{fig:qsz_T_L0}
\end{center} 
\end{figure*}

\begin{figure*}[tbp]
\begin{center}
        	\includegraphics[scale = 0.45]{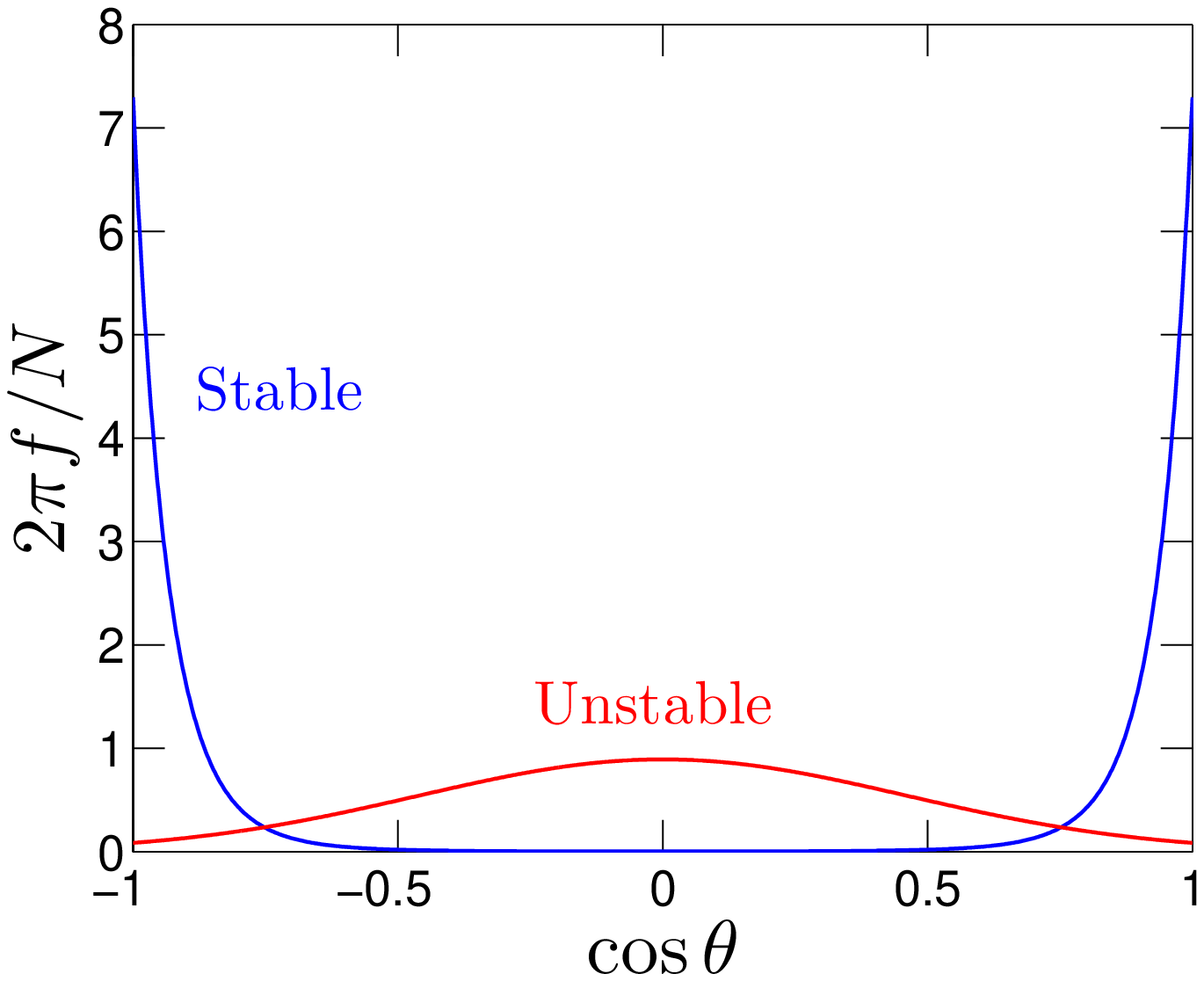} 
        	\includegraphics[scale = 0.45]{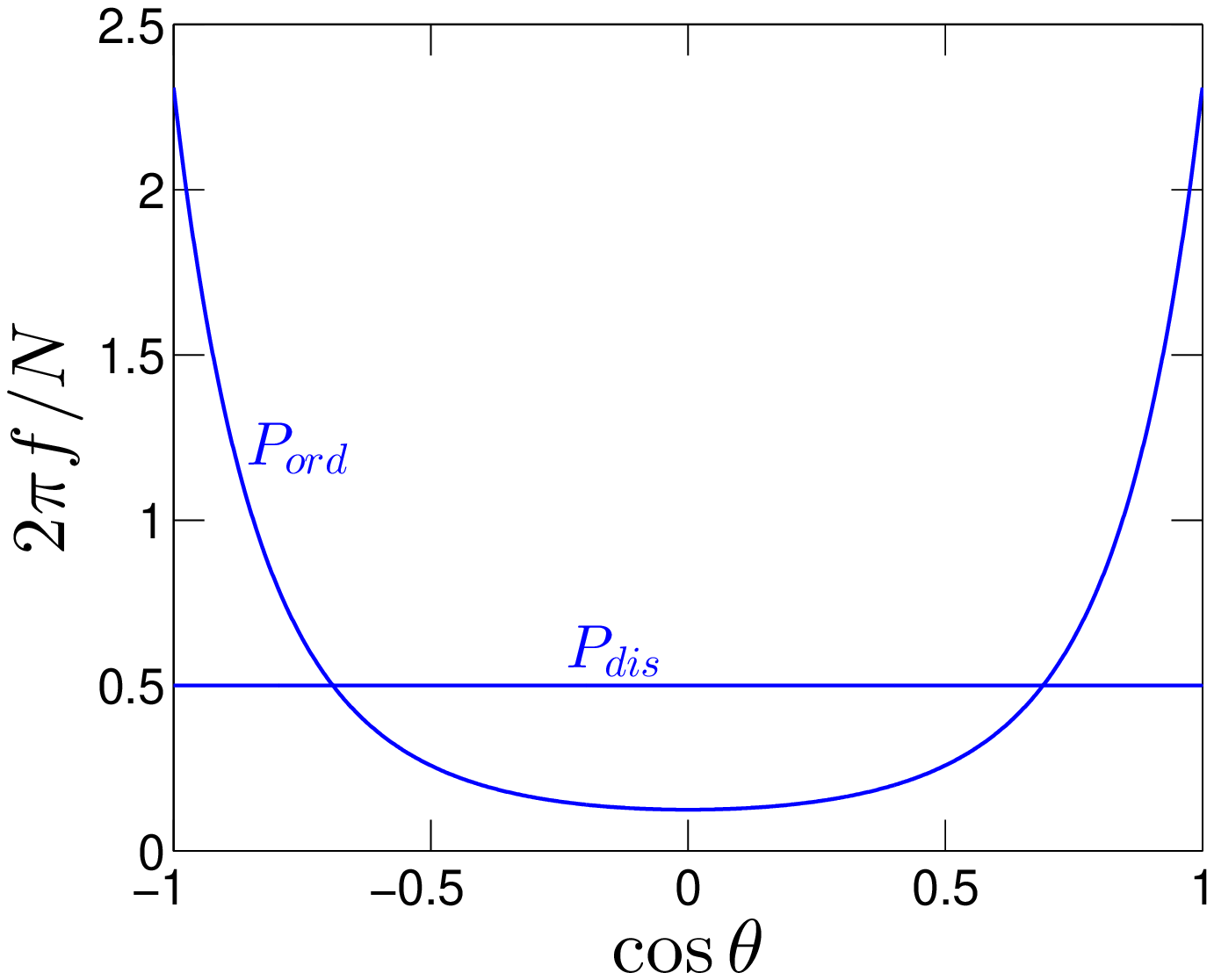} 
	\caption{The probability density of the $z$-coordinate of the normalized angular-momentum vector ($s=\cos\theta$) for equilibrium systems with zero total angular momentum at two different temperatures. \textit{Left: } The stable ordered phase (branch $O_1P_{\rm ord}$ of Figure \ref{fig:qsz_T_L0}) and an unstable phase (branch $O_2B$) at temperature $kT/(JN) = 0.1$ (the states $S_1$ and $S_2$ in Figure~\ref{fig:F_q_w_L0_T5e-2} lie on these branches, but at a lower temperature $kT/(JN)=0.05$). \textit{Right: }The two phases at the transition temperature $kT_P/(JN) = 0.146796$ (see Figure~\ref{fig:F_T_ABmeta_L0}). Phase $P_{\rm ord}$ is the ordered phase, which presents a disk+halo structure with an equal number of prograde and retrograde orbits. Phase $P_{\rm dis}$ is the isotropic disordered phase, which is independent of temperature.}
\label{fig:f_s_L0}
\end{center} 
\end{figure*}

The left panel of Figure \ref{fig:qsz_T_L0} shows the order parameter $Q$ for a series of axisymmetric equilibria as a function of temperature (only positive temperatures are shown). There are three zero-temperature states $O_1$, $O_2$ and $O_3$ with $Q = \frac{2}{3}$, $-\frac{1}{3}$, and 0. The state $O_3$ corresponds to the isotropic distribution function
\begin{equation}\label{eq:iso_df}
f(\bm{n}) = \frac{N}{4\pi}
\end{equation}
which satisfies the self-consistency conditions (\ref{eq:Lav}), (\ref{eq:w_order}), and (\ref{eq:q_order}) when $Q=W=\gamma=0$ for any inverse temperature $\beta$; thus $O_3$ is identical to $S_3$. The state $O_1$ has $Q=\frac{2}{3}$, and since this is the maximum allowed value of $q$ by Eq.\ (\ref{eq:allowed}) all bodies in the system must have $q=\frac{2}{3}$. In other words all of the orbits have unit normals $\bm{n}=(0,0,\pm1)$ so they form a razor-thin disk in the $x$-$y$ plane with equal numbers of prograde and retrograde orbits and energy $E=-\frac{1}{3}JN^2$. 
This is the most extreme configuration of what we call the \textit{ordered phase}. The state $O_2$ has $Q=-\frac{1}{3}$ and since this is the minimum allowed value of $q$ all bodies must have $q=-\frac{1}{3}$, that is, their orbit normals lie in the equatorial plane $\theta=\frac{1}{2}\pi$. The azimuths of the orbit normals are uniformly distributed. These three states are respectively analogous to $S_1$, $S_2$, and $S_3$ in Figure \ref{fig:F_q_w_L0_T5e-2} for zero temperature, which indicates that $O_1$ is stable, while $O_2$ and $O_3$ are unstable. A mathematical description of these states is given in Appendix \ref{s:app:partition}.
The right panel of Figure \ref{fig:qsz_T_L0} shows the mean angular momenta of the clockwise and counterclockwise orbits respectively, $L_\pm=Nl\LA s\RA_{\pm}$ as defined in Eq.~(\ref{eq:X_pm}). As we have already shown, the stable zero-temperature state $O_1$ corresponds to a razor-thin disk with equal numbers of bodies orbiting clockwise ($s=-1$) and counterclockwise ($s=1$). 
At higher temperatures most of the bodies remain in a disk, which thickens as the temperature increases. In addition a few bodies are found at higher inclinations forming a ``halo'' as shown in Figure \ref{fig:f_s_L0}. 
As Eq.\ (\ref{eq:iso_df}) implies, the disordered phase corresponding to $O_3$ is isotropic at all temperatures for zero total angular momentum ($Q = W = \LA s\RA = 0$). 

Figure \ref{fig:qsz_T_L0} marks the states $P_{\rm ord}$, $A$, $B$, and $P_{\rm dis}$, where stability properties change in the canonical ensemble, as shown in Figures~\ref{fig:F_T_ABmeta_L0} and \ref{fig:Fbeta_L0} and in Appendix \ref{s:app:partition:L0}\footnote{For the case of zero total angular momentum discussed here, $B$, $P_{\rm dis}$, $D$, and $O_3$ are identical: they all represent the isotropic configuration, $f(\bm n)=N/(4\pi)$. However, these states are distinct for nonzero angular momentum $L$.
 The isotropic disordered state is unstable at low temperature, but in the canonical ensemble it becomes metastable for temperatures in the range $T_B<T<T_P$ where $kT_B/(JN)=\frac{2}{15}=0.133333$ and $kT_P/(JN) = 0.146796$. These states become stable for $T>T_P$. The ordered phase (branch $O_1P_{\rm ord}$) becomes metastable in the canonical ensemble for temperatures in the range $T_{P}<T<T_{A}$, where $kT_A/(JN)=0.148556$. At $T_P$, the system suffers a first-order phase transition from the ordered phase $P_{\rm ord}$ to the disordered phase $P_{\rm dis}$. At the transition point $P_{\rm ord}$,  $Q=0.286014$, $L_{\pm}/(Nl)=\pm 0.747913$ and $E/(JN^2)=0.0613553$, while at $P_{\rm dis}$, $Q = 0$, $L_{\pm}/(Nl) = \pm \frac12$, and $E/(JN^2)=-\frac{1}{3}$. The distribution functions at $P_{\rm ord}$ and $P_{\rm dis}$ are shown in Figure \ref{fig:f_s_L0}. }

\begin{figure*}[tbp]
\begin{center}
		\includegraphics[scale = 0.45]{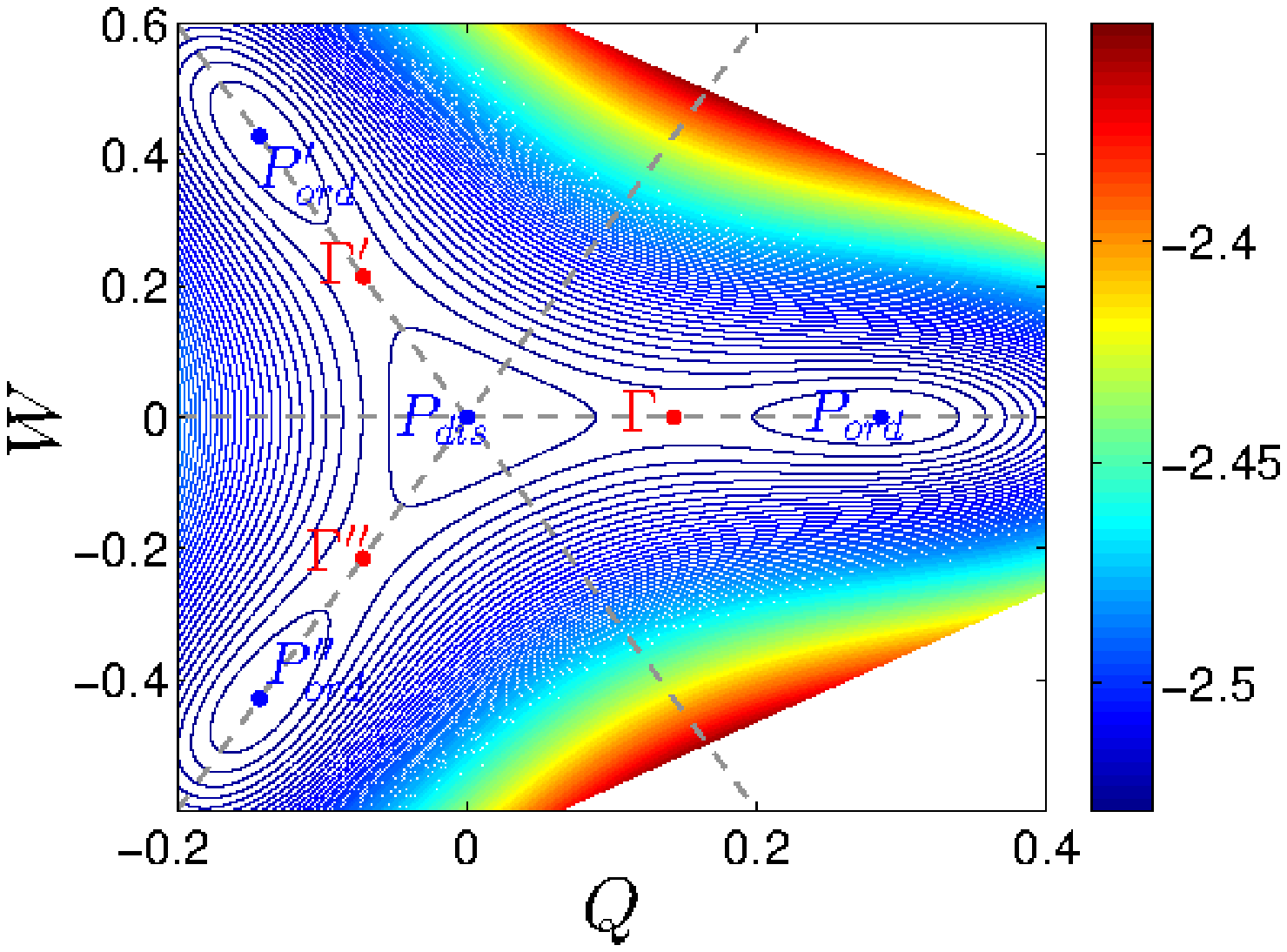}
		\includegraphics[scale = 0.45]{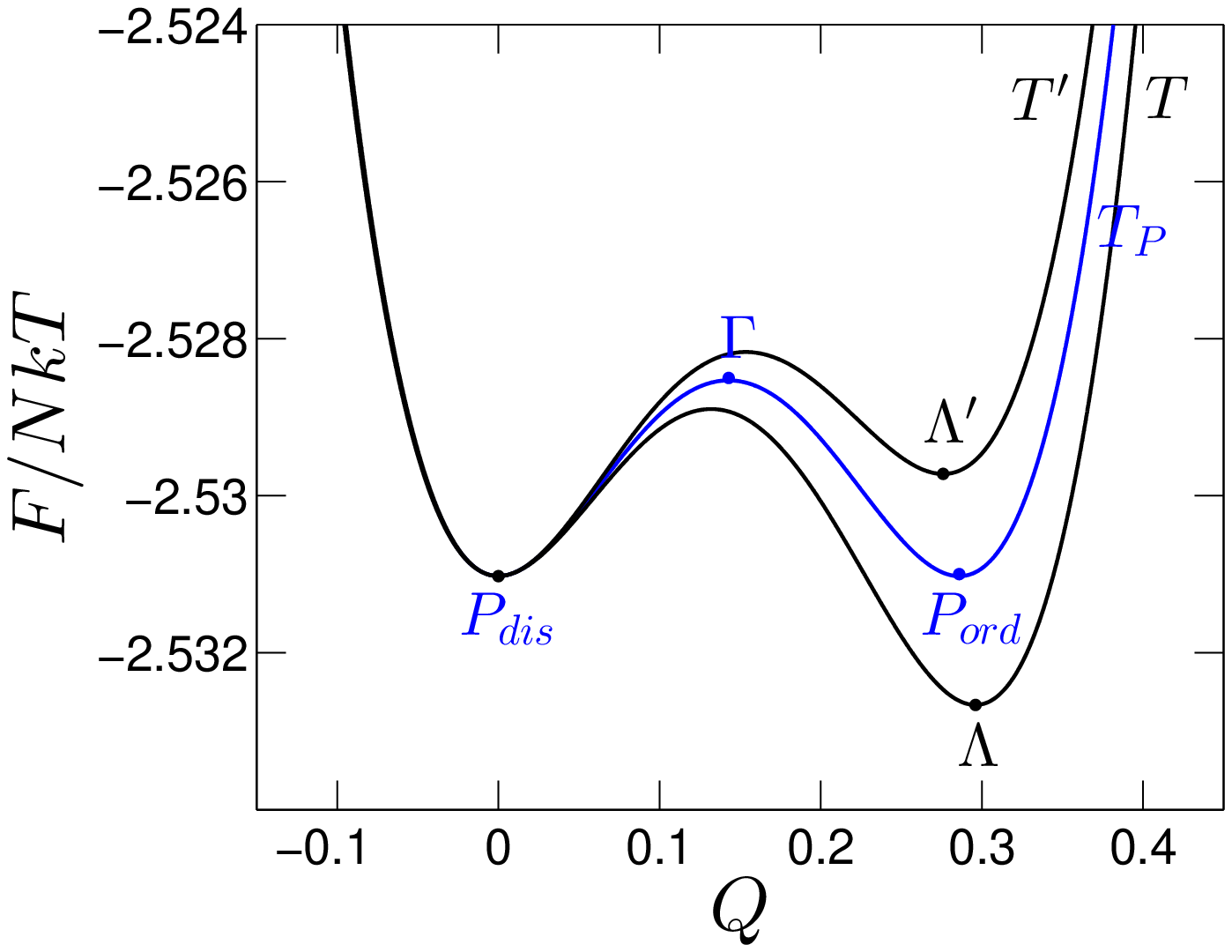} 
	\caption{
    The free energy $F/(NkT)$ with respect to the order parameters at a fixed temperature near the gravitational phase transition $kT_P/(JN) = 0.146796$, for systems with zero total angular momentum $L = 0$. \textit{Left:} Same as Figure~\ref{fig:F_q_w_L0_T5e-2}, but for temperature $T=T_P$. \textit{Right:} Free energy of axisymmetric configurations ($W = 0$) for three different temperatures, as a function of $Q$. The middle curve corresponds to the transition temperature $T_P$. The bottom curve corresponds to a slightly lower temperature ($T_B<T<T_P$), and the top curve to a higher temperature ($T_P<T'<T_A$). The temperatures $T_A$ and $T_B$ are defined in Figure~\ref{fig:qsz_T_L0}. The ordered state ($Q_{P_{\rm ord}} =0.286014$) changes from stable to metastable as the temperature increases past $T_P$, and the isotropic disordered state ($Q_{P_{\rm dis}} =0$) changes from metastable to stable. Therefore the branches $BP_{\rm dis}$ and $P_{\rm ord} A$ in Figure~\ref{fig:qsz_T_L0} are metastable in the canonical ensemble. 
    The states $\Lambda$, $P_{\rm ord}$, and $\Lambda'$ may be obtained by slow heating of $S_1$ or $O_1$ shown in Figures~\ref{fig:F_q_w_L0_T5e-2} and \ref{fig:qsz_T_L0} respectively, while $\Gamma$ is obtained by heating $S_2$ or $O_2$.}
	\label{fig:F_T_ABmeta_L0}
\end{center} 
\end{figure*}

\begin{figure*}[tbp]
\begin{center}
		\includegraphics[scale = 0.5]{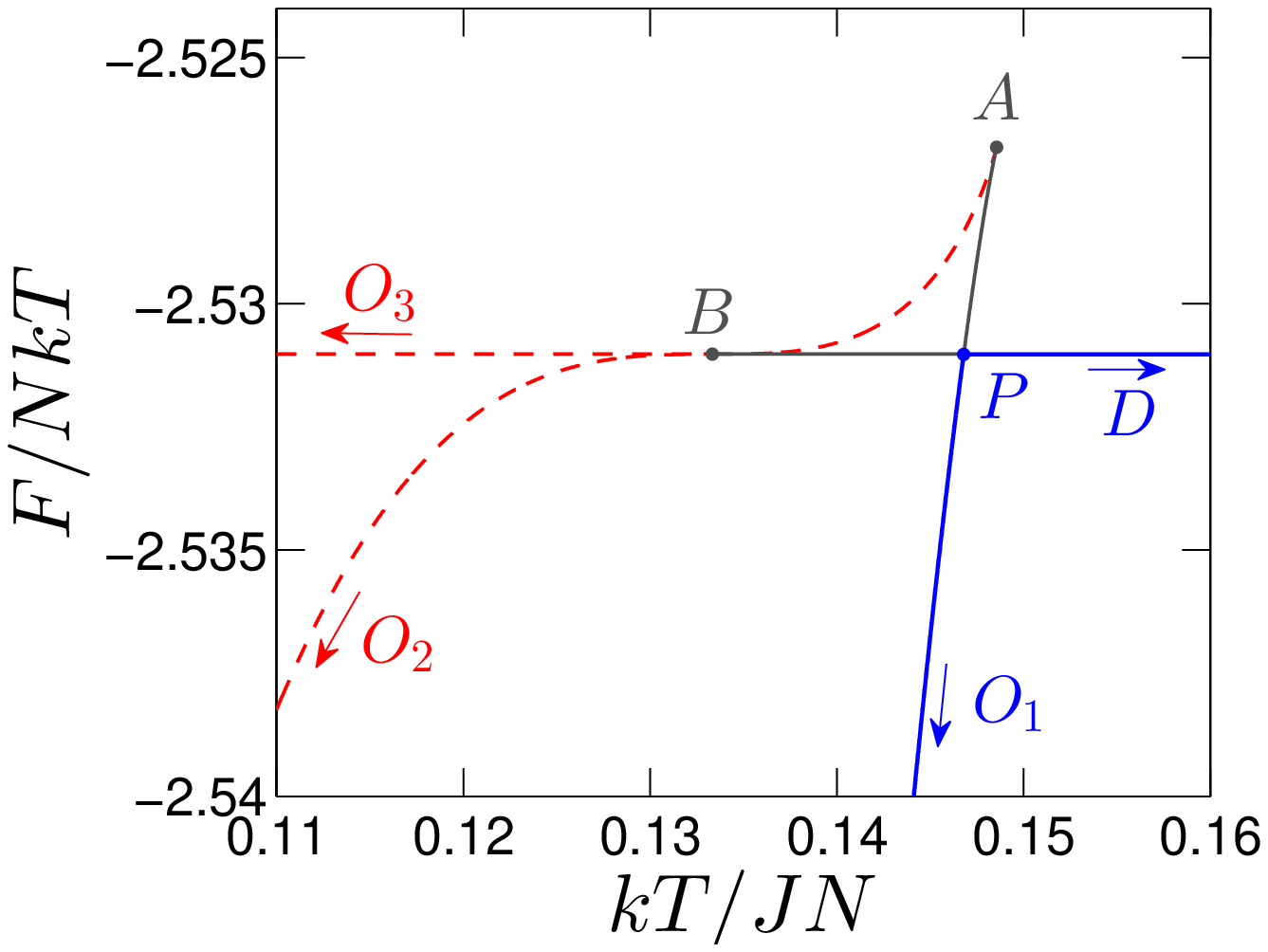} 
		\includegraphics[scale = 0.5]{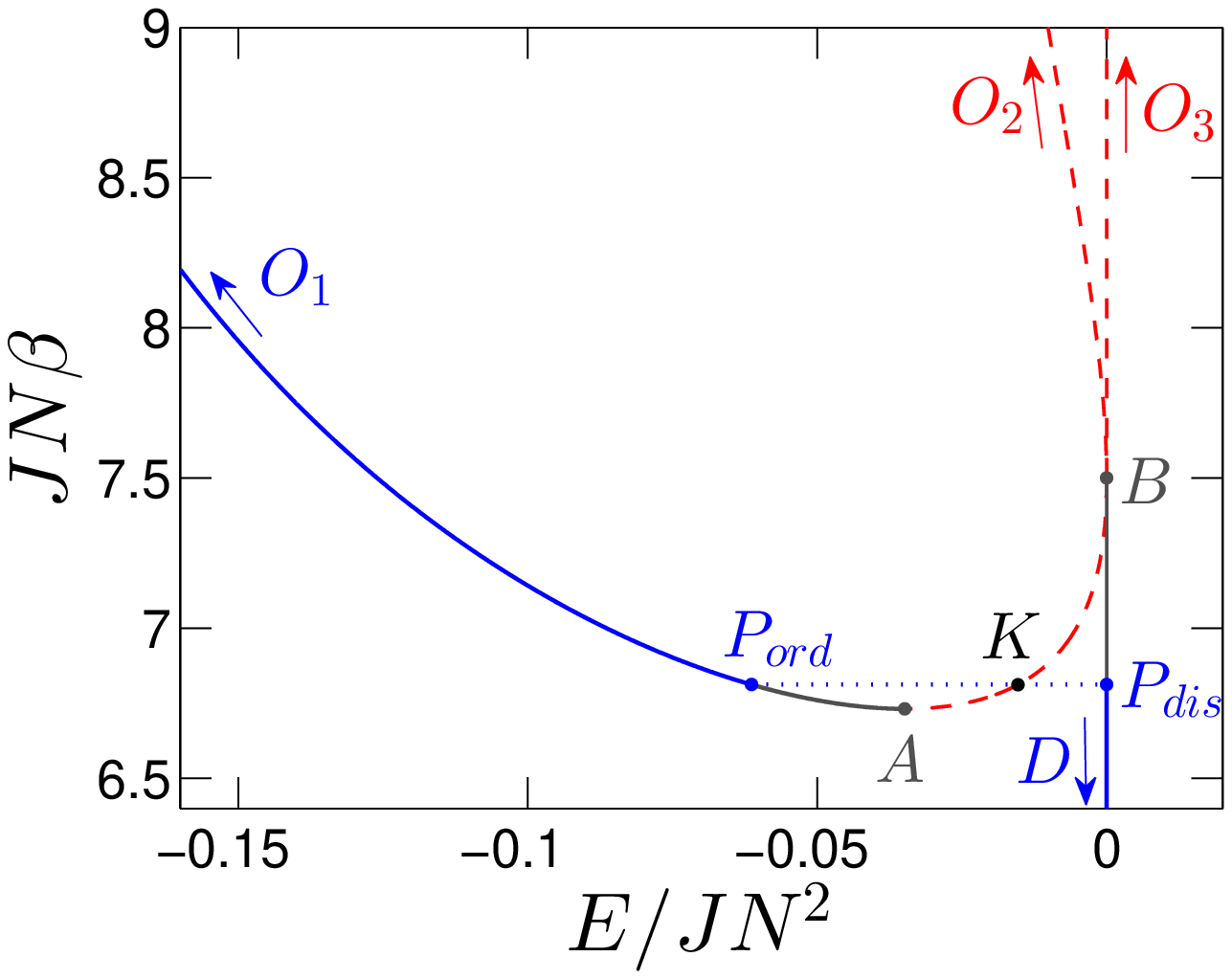} 
	\caption{Properties of equilibrium systems with zero total angular momentum near the first-order phase transition at $T_P=0.146796$. Blue solid and red dashed lines denote respectively stable and unstable equilibria. \textit{Left: } Free energy vs.\ temperature. \textit{Right:} Inverse temperature vs.\ energy (the caloric curve). The phase transition of the canonical ensemble occurs at $P$, where the two phases $P_{\rm ord}$ and $P_{\rm dis}$ (see Figure~\ref{fig:qsz_T_L0} and the right panel of Figure~\ref{fig:f_s_L0}) have the same free energy. The right panel shows the Maxwell construction: $P_{\rm ord} AK$ and $P_{\rm dis} B K$ have equal area. The phase transition involves a latent heat $E_{P_{\rm dis}}-E_{P_{\rm ord}} = 0.0613553 J N^2$. The gray branches $P_{\rm ord}A$, $P_{\rm dis} B$ are metastable as shown by Figure~\ref{fig:F_T_ABmeta_L0}, and the red dashed branch $AB$ is unstable in the canonical ensemble. Two phases $P_{\rm ord}$ and $P_{\rm dis}$ cannot coexist in a one-component system, hence the dotted line $P_{\rm ord}P_{\rm dis}$ does not represent equilibria in that case (the situation can be different in a multi-component system; see Section \ref{sec:multi}). In the canonical ensemble, the system passes abruptly from $P_{\rm ord}$ to $P_{\rm dis}$. In the microcanonical ensemble, all states along the branch $P_{\rm ord}AKBP_{\rm dis}$ are stable, and a continuous second-order phase transition takes place at $B$. Branches $BO_2$ and $BO_3$ are unstable, $O_1P_{\rm ord}$ and $P_{\rm dis} D O_3$ are stable in both the canonical and microcanonical ensembles.}  
	\label{fig:Fbeta_L0}
\end{center} 
\end{figure*}

To understand the nature of the phase transition, let us imagine slowly heating a one-component system from low temperature across the phase transition in the canonical ensemble\footnote{This would occur, for example, if the system interacts with massive distant objects with a nearly isotropic angular-momentum distribution.}. At low temperatures, the equilibria along the sequence starting with $O_1$ are minima of the free energy and therefore stable, while those along the sequences starting with $O_2$ and $O_3$ are saddle points and maxima, respectively, and therefore unstable.
A comparison of Figures \ref{fig:F_q_w_L0_T5e-2}, \ref{fig:qsz_T_L0}, and \ref{fig:F_T_ABmeta_L0} shows how the equilibria along these sequences change as the temperature increases: the sequences starting at $O_1$ and $O_2$ move toward the isotropic configuration, in the sense that the disk thickens and $|Q|$ decreases for both. At $k T_B/(JN)= 2/15=0.133333$, the unstable sequences starting from $O_2$ and $O_3$ intersect. As the temperature increases across $T_B$, the equilibria on the isotropic sequence that started at $O_3$ change from maxima to minima of the free energy so the sequence becomes metastable, while the sequence starting from $O_2$ remains unstable. The stable ordered sequence that begins at $O_1$ remains stable until the temperature increases to $T_P=0.146796\,JN/k$. At $T_P$, the free energies of the ordered and disordered states, $P_{\rm ord}$ and $P_{\rm dis}$ in Figure~\ref{fig:qsz_T_L0}, become equal. When the temperature is increased above  $T_P$, a first-order phase transition occurs in the canonical ensemble, in which the isotropic disordered state becomes stable and the ordered state becomes metastable. As the temperature continues to increase, the sequences that began at $O_1$ and $O_2$ approach the same state, and they eventually coincide at $T_A=0.148556\,JN/k$, For $T>T_A$ there is no equilibrium other than the isotropic disordered state, which is stable. 

The behavior near the phase transition is explored further in Figure 
\ref{fig:F_T_ABmeta_L0}. The left panel shows the free energy as a
function of the order parameters $Q$ and $W$ at the temperature of the phase transition, $T_P$. Since the equilibria are all axisymmetric, we can set $W=0$ and plot the free energy as a function of one order parameter $Q$, which we do in the right panel. 
The free energy is shown for three temperatures. The bottom curve is for temperatures between $T_B$ and $T_P$, where the minimum at $P_{\rm ord}$ corresponds to the stable ordered sequence starting at $O_1$, the minimum at $P_{\rm dis}$ corresponds to the metastable disordered sequence starting at $O_3$, and the intervening maximum corresponds to the unstable sequence starting at $O_2$. The middle curve is at the temperature $T_P$ of the phase transition, where the free energies at $P_{\rm ord}$ and $P_{\rm dis}$ are equal, and the upper curve is for temperatures between $T_P$ and $T_A$. 

Figure \ref{fig:Fbeta_L0} shows $F/(NkT)$ near the phase transition. At $T_P$, the free energy is the same for the ordered and disordered phases ($P$ in the left panel), but the energies of these states are different ($P_{\rm ord}$ and $P_{\rm dis}$ in the right panel). This panel shows the caloric curve and Maxwell's construction for the first-order phase transition. In a nonadditive system multiple phases cannot coexist in equilibrium 
(see Section \ref{s:E_In}) and so the dotted line $P_{\rm ord}P_{\rm dis}$ in the right panel of Figure \ref{fig:Fbeta_L0} is unphysical. However, phase separation can occur in separable multi-component systems, as described in Section \ref{sec:multi}; in this case individual components are either in $P_{\rm ord}$ or in $P_{\rm dis}$ but the full system can lie anywhere along $P_{\rm ord}P_{\rm dis}$. 

This analysis is different in the microcanonical ensemble for a one-component system, i.e., an isolated system under conditions of constant energy. In that case, we find that the series of equilibria $P_{\rm ord} A BP_{\rm dis}$ represent the highest entropy states at fixed energy, while the states along $O_2B$ and $O_3B$ at fixed energy have an entropy minimum. Thus, the branch $P_{\rm ord} A BP_{\rm dis}$ is stable in the microcanonical ensemble, and a continuous second-order phase transition takes place at point $B$ in the microcanonical ensemble at temperature $k T_B/(JN)= 2/15=0.133333$ (see Figure~\ref{fig:Fbeta_L0}).
Branches $BO_2$ and $BO_3$ are unstable in both canonical and microcanonical ensembles. 

\begin{figure*}[tbp]
\begin{center}
		\includegraphics[scale = 0.5]{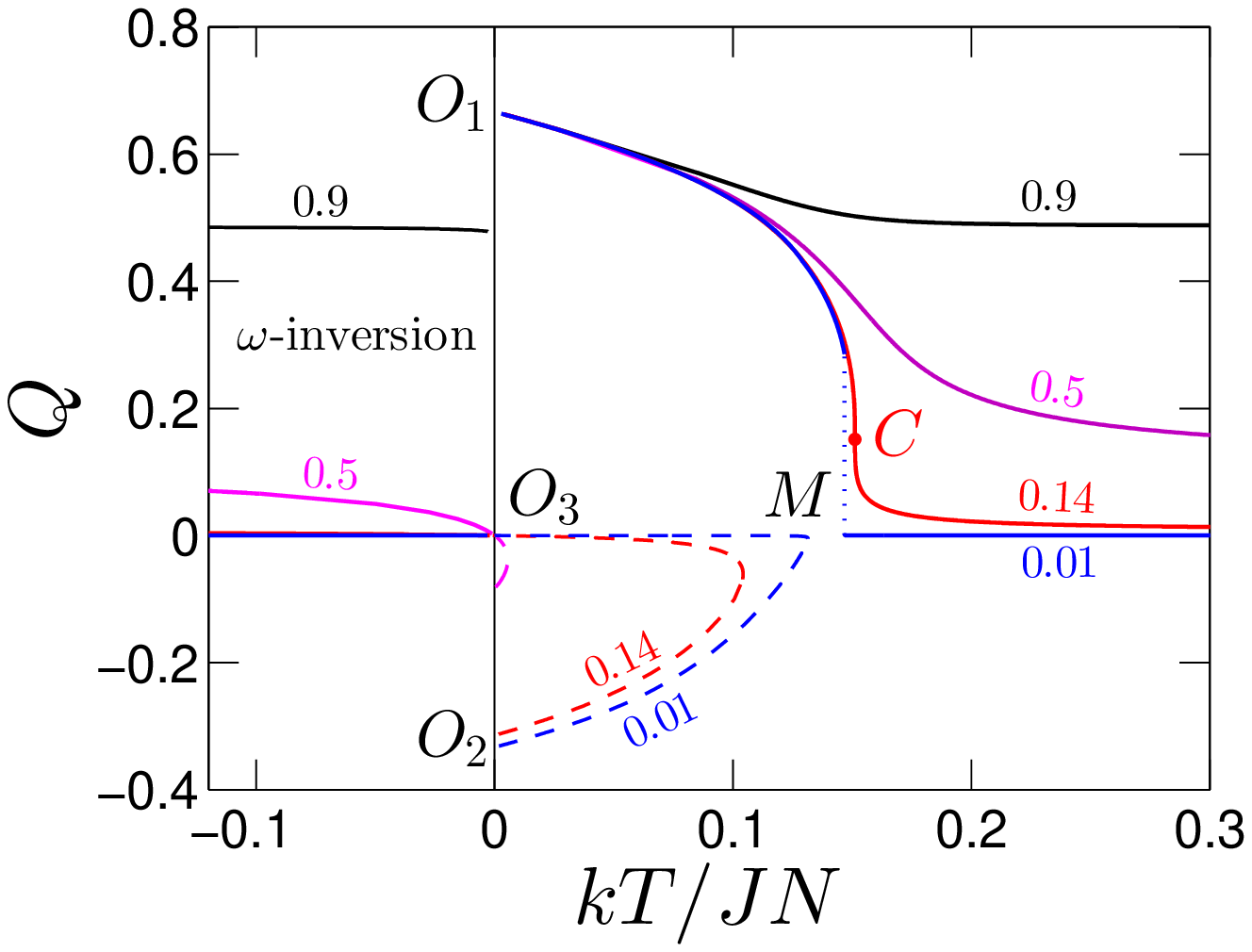} 
		\includegraphics[scale = 0.5]{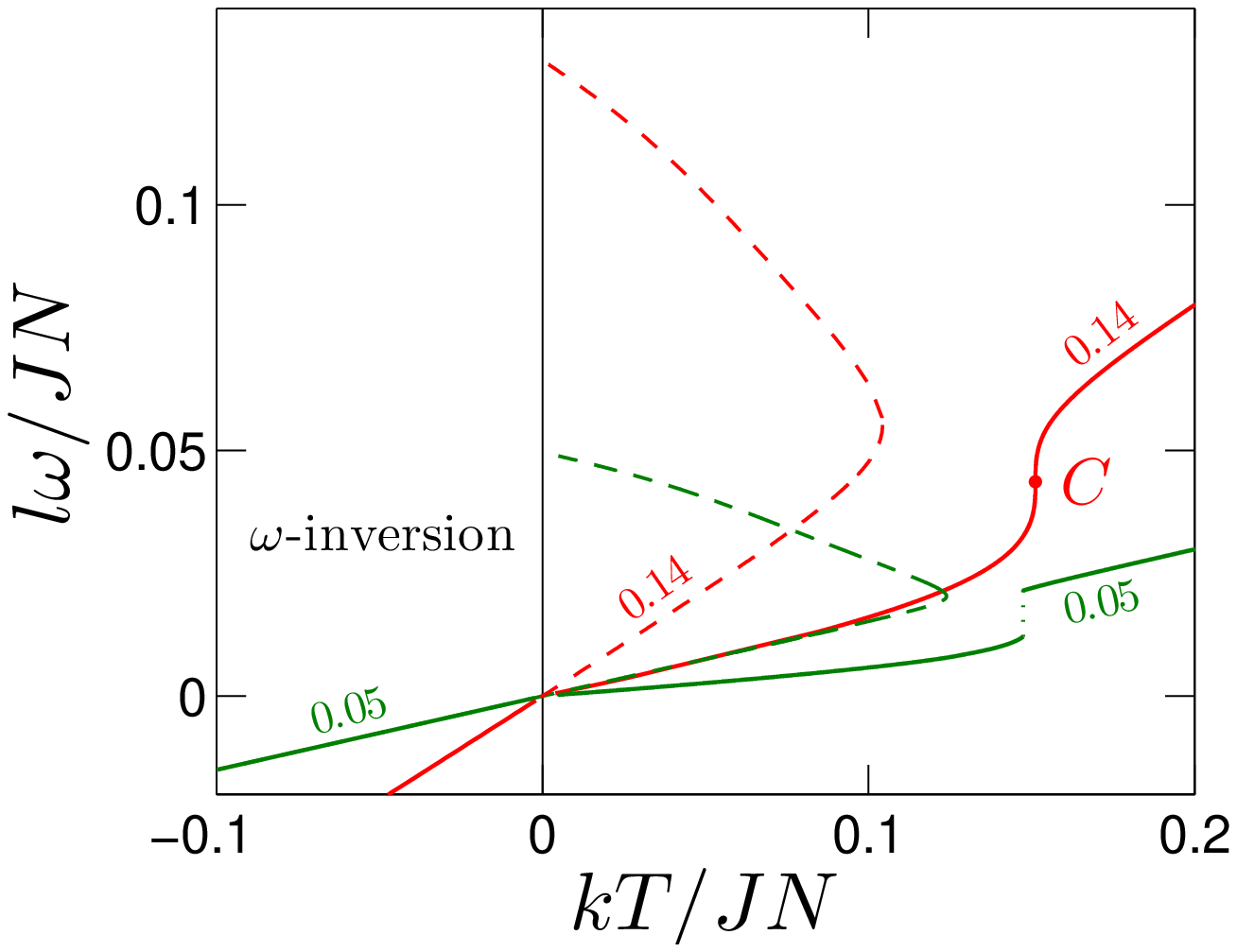} 
\caption{\textit{Left:} The order parameter $Q$ of axisymmetric equilibria in the canonical ensemble, plotted with respect to temperature for several values of the total angular momentum, $L/(Nl)=0.01$ (blue), $0.14$ (red), $0.5$ (magenta), $0.9$ (black). Figure~\ref{fig:qsz_T_L0} shows the analogous plot for $L=0$. Stable equilibria are denoted by solid lines and unstable equilibria by dashed lines. For angular momentum less than a critical value $L_{\rm cr}/(Nl) = 0.13714$, a first-order phase transition takes place in the canonical ensemble, as in the non-rotating case. At $L_{\rm cr}$, the phase transition becomes second-order, marked by the critical point $C$ (see Figure~\ref{fig:phase_trans_L} for an expanded view). For $L > L_{\rm cr}$ there is no phase transition.  The stable branches $O_2 M$ and $O_3 M$ contain the equilibria $\Sigma_2$ and $\Sigma_3$ of Figure \ref{fig:F_q_w}, respectively. The two ground states $O_2$ and $O_3$ merge for $L/(Nl) \geq 1/\sqrt{3}$ and no unstable axisymmetric branch exists for higher $L$. Negative-temperature equilibria exist for all $L$ and are stable in both canonical and microcanonical ensembles (see appendix \ref{s:app:S2}).
\textit{Right:} The rotation parameter $\omega$, defined in Eq.~(\ref{eq:omega}), is shown with respect to temperature for equilibria with $L/(Nl) = 0.05$ (green curves) 
and $L_{\rm cr}$ (red curves). 
At negative temperature, $\omega=\gamma/\beta$ becomes negative, so there is an 
``$\omega$-inversion'' in the sense that more bodies counter-rotate with respect to $\bm{\omega}$, since the coordinate system is aligned with $\bm{L}$. In other words $\bm{\omega}$ is anti-parallel to $\bm{L}$ at negative temperatures and parallel to it at positive ones.}
	\label{fig:qomega_T_L}
\end{center} 
\end{figure*}

\begin{figure*}[tbp]
\begin{center}
		\includegraphics[scale = 0.5]{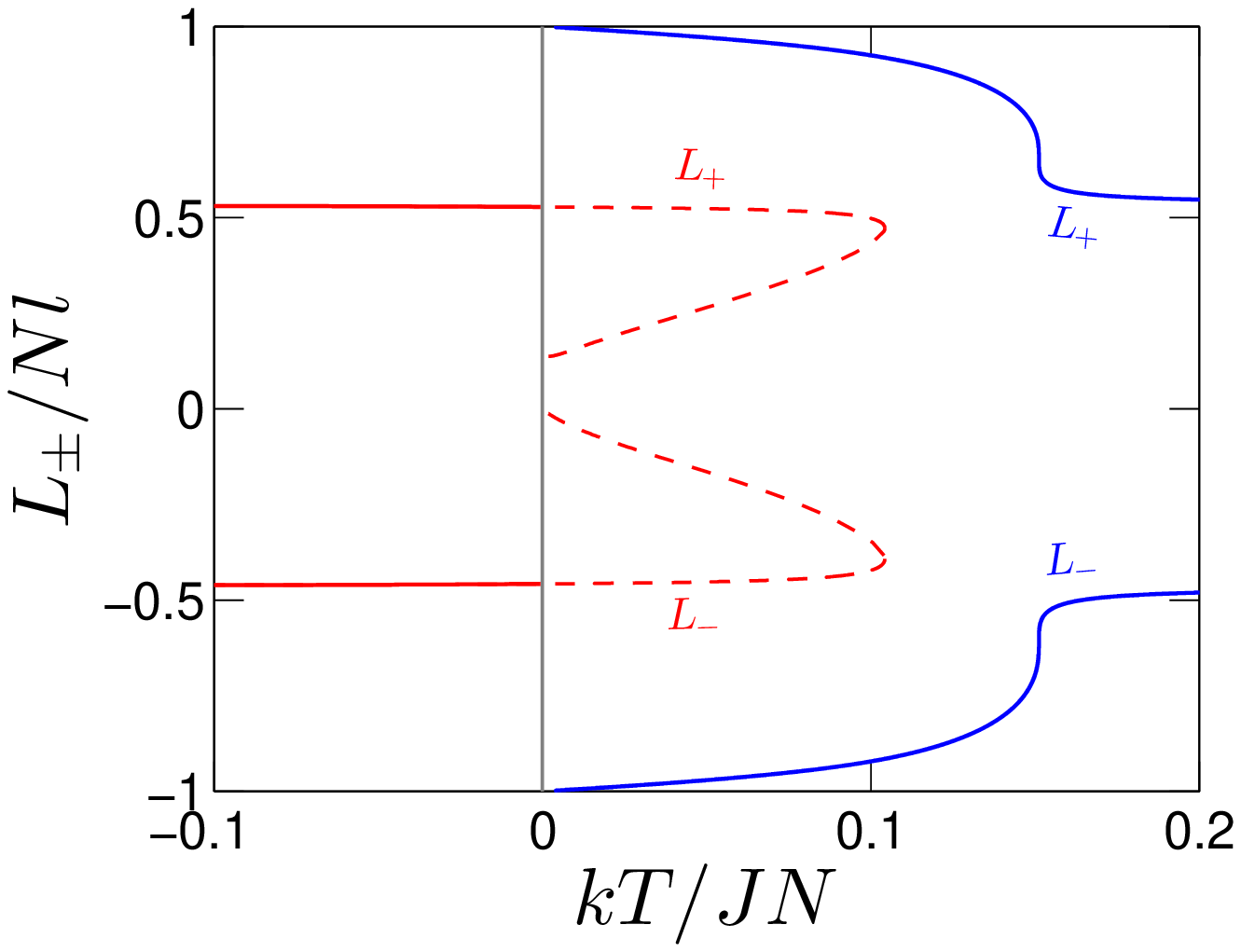} 
		\includegraphics[scale = 0.5]{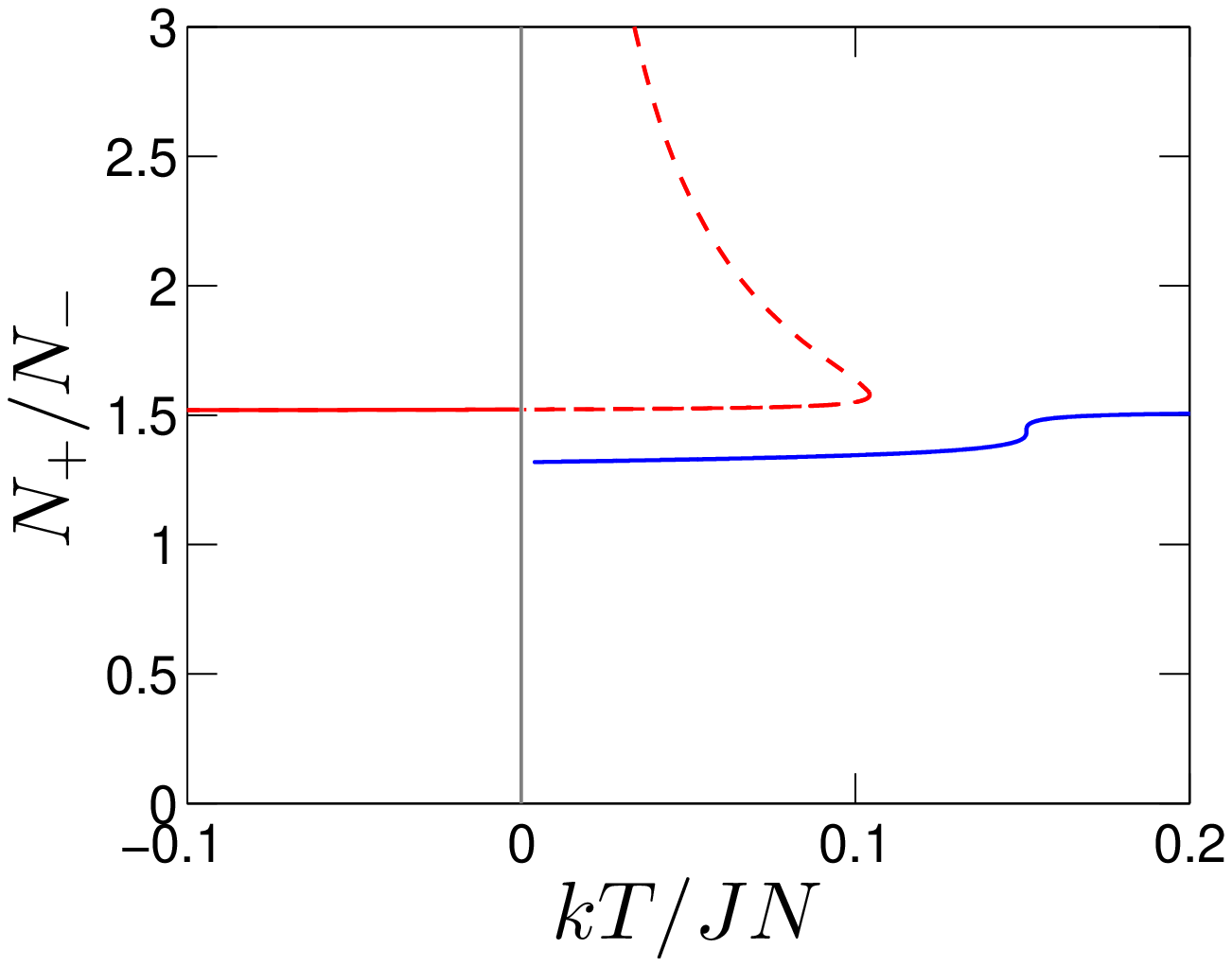} 
	\caption{ \textit{Left:} The mean angular momentum of prograde and retrograde bodies, $L_{\pm}$, for the critical total angular momentum $L_{\rm cr} = 0.13714$ (cf.\ Figure~\ref{fig:qsz_T_L0}).
    \textit{Right:} The ratio of the number of prograde and retrograde orbits. In both panels solid lines denote stable equilibria and dashed lines unstable equilibria.
    }
	\label{fig:szNpm_T_L}
\end{center} 
\end{figure*}

\section{Non-zero angular momentum}

Let us turn to the more general case of a one-component system with non-zero total angular momentum, $L \neq 0$ in Eq.~(\ref{eq:Lav}). In this case, the angular momentum constraint can only be satisfied if the Lagrange multiplier $\bm{\gamma}=\beta\bm{\omega}$ is non-zero, which introduces a factor $\exp(l \beta\bm{\omega}\cdot\bm{n})$ in the distribution function (Eq.~\ref{eq:f_qw}). This is similar to the effect of a paramagnetic term in the Hamiltonian, and hence the distribution function, of a spin system in a magnetic field, where the role of magnetic field is played by $\bm{\omega}$, the magnetic moment is replaced by $l \bm{n}$ and the spin is replaced by $\bm{n}$.\footnote{This is different in the Maier-Saupe model of liquid crystals, where an external magnetic field $\bm{B}$ gives rise to a term proportional to $(\bm{B}\cdot \bm{n})^2$ where $\bm{n}$ describes the orientation of the molecules \citep{Wojtowicz_1974,Muhoray_1982}.}.  

We work in a coordinate system in which the total angular momentum is parallel to the positive $z$-axis.  The equilibrium distribution function $f(\bm{n})$ may be either axisymmetric around the $z$-axis (e.g., $\Sigma_1$, $\Sigma_2$, $\Sigma_3$ in Figure~\ref{fig:F_q_w}) or else non-axisymmetric, with three distinct eigenvalues (i.e., $Q\neq \pm \frac{1}{3} W$ and $W \neq 0$ as in $\Sigma_4$, $\Sigma_5$ in Figure~\ref{fig:F_q_w}). We study these cases separately.

\subsection{Axially symmetric equilibria}\label{s:ax-sym}

Let us consider first axisymmetric configurations (e.g., $\Sigma_1$, $\Sigma_2$, $\Sigma_3$ in Figure~\ref{fig:F_q_w}). These constitute the most important cases, since global free energy minima at fixed temperature and global entropy maxima at fixed total energy are always axisymmetric as we will see below. We derive analytic expressions for the order parameters valid at any temperature and angular momenta in Appendix~\ref{s:app:partition:rotating}.

\begin{figure*}[tbp]
\begin{center}
		\includegraphics[scale = 0.5]{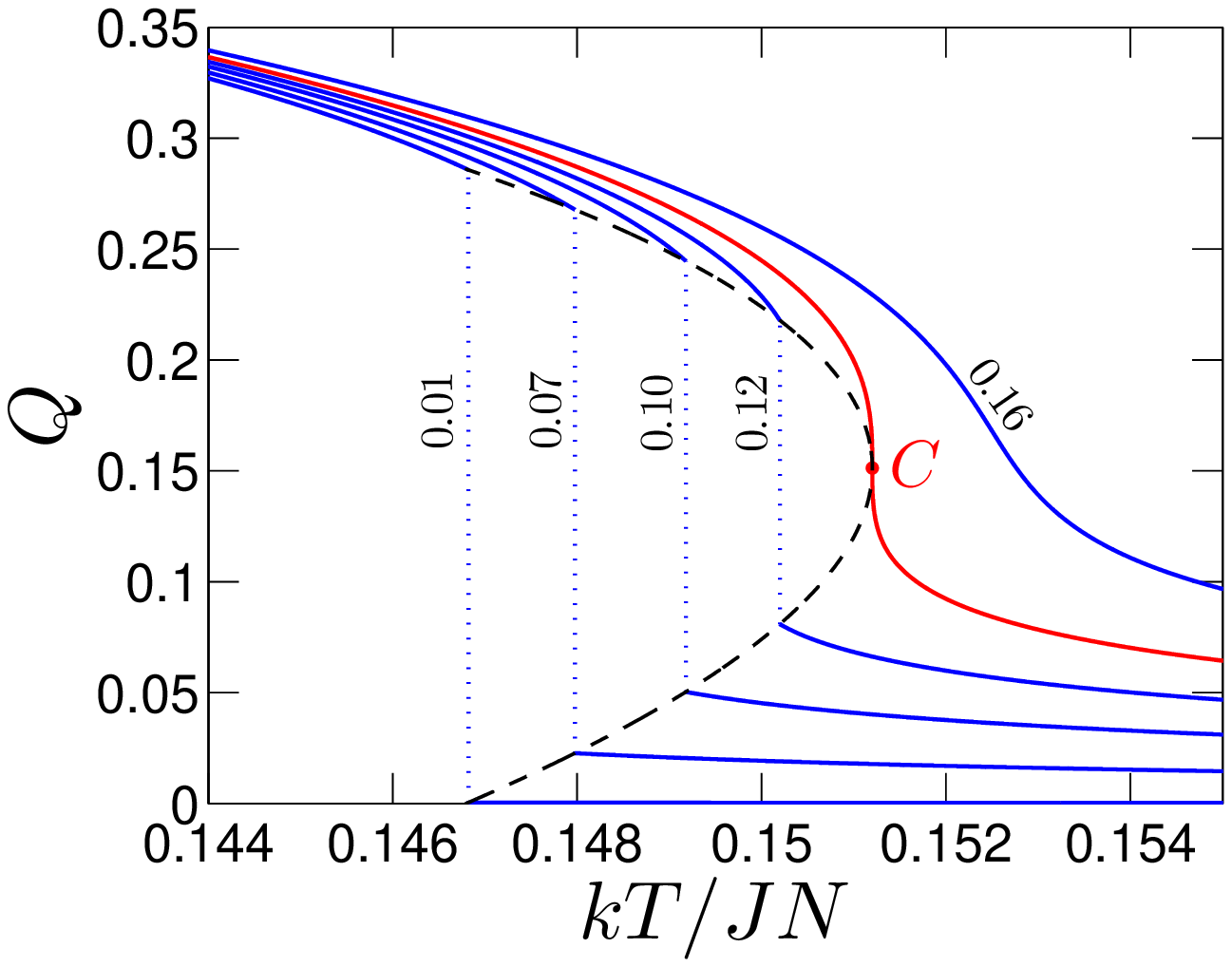} 
		\includegraphics[scale = 0.5]{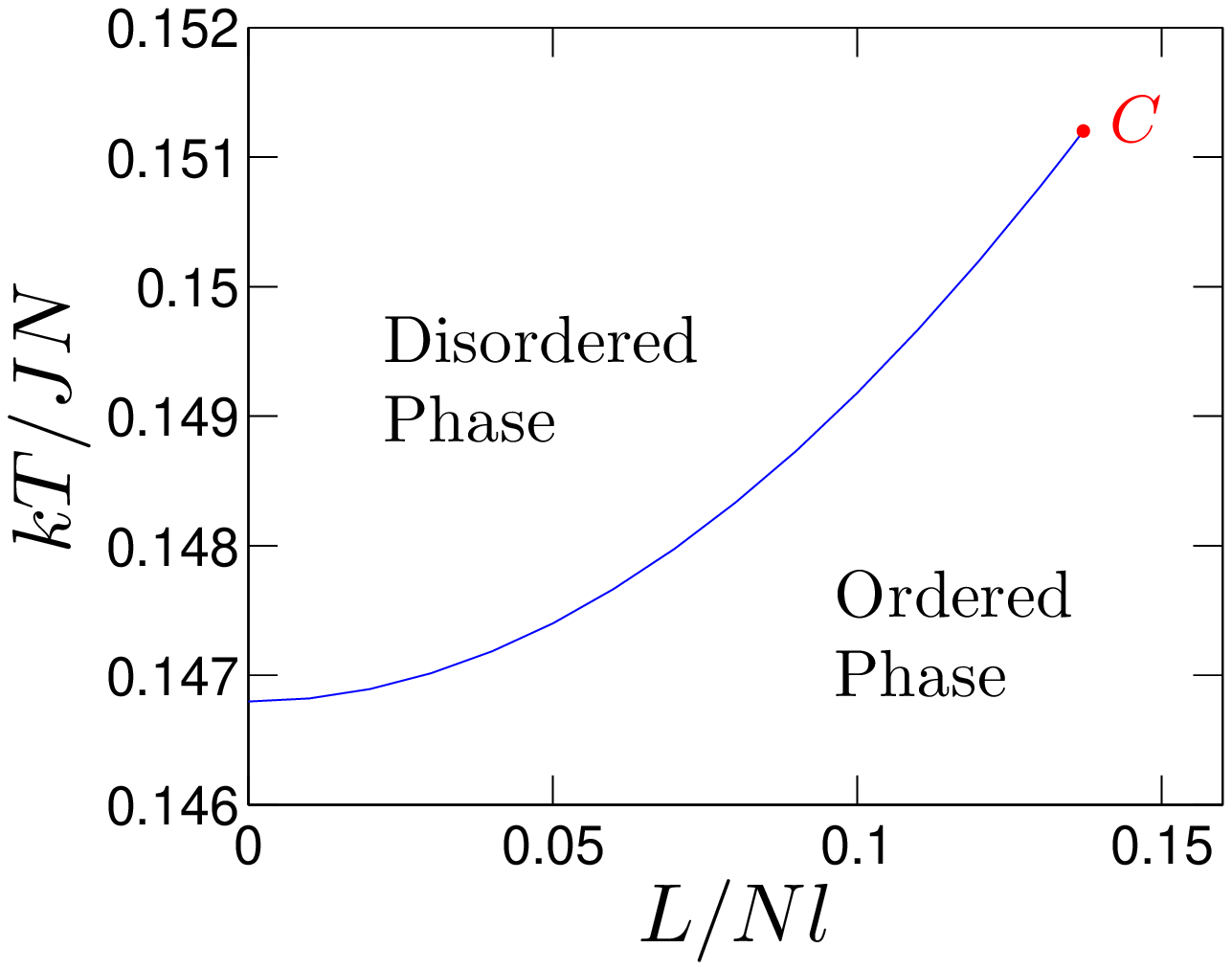} 
	\caption{\textit{Left:} Zoom-in on the phase transition of the canonical ensemble (Figure~\ref{fig:qomega_T_L}) for various values of the total angular momentum (labeled) below and above the critical value $L_{\rm cr}$. The dashed line is parabolic (Eq.~\ref{eq:phasetransition-q}). For $L < L_{\rm cr}$ the phase transition is first-order in the canonical ensemble, and it is second-order at the critical point $C$, with $L_{\rm cr}=0.13714 Nl$, $Q_C = 0.151201$, $kT_C = 0.1512 JN$, $l \omega_C = 0.0291\,JN$. 
    \textit{Right:} The phase transition temperature as a function of the dimensionless total angular momentum.}
	\label{fig:phase_trans_L}
\end{center} 
\end{figure*}

\begin{figure}[tbp]
\begin{center}
		\includegraphics[scale = 0.43]{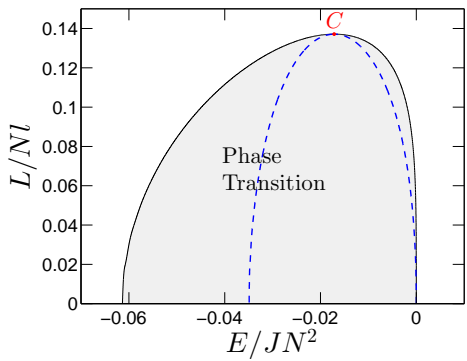} 
\caption{
The total dimensionless angular momentum $L/(Nl)$ with respect to the energy at the phase transition points of the canonical ensemble where $L$ is fixed (cf. Figure~\ref{fig:gensemble_Omega} in which $L$ may change). 
The boundary of this region corresponds to the series of points $P_{\rm ord}$ and $P_{\rm dis}$ for energies lower and higher than the energy at the critical point, respectively.
The region inside the blue dashed curve corresponds to equilibria that have negative specific heat, which are unstable in the canonical ensemble and stable in the microcanonical ensemble.}
	\label{fig:L_E_Phase_Transition}
\end{center} 
\end{figure}

\begin{figure*}[tbp]
\begin{center}
		\includegraphics[scale = 0.5]{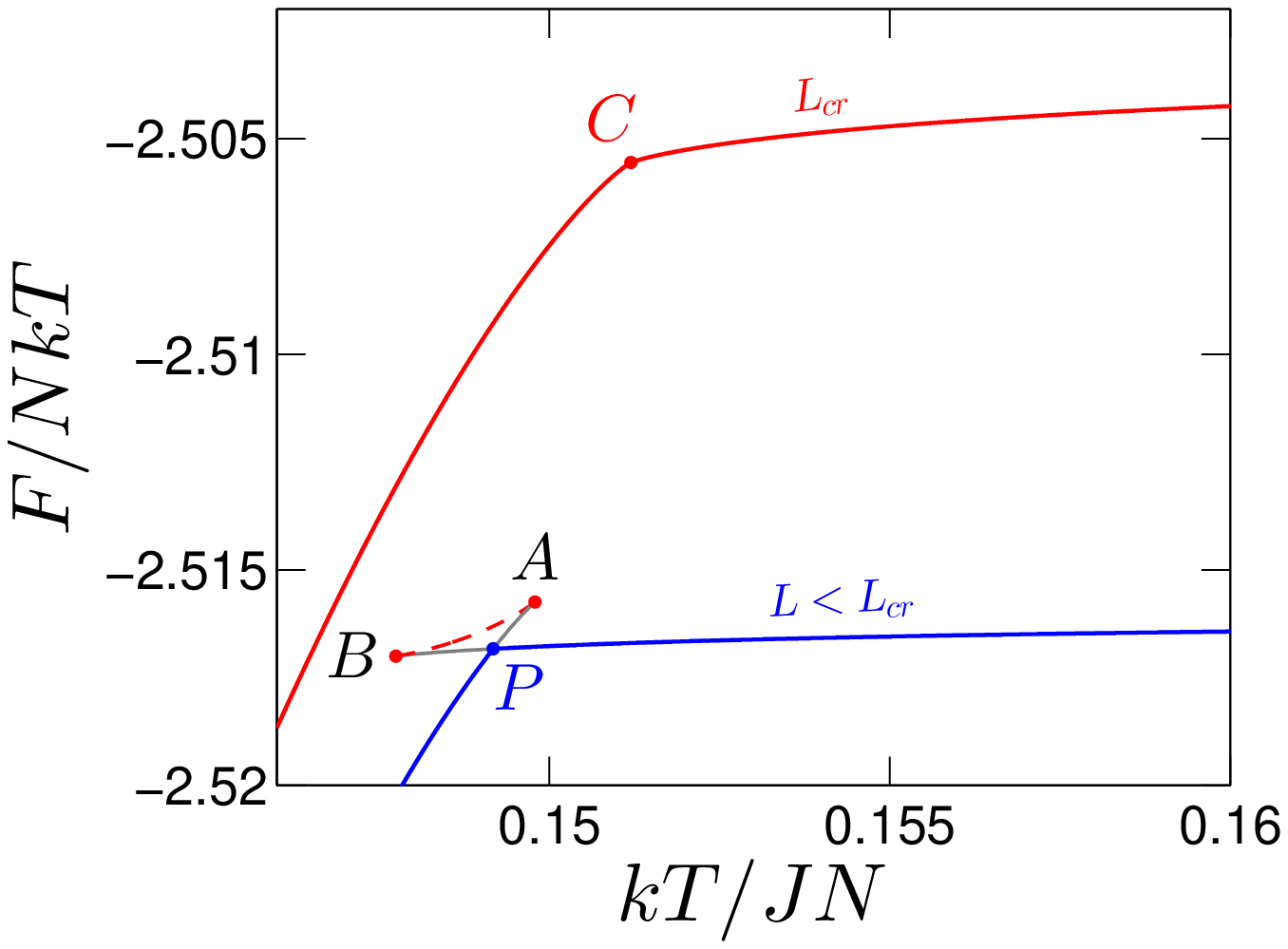} 
		\includegraphics[scale = 0.5]{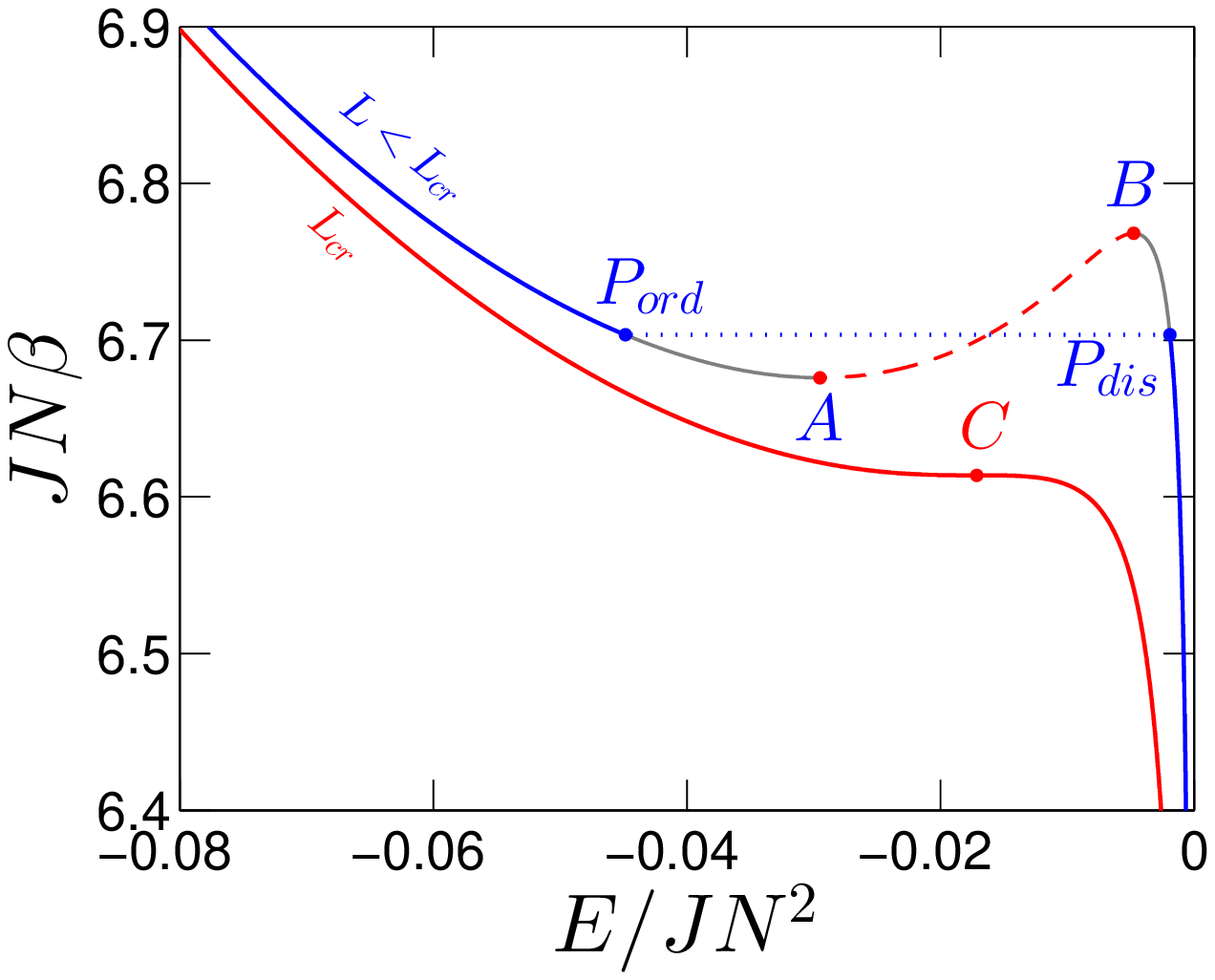} 
	\caption{ Same as Figure~\ref{fig:Fbeta_L0} but for $L = 0.1 Nl < L_{\rm cr}$ and $L=L_{\rm cr}$. Point $C$ is a critical point, where the first-order phase transition (point $P$) becomes a continuous second-order phase transition in the canonical ensemble (see Figure~\ref{fig:phase_trans_L}). Curves $P_{\rm ord}A$ and $BP_{\rm dis}$ are metastable in the canonical ensemble, similar to the case of no net rotation (see Figure~\ref{fig:F_T_ABmeta_L0}). The dashed branch $AB$ corresponds to unstable equilibria. $P_{\rm ord}$ represents the ordered disk+halo phase, while $P_{\rm dis}$ is the disordered, nearly isotropic, phase whose distribution function is shown in Figure~\ref{fig:f_s_rot}. The $P_{\rm ord}P_{\rm dis}$ dotted line does not represent a series of equilibria for a non-additive long-range interaction system and the system passes discontinuously from one phase to the other in the canonical ensemble. In the microcanonical ensemble there is no phase transition and the branches $P_{\rm ord} ABP_{\rm dis}$ are stable even though the branch $AB$ has negative heat capacity. 
All of the equilibria along the red critical curve are stable.
    }
	\label{fig:Fbeta_L} 
\end{center} 
\end{figure*}

Figure \ref{fig:qomega_T_L} shows the order parameter $Q$ with respect to temperature for axisymmetric equilibria, at several values of the dimensionless total angular momentum. For small values of $L$, three axisymmetric ground states exist at zero temperature, similar to the case with $L=0$ shown in Figure \ref{fig:qsz_T_L0}. The stable ordered state $O_1$ has $Q_{O_1} = \frac{2}{3}$ at $T\rightarrow 0^+$, for any $L$.
This state is a razor-thin disk in physical space, with a fraction of prograde orbits equal to $\frac{1}{2}(1+L/(Nl))$. 
As $L$ increases, $Q_{O_2}$ at $T=0^+$ moves toward $Q_{O_3}=0$ and they merge at $L/(Nl) = 1/\sqrt{3}$. This behavior is explained using the asymptotics of the partition function in Appendix~\ref{s:app:partition}. Figure \ref{fig:szNpm_T_L} shows the mean angular momentum $L_{\pm}$ and the ratio of prograde or retrograde orbits $N_+/N_-$ at the critical total angular momentum, $L_{\rm cr}/(Nl) = 0.13714$. This is the maximum total angular momentum for which a phase transition occurs. 

\begin{figure*}[tbp]
\begin{center}
		\includegraphics[scale = 0.45]{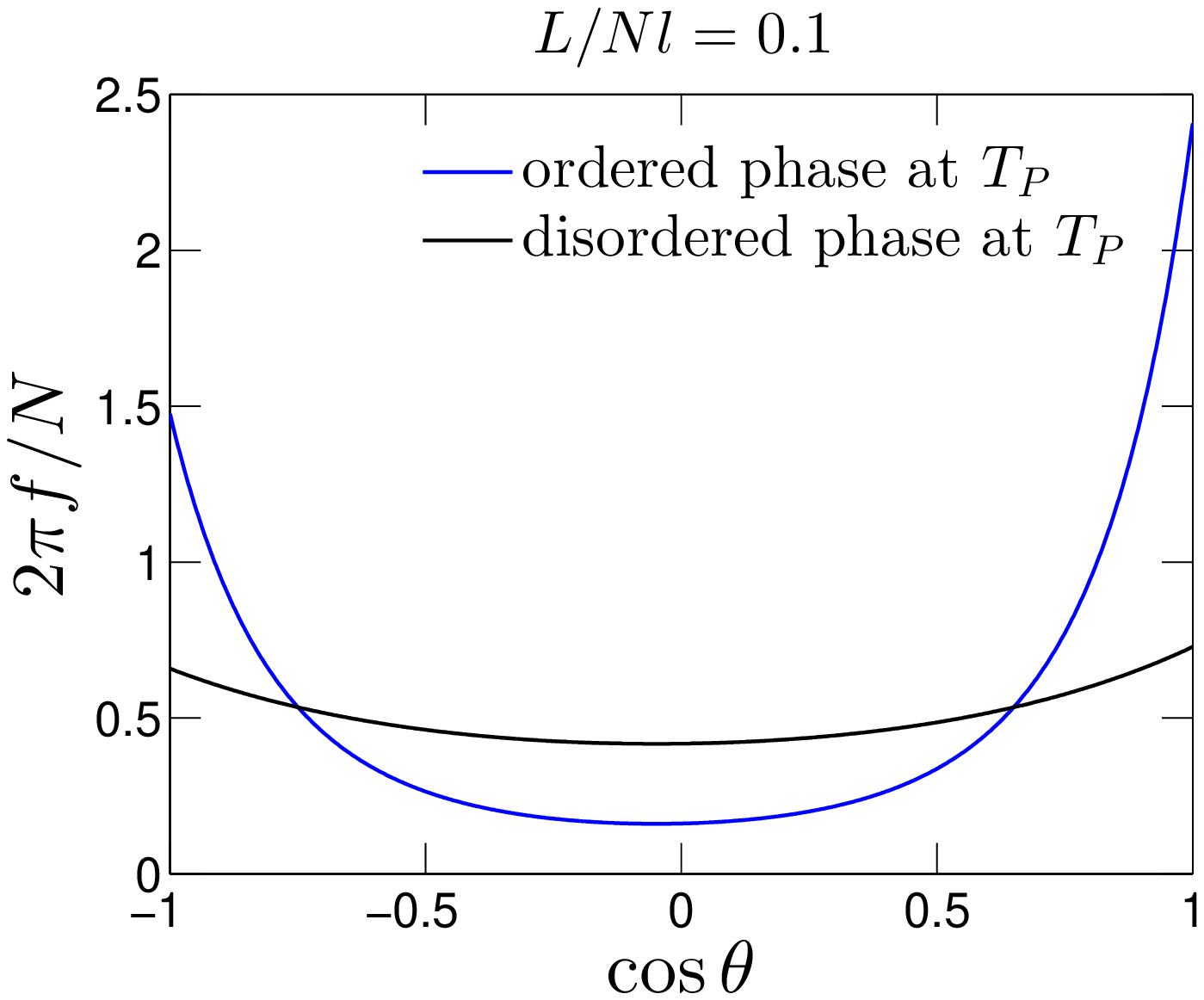} 
		\includegraphics[scale = 0.45]{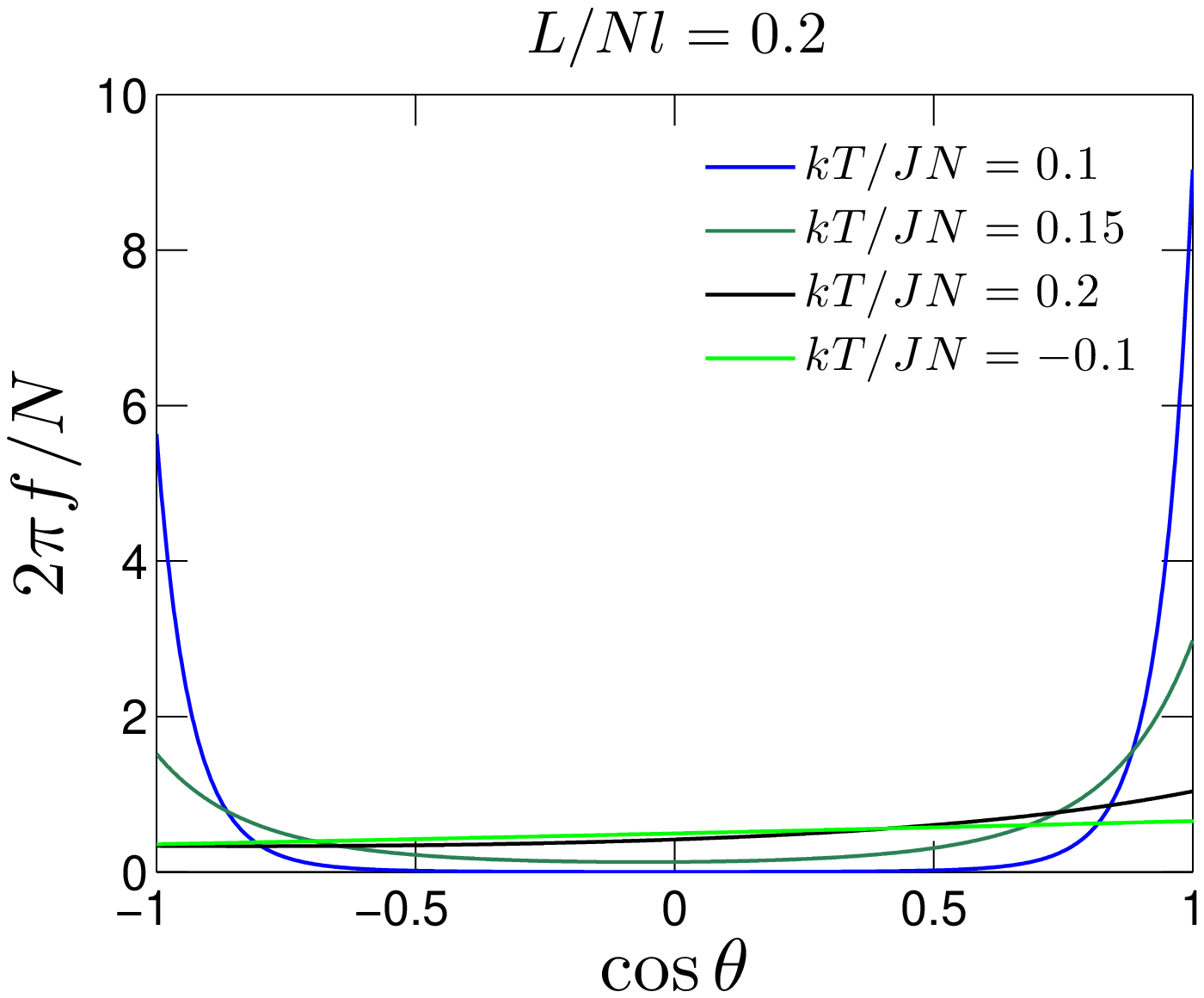} 
\caption{The probability density of the $z$-coordinate of the normalized angular-momentum vector ($s = \cos\theta$) for axisymmetric stable equilibria (cf.\  Figure~\ref{fig:f_s_L0}). \textit{Left:} The two phases at the phase-transition temperature $kT_P/(JN) = 0.1512$  for 
$L/(Nl) = \LA s \RA = 0.1$. 
The ordered phase consists of a disk containing both prograde and retrograde bodies and a dilute halo. The disordered phase is a nearly isotropic halo.
\textit{Right:} The distribution at different temperatures for $L = \LA s\RA = 0.2$. Although there is no phase transition here since $L > L_{\rm cr} = 0.137 Nl$ (see Figure~\ref{fig:phase_trans_L}), the qualitative characteristics of the disk+halo structure and nearly isotropic configurations at low and high temperatures are similar.}
	\label{fig:f_s_rot}
\end{center} 
\end{figure*}

\begin{figure*}[tbp]
\begin{center}
		\includegraphics[scale = 0.45]{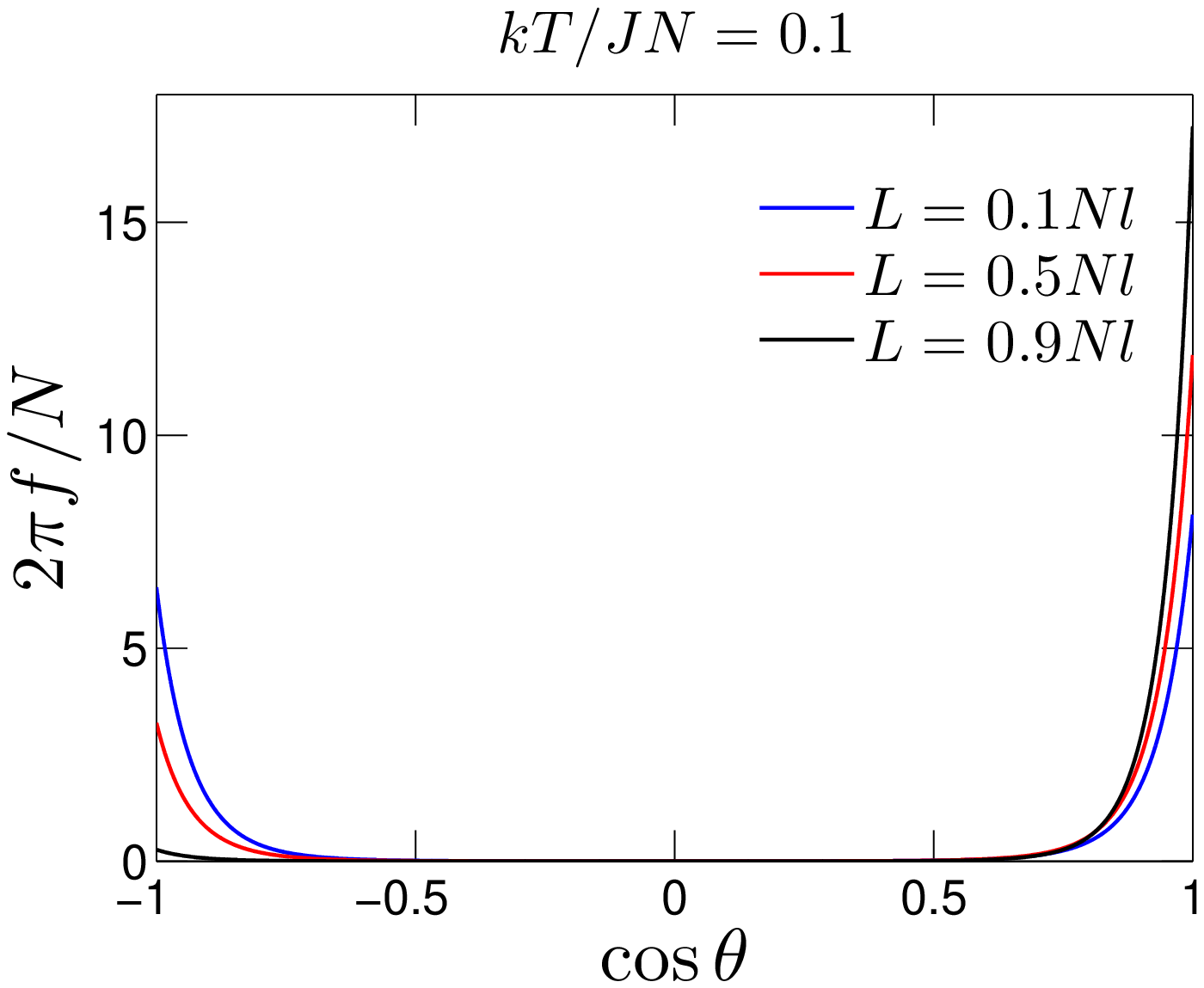} 
		\includegraphics[scale = 0.45]{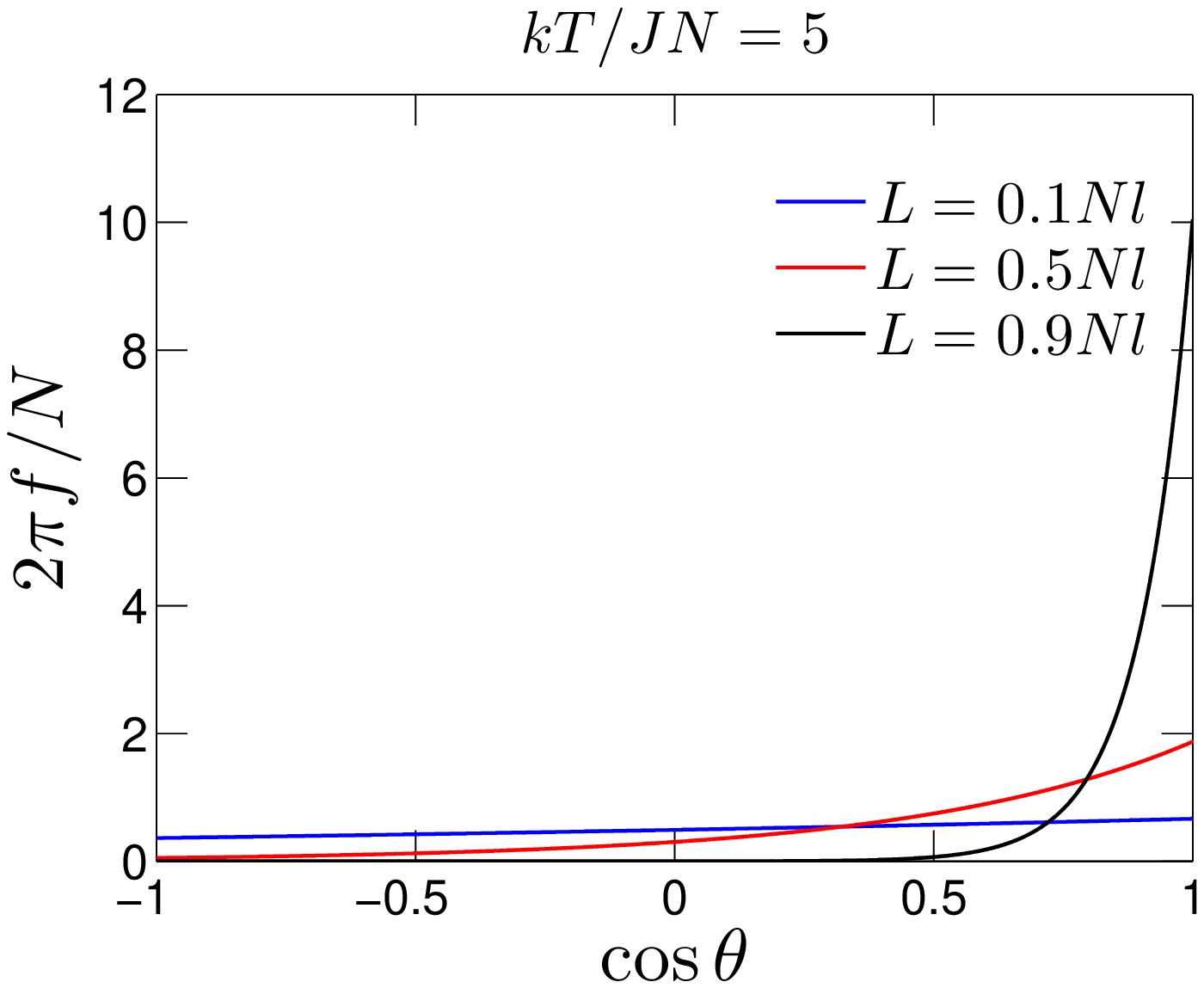} 
\caption{Same as Figure~\ref{fig:f_s_rot}, but for fixed temperatures (low in the left panel, and high in the right panel as labeled) and different total angular momentum (different curves as labeled by $L$). For low $L$, the angular momenta are oriented along $\cos \theta\sim \pm1$ at low temperature, representing a thin disk with both prograde and retrograde orbits, and they are nearly isotropically distributed at high temperature. For high $L$, almost all bodies orbit in the same direction in a disk for either low or high temperature.}
	\label{fig:f_s_var_rot}
\end{center} 
\end{figure*}

\begin{figure}[tbp]
\begin{center}
		\includegraphics[scale = 0.5]{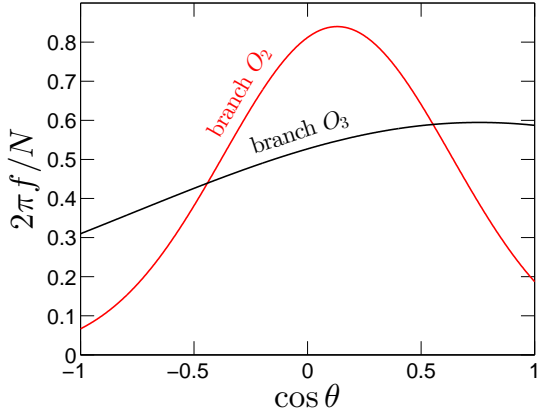} 
	\caption{Same as Figures~\ref{fig:f_s_rot} and \ref{fig:f_s_var_rot}, but for the \textit{unstable} axisymmetric equilibria with $L/(Nl) = 0.1$. Here $kT/(JN) = 0.1$ for the curves representing branches $O_2$ and $O_3$ in Figure~\ref{fig:qomega_T_L} (see $\Sigma_2$ and $\Sigma_3$ in Figure~\ref{fig:F_q_w}).}
\label{fig:f_s_1e-1_T1e-1_unst}
\end{center} 
\end{figure}

\begin{figure}[tbp]
\begin{center}
		\includegraphics[scale = 0.45]{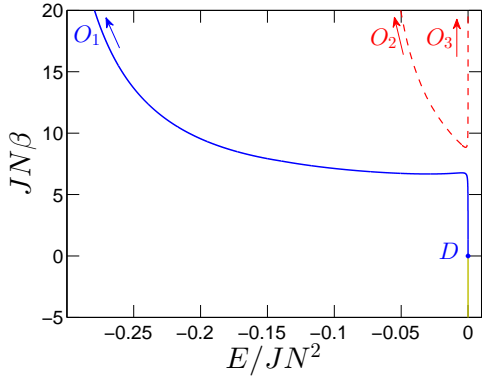} 
\caption{The caloric curve (inverse temperature vs.\ VRR energy) for axisymmetric equilibria in the microcanonical ensemble with $L/(Nl) = 0.1$. States $O_1$, $O_2$, $O_3$ are the ground states referring to the limit $\beta \rightarrow \infty$. The branch that starts at $O_1$ is stable for a one-component system in the microcanonical ensemble all the way down to point $D$ (including the phase transition region highlighted in Figure~\ref{fig:Fbeta_L}), while branches $O_2$ and $O_3$ are unstable. The beige negative-temperature branch $\beta < 0$ that starts at $D$ is stable (see Appendix \ref{s:app:S2}). 
            } 
	\label{fig:caloric_S_micro}
\end{center} 
\end{figure}

\begin{figure}[tbp]
\begin{center}
		\includegraphics[scale = 0.45]{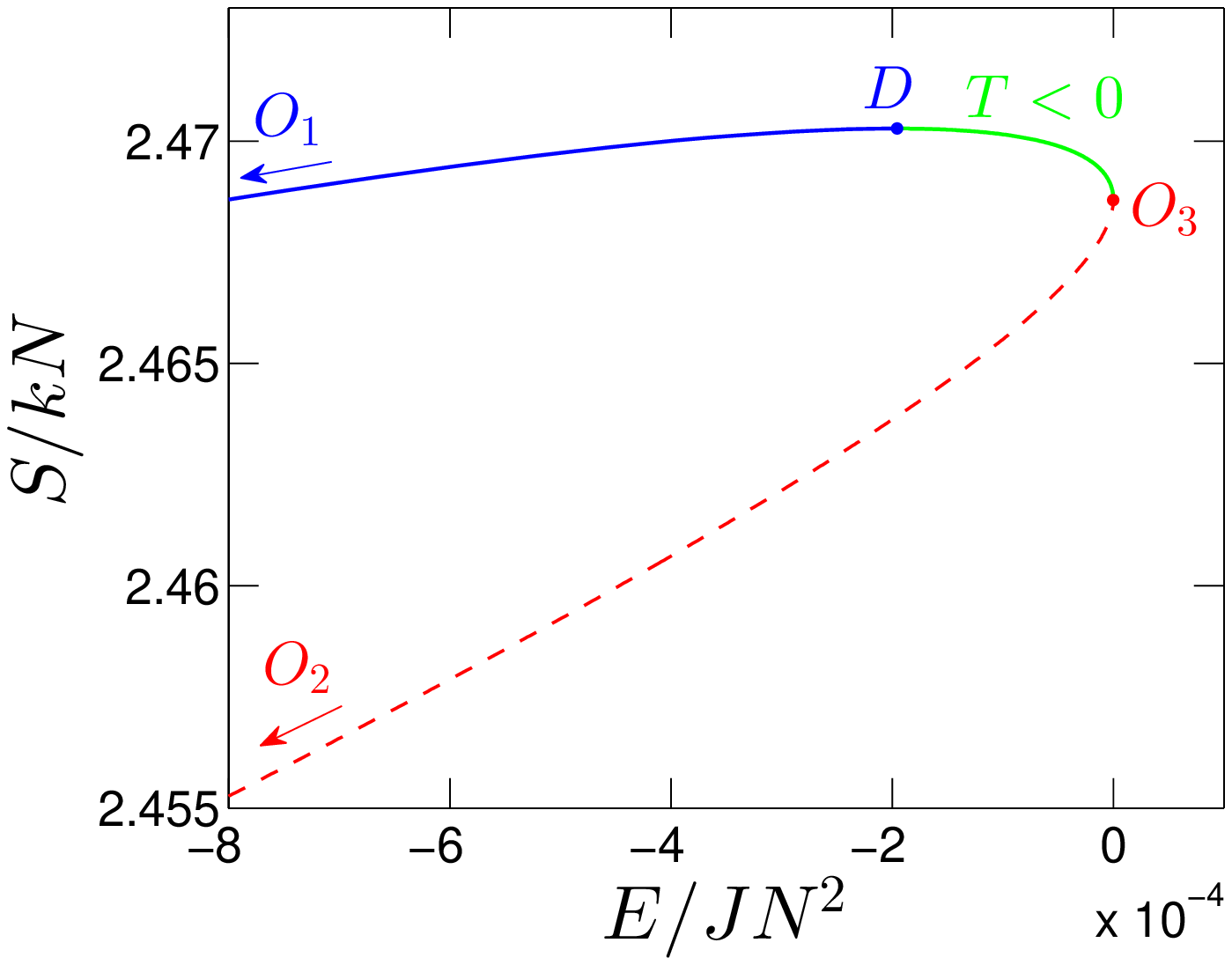}    
		\includegraphics[scale = 0.43]{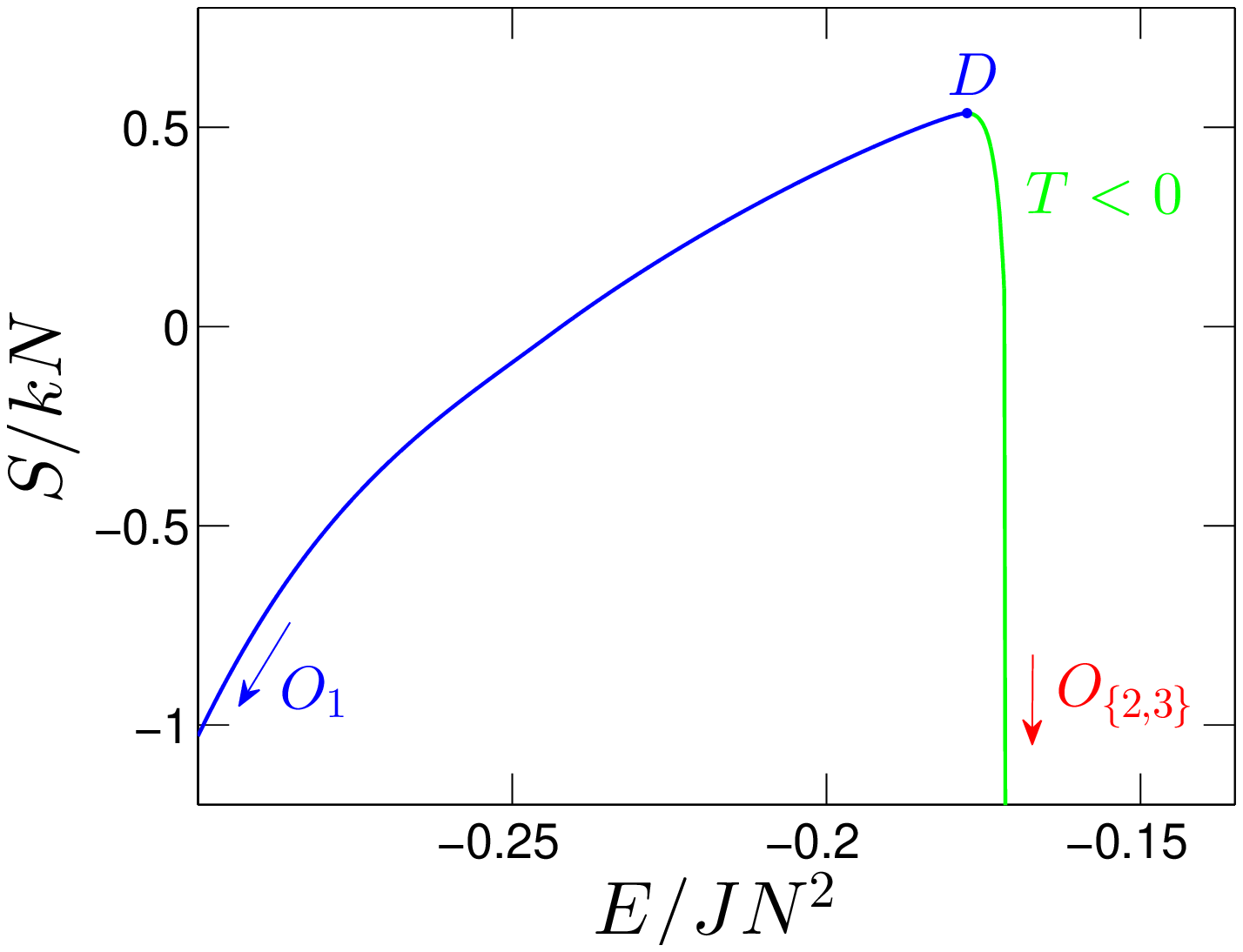} 
\caption{ The entropy with respect to VRR energy for a series of equilibria with total angular momenta $L/(Nl) = 0.2$ and $L/(Nl) = 0.9$ (top and bottom panels, respectively). Its derivative defines the absolute temperature ($\beta=1/kT=\partial S/\partial E|_{L}$) which is positive and negative to the left and right of $D$, shown by blue and green curves respectively. The states $O_3$ and $D$ are the nearly isotropic states at different temperature extremes, $T\rightarrow 0^\pm$ and $T\rightarrow \pm \infty$, respectively (see also Figure \ref{fig:caloric_S_micro}). For $L/(Nl)\geq 1/\sqrt{3}$, $O_3$ coincides with $O_2$, but for $L/(Nl)<1/\sqrt{3}$ they are inequivalent (Appendix~\ref{s:app:partition:rotating}). However, at such high total angular momentum, as in the bottom panel, ($L/(Nl) >1/\sqrt{3}$) there is no unstable branch at positive temperature and the states $O_2$, $O_3$ merge. The green negative temperature branch $DO_3$ is stable (see Appendix \ref{s:app:S2}), while the blue $O_1 D$ branch is stable and the dashed red $O_3O_2$ branch is unstable. 
} 
	\label{fig:S_E_L_9e-1}
\end{center} 
\end{figure}

The first-order phase transition in the canonical ensemble presented in Figure~\ref{fig:F_T_ABmeta_L0} for $L = 0$ is similar for any $0 < L < L_{\rm cr} $. As $L$ is increased, the jumps $Q_{P_{\rm ord}}-Q_{P_{\rm dis}}$ and $E_{P_{\rm dis}}-E_{P_{\rm ord}}$ shrink, and above $L_{\rm cr}$ there is no phase transition. 
This is illustrated in Figure \ref{fig:L_E_Phase_Transition}, which shows 
the region in energy-angular momentum space that corresponds to the phase transition.

On the critical curve $L = L_{\rm cr}$, the phase transition in the canonical ensemble becomes second-order. Figure \ref{fig:phase_trans_L} shows the order parameter with respect to temperature for several values of angular momentum below and above the critical value, and the free energy and the caloric curve are shown in Figure \ref{fig:Fbeta_L}. The \textit{critical point} is labeled with $C$ on the $L_{\rm cr}$ critical curve. In the left panel of Figure \ref{fig:phase_trans_L} we also plot the curve defined by the phase transition points  (dashed line), the ``coexistence curve''\footnote{Although in the one-component model there cannot be phase coexistence in equilibrium, see Section~\ref{s:E_In}.}, which is found empirically to follow a parabolic rule 
\begin{equation}\label{eq:phasetransition-q}
	(Q_{P_{\rm ord}}- Q_{P_{\rm dis}} )^2 = 99 \frac{k}{JN}(T_C-T_P)\,,
\end{equation}
where 
$T_P$ is the temperature of the corresponding first-order phase transition and $T_C$ the temperature at the critical point. The right panel of Figure \ref{fig:phase_trans_L} shows the temperature-angular momentum phase diagram.
In the vicinity of the critical point and on the critical curve, where the phase transition becomes second-order, we get the scaling 
\begin{equation}
	(Q- Q_C )^3 = 
	\frac{k}{JN}\times\left\lbrace
	\begin{array}{ll} 
	2.42 (T_C-T), & Q > Q_{C} \\
	2.18 (T_C-T), & Q < Q_{C} \\
	\end{array}
	\right.
\end{equation}
These scalings are the ones predicted by mean-field theory \citep{stanley1987introduction,papon2002the}
and the critical point as described above is completely analogous to the critical points displayed not only by liquid crystals, but also by ferromagnetic and liquid-vapor systems \citep{stanley1987introduction,Wojtowicz_1974,Muhoray_1982}. The thermodynamic quantities $Q - Q_C$, $T$, $L$ of the one-component quadrupole VRR system discussed here correspond to $S-S_C$, $T$ and $H^2$, respectively, for liquid crystals in a magnetic field (see \citealt{Muhoray_1982} for the definitions), or $M$, $H$ and $T$ for ferromagnets, or $(\rho-\rho_C)$, $P$ and $T$ for liquid-vapour systems (see \citealt{stanley1987introduction}).

The first-order phase transition shown for $L \neq 0$ in Figure~\ref{fig:Fbeta_L} is qualitatively similar to that for $L = 0$ shown in Figure~\ref{fig:Fbeta_L0}. Figure \ref{fig:f_s_rot} (left panel, black curve) shows that the disordered phase is nearly isotropic (for all $L < L_C$ for which the phase transition is defined), while the ordered phase corresponds again to a disk+halo structure in physical space (cf.\ Figure~\ref{fig:f_s_L0} and see Section~\ref{s:VRRequilibrium}). 
In the ordered phase the system consists of a disk with two components, rotating in opposite directions, and a dilute halo. Due to our definition of the coordinate system, more bodies have $L_z > 0$ than $L_z < 0$ as shown by the asymmetry of the curve in the left panel of Figure \ref{fig:f_s_rot}. For high values of $L$ the system is a thin disk rotating in practically one direction only as depicted in Figure \ref{fig:f_s_var_rot}. Figure \ref{fig:f_s_1e-1_T1e-1_unst} shows the distribution function for states along the unstable branches. 

\begin{figure}[tbp]
\begin{center}
		\includegraphics[scale = 0.45]{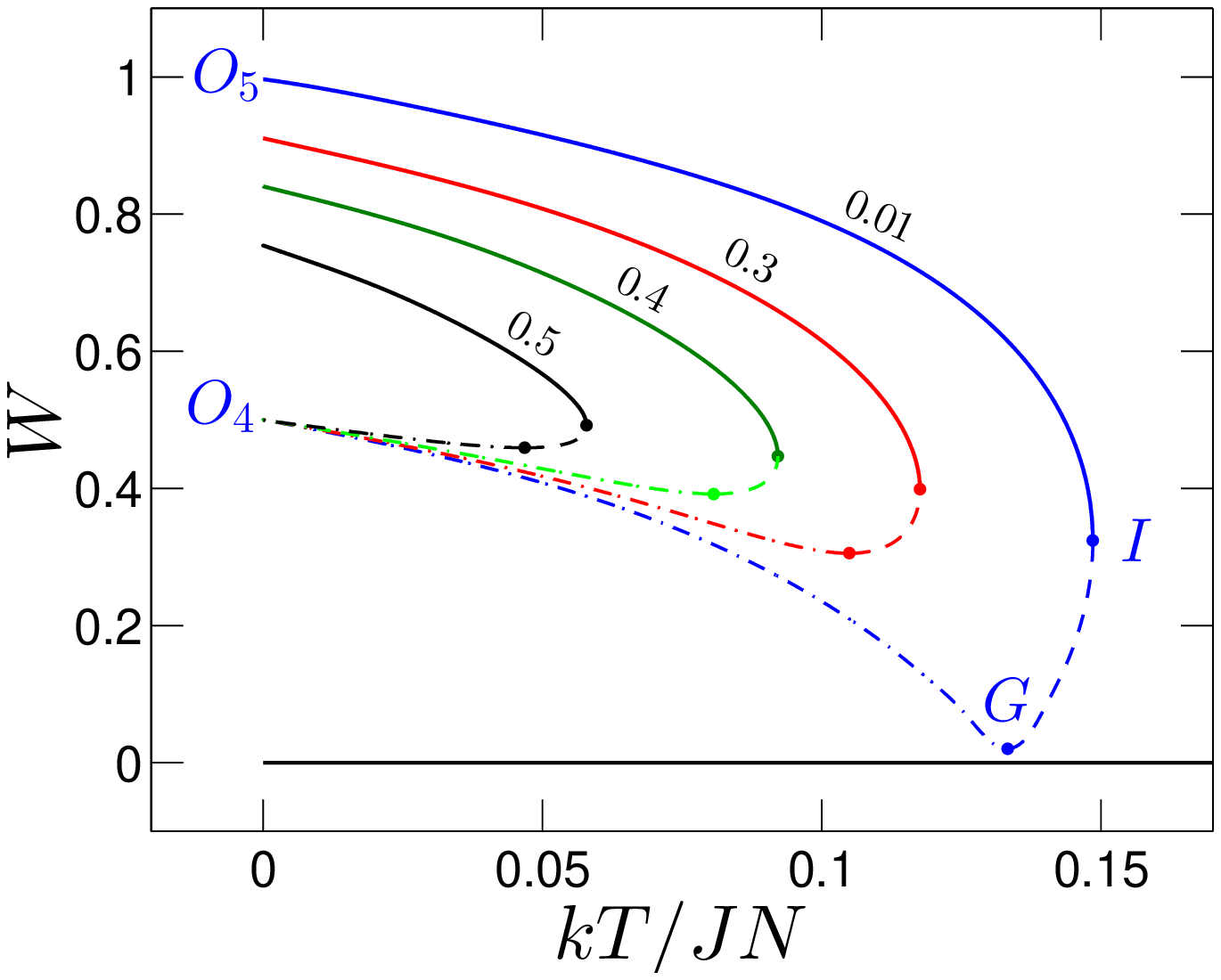} 
		\includegraphics[scale = 0.45]{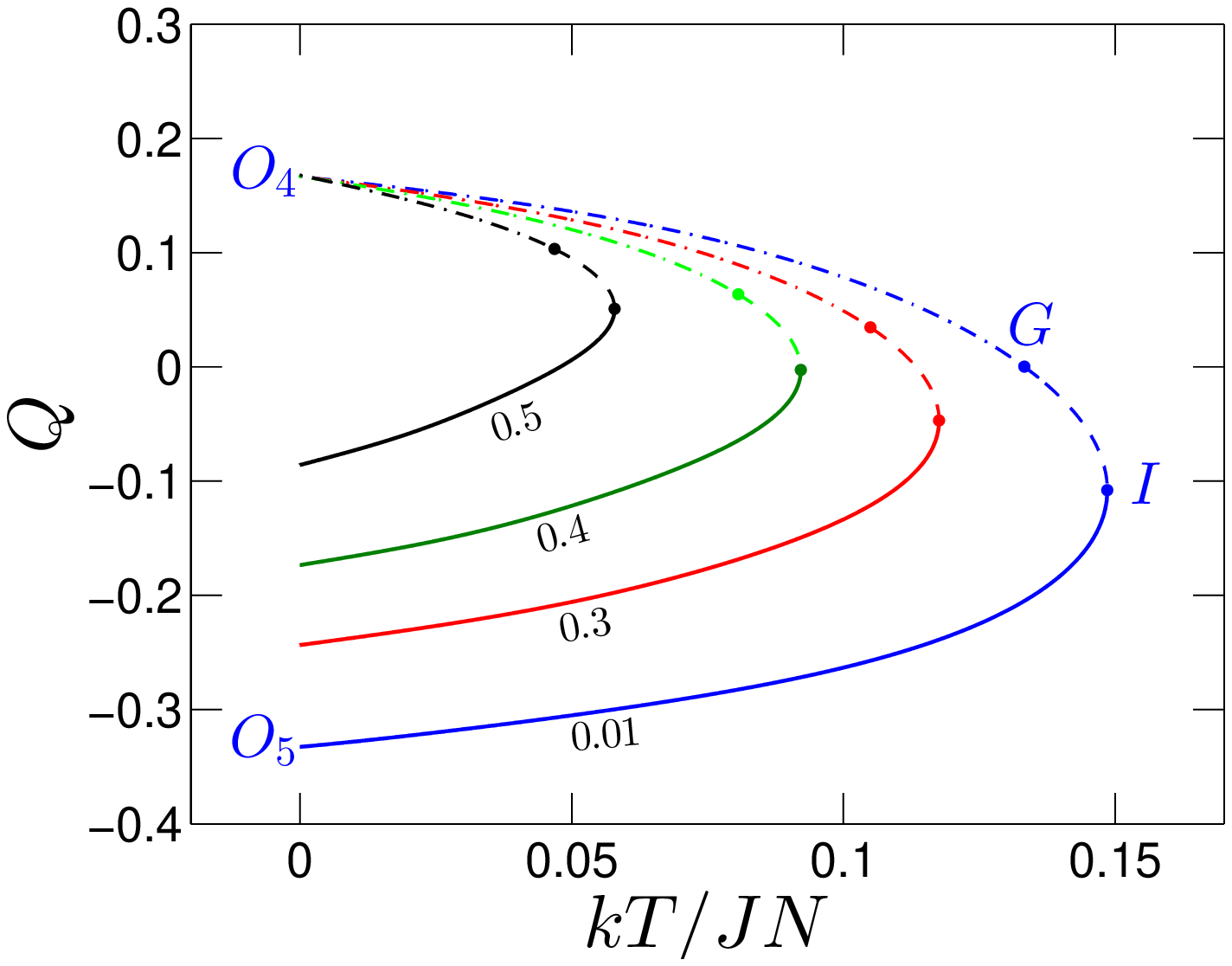} 
	\caption{The $W$  and $Q$ order parameters with respect to temperature for four values of the  dimensionless total angular momentum $\LA s\RA=L/(Nl)$ for the biaxial states as labeled. The equilibria shown are non-axisymmetric, that is, $W \not=0$ and $W \not=\pm3\,Q$ (cf.\ the axisymmetric equilibria shown in Figure~\ref{fig:qomega_T_L}). Solid lines represent metastable equilibria (analogous to $\Sigma_5$ in Figure~\ref{fig:F_q_w}), while dashed curves represent equilibria that are unstable in the canonical ensemble (analogous to $\Sigma_4$ in Figure~\ref{fig:F_q_w}). An analysis similar to that of Figure~\ref{fig:Fbeta_L} shows that in the microcanonical ensemble, the dashed branch $IG$ is metastable, while the dashed-dotted branch $O_4G$ is unstable. The stable axisymmetric equilibria have lower free energy at any temperature and angular momentum as shown in Figure \ref{fig:caloric_F_tr}. At points $I$ and $G$ instabilities arise in the canonical and microcanonical ensembles respectively, and the system transitions to the axisymmetric phase. 
    }
    \label{fig:QW_T_tr}
\end{center} 
\end{figure}

For non-zero angular momentum, there is no phase transition in the microcanonical ensemble, but the system passes continuously from the ordered to the disordered phase as the total VRR energy is slowly increased. The caloric curve is shown in Figure \ref{fig:caloric_S_micro} for the microcanonical ensemble, while the entropy with respect to energy is presented in Figure \ref{fig:S_E_L_9e-1}. 
The equilibrium configurations are identical in the microcanonical ensemble to those in the canonical ensemble apart from the phase-transition region of the canonical ensemble. In both ensembles, the system presents a disk+halo structure at low temperature and a nearly isotropic structure at high temperature for low total angular momentum. However, there are no stable configurations in the phase-transition region in the canonical ensemble between $P_{\rm ord}$ and $P_{\rm dis}$ and the system passes discontinuously from the one phase to the other, while in the microcanonical ensemble the corresponding configurations (branch $P_{\rm ord}ABP_{\rm dis}$ of Figure \ref{fig:Fbeta_L}) are stable.

\subsection{Non-axisymmetric equilibria}\label{s:biax}

In Figure~\ref{fig:F_q_w}, we have seen that VRR free energy extrema exist with $W \neq 0$ and $W\not=\pm3Q$ ($\Sigma_4$, $\Sigma_5$, and their rotations $\Sigma'_4$, $\Sigma'_5$). For $L\not=0$, these states represent non-axisymmetric configurations in the sense that the tensor $\LA n_{\mu}n_{\nu}\RA$ has three distinct eigenvalues (see Eq.~\ref{eq:Qmatrix}). 

\begin{figure*}[tbp]
\begin{center}
    	\includegraphics[scale = 0.45]{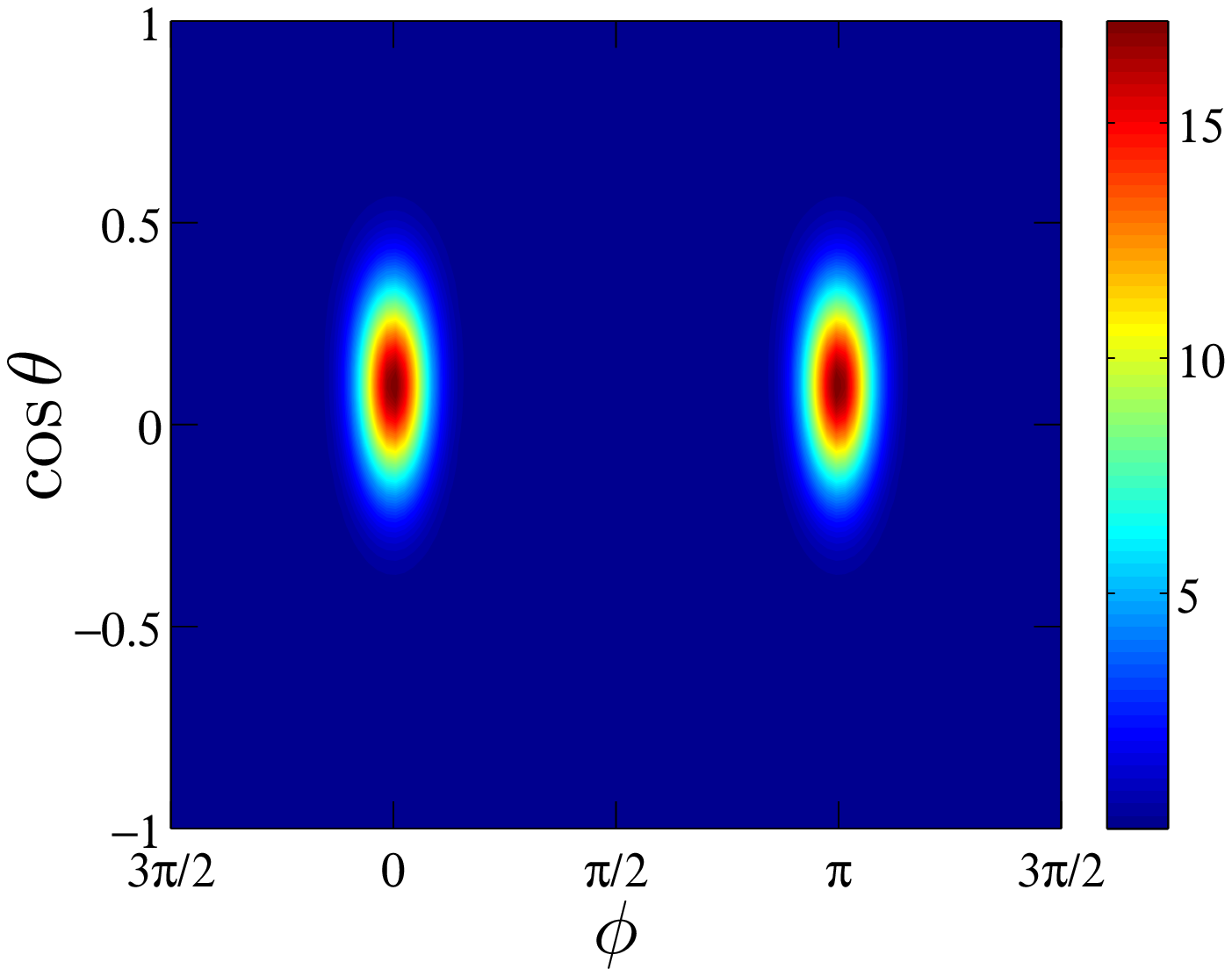} 
    	\includegraphics[scale = 0.45]{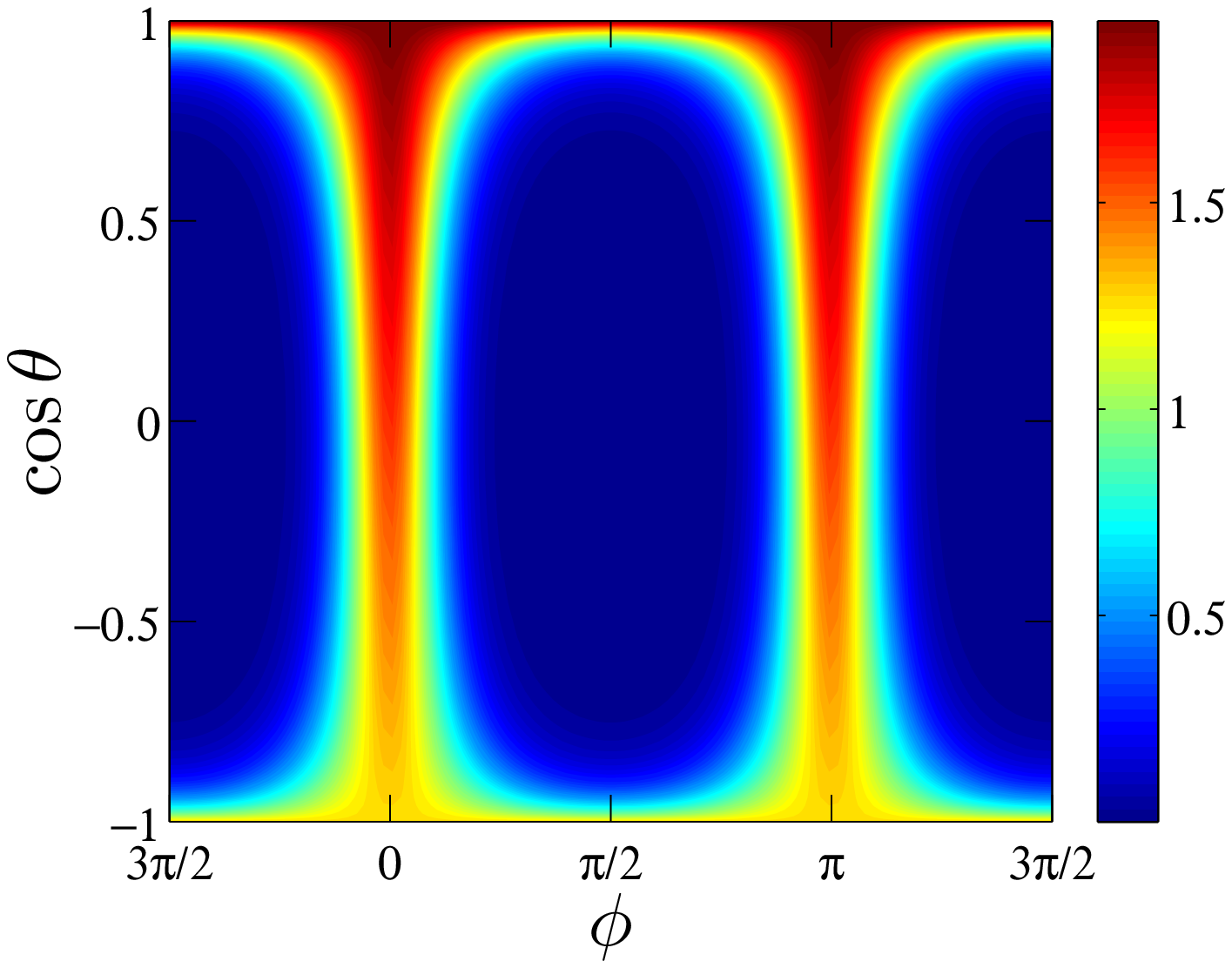} 
    \\
    	\includegraphics[scale = 0.45]{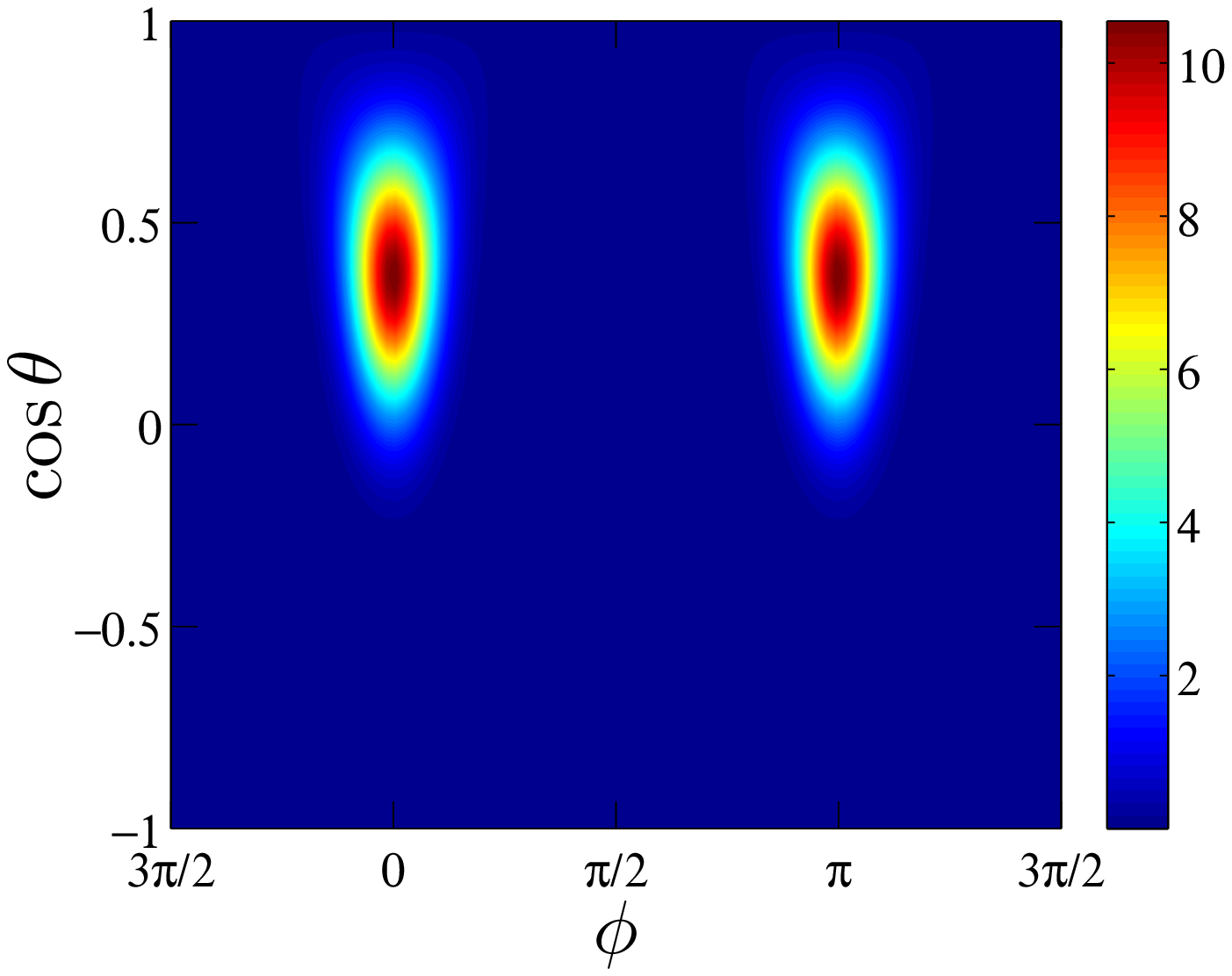} 
    	\includegraphics[scale = 0.45]{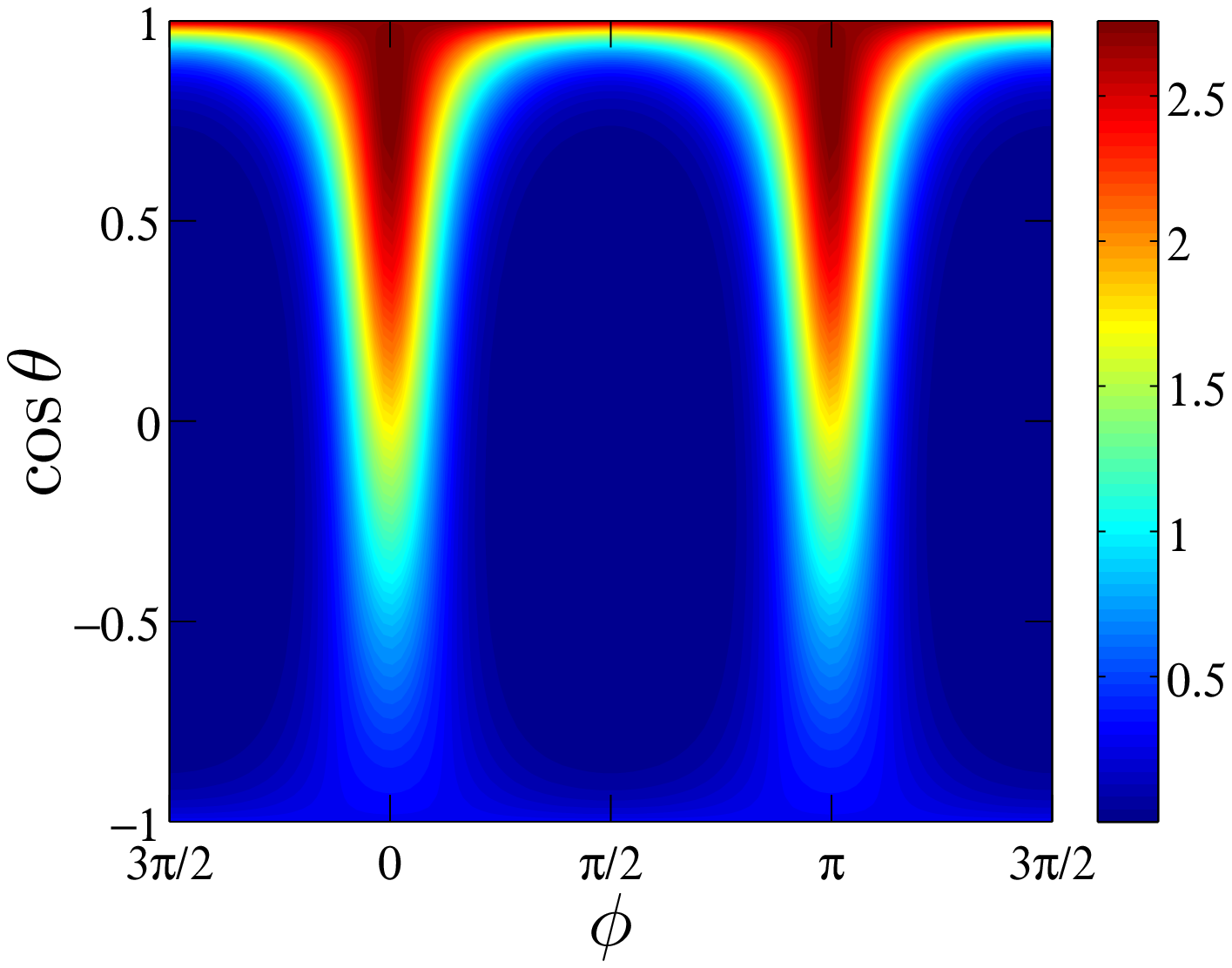} 
    \caption{The probability density distribution function $f(\theta,\phi)/N$ of the orbital angular-momentum vectors for non-axisymmetric equilibria at temperature $kT/(JN) = 0.05$. The left/right panels show the metastable/unstable states which correspond to $\Sigma_5$/$\Sigma_4$ in Figure~\ref{fig:F_q_w}, respectively. The upper/lower panels show equilibria with $\LA s\RA = 0.1$ and $0.4$, respectively. For the metastable state \textit{(left panels)}, the angular-momentum vector directions are distributed around two directions centered at $\cos \theta\sim \LA s\RA$ and $\phi=0$ or $\pi$. These represent two axisymmetric disks in physical space which counterrotate with mutual inclination $\cos \iota \sim 2\LA s\RA^2 -1$. For the unstable states \textit{(right panels)}, the angular-momentum vector directions form a ring or arc along a line of longitude, with maximum density at the north pole (the direction of the total angular-momentum vector).}
    \label{fig:f_QW}
\end{center} 
\end{figure*}

Figure \ref{fig:QW_T_tr} shows the order parameters of the non-axisymmetric equilibria as a function of temperature, for four values of the total angular momentum. For $L/(Nl)\leq 1/\sqrt{2}$ and sufficiently low temperature, there are two equilibria shown by solid and dashed lines, which correspond respectively to $\Sigma_5$ and $\Sigma_4$ in Figure~\ref{fig:F_q_w}. The solid lines represent local free-energy minima which are therefore metastable, and the dashed lines represent free-energy saddle points which are unstable (see Figure~\ref{fig:F_q_w}). 

\begin{figure*}[tbp]
\begin{center}
    	\includegraphics[scale = 0.45]{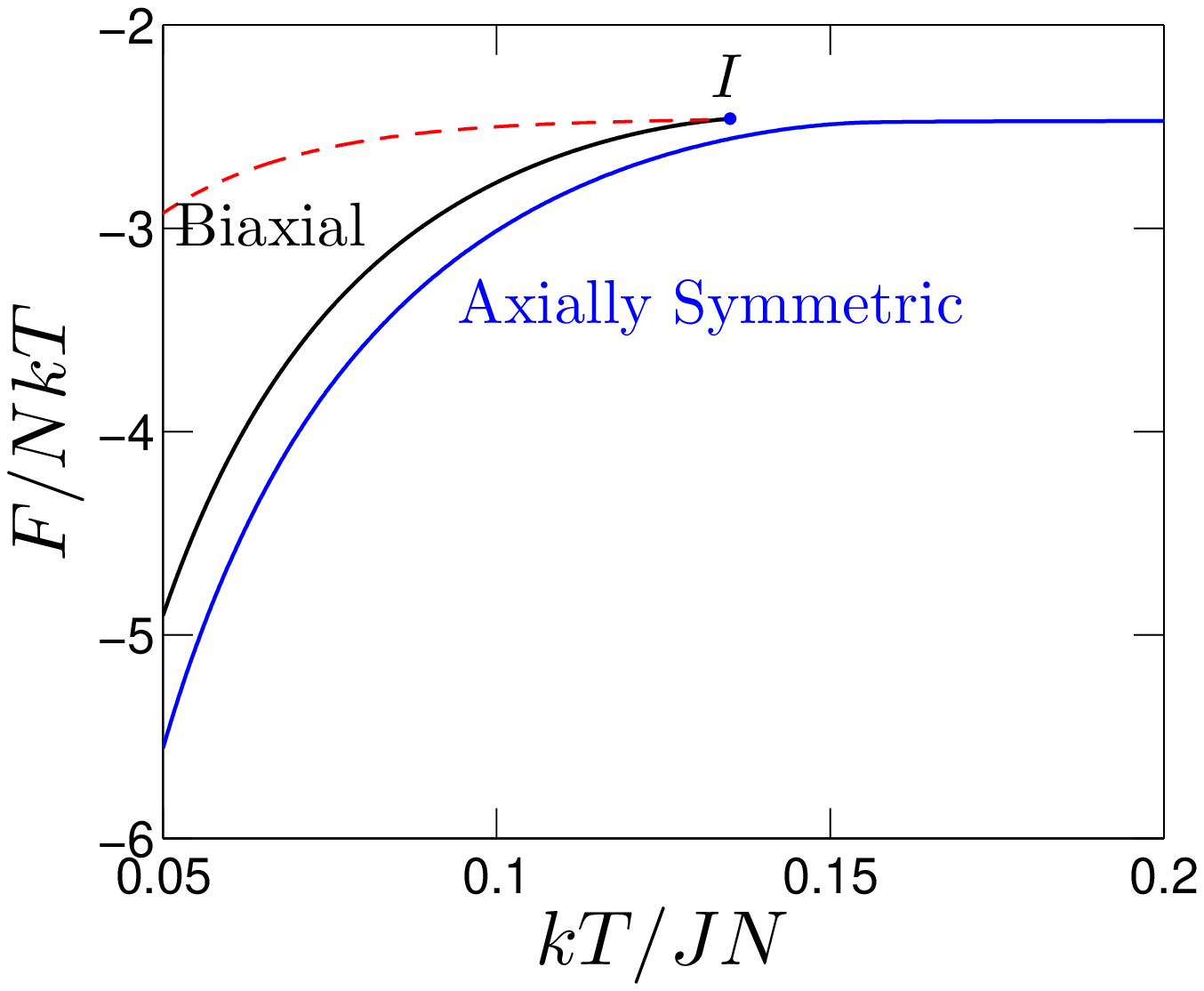} 
    	\includegraphics[scale = 0.45]{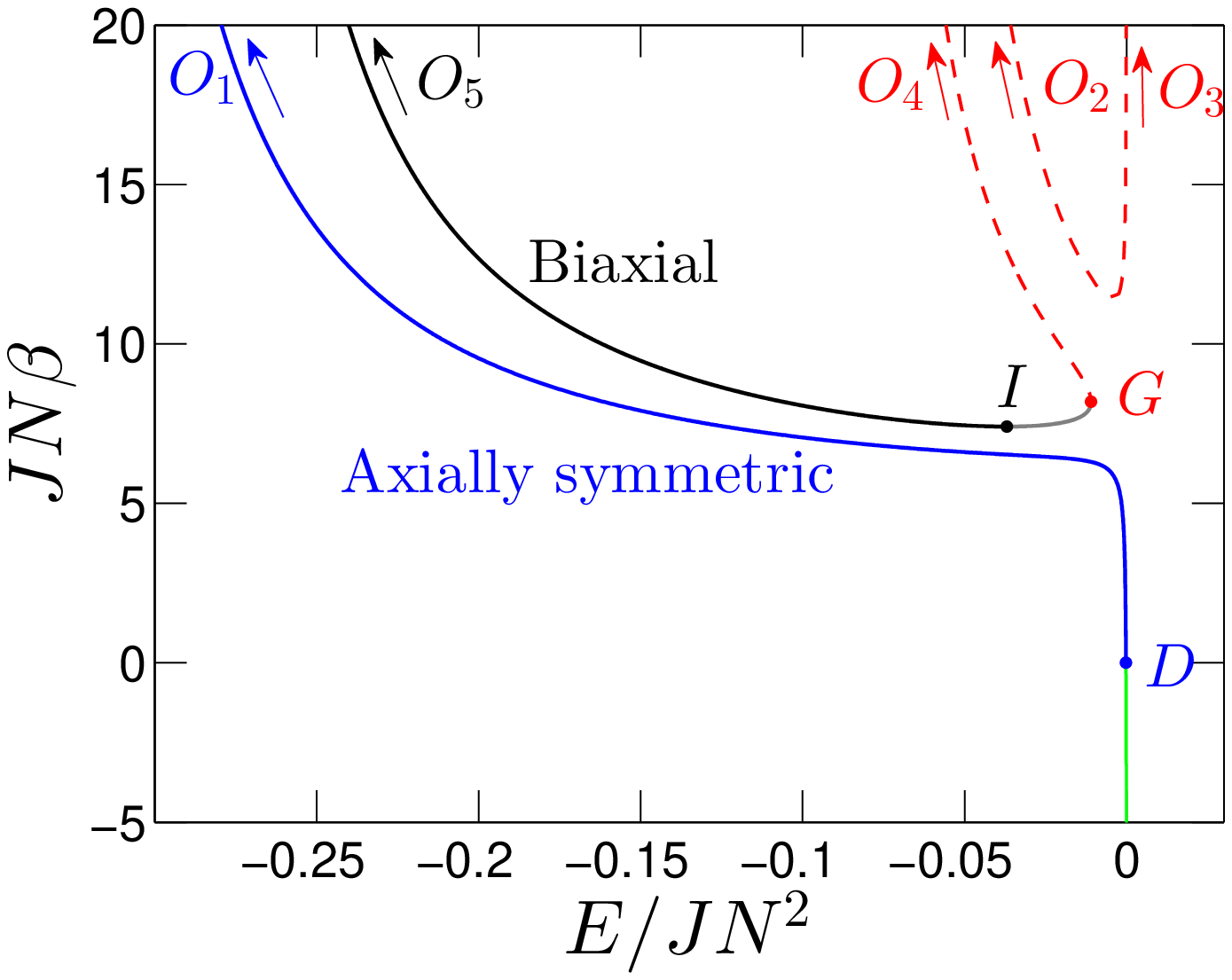} 
    \caption{\textit{Left:} 
    The VRR free energy vs.\ temperature showing non-axisymmetric (biaxial) and axisymmetric series of equilibria with $L/(Nl) = 0.2$ (cf. Figures~\ref{fig:Fbeta_L0} and \ref{fig:Fbeta_L} for axisymmetric equilibria for other $L$ values). At low $T$, the solid black and red dashed curves correspond to $\Sigma_5$ and $\Sigma_4$ in Figure~\ref{fig:F_q_w}, and the solid blue curve corresponds to $\Sigma_1$ (the unstable branches corresponding to $\Sigma_2$ and $\Sigma_3$ in Figure~\ref{fig:F_q_w} are not shown for clarity). There are only axisymmetric equilibria above temperature $T_I$ (see also Figures~\ref{fig:QW_T_tr} and \ref{fig:tr_meta}). 
    \textit{Right:} The caloric curve showing all equilibria for $L/(Nl)= 0.2$ as labeled. Solid curves denote stable or metastable equilibria and dashed curves denote unstable equilibria in both the canonical and microcanonical ensembles, with only exception the part $IG$. For biaxial equilibria, an instability sets in at point $I$ in the canonical ensemble where $|\partial E/\partial \beta|_{\bm{L},N} = \infty$. In the microcanonical ensemble, however, the instability sets in at point $G$ where $|\partial \beta/\partial E|_{\bm{L},N} = \infty$, and the branch $IG$ which has negative specific heat remains metastable. 
    }
    \label{fig:caloric_F_tr}
\end{center} 
\end{figure*}

\begin{figure}[tbp]
\begin{center}
		\includegraphics[scale = 0.45]{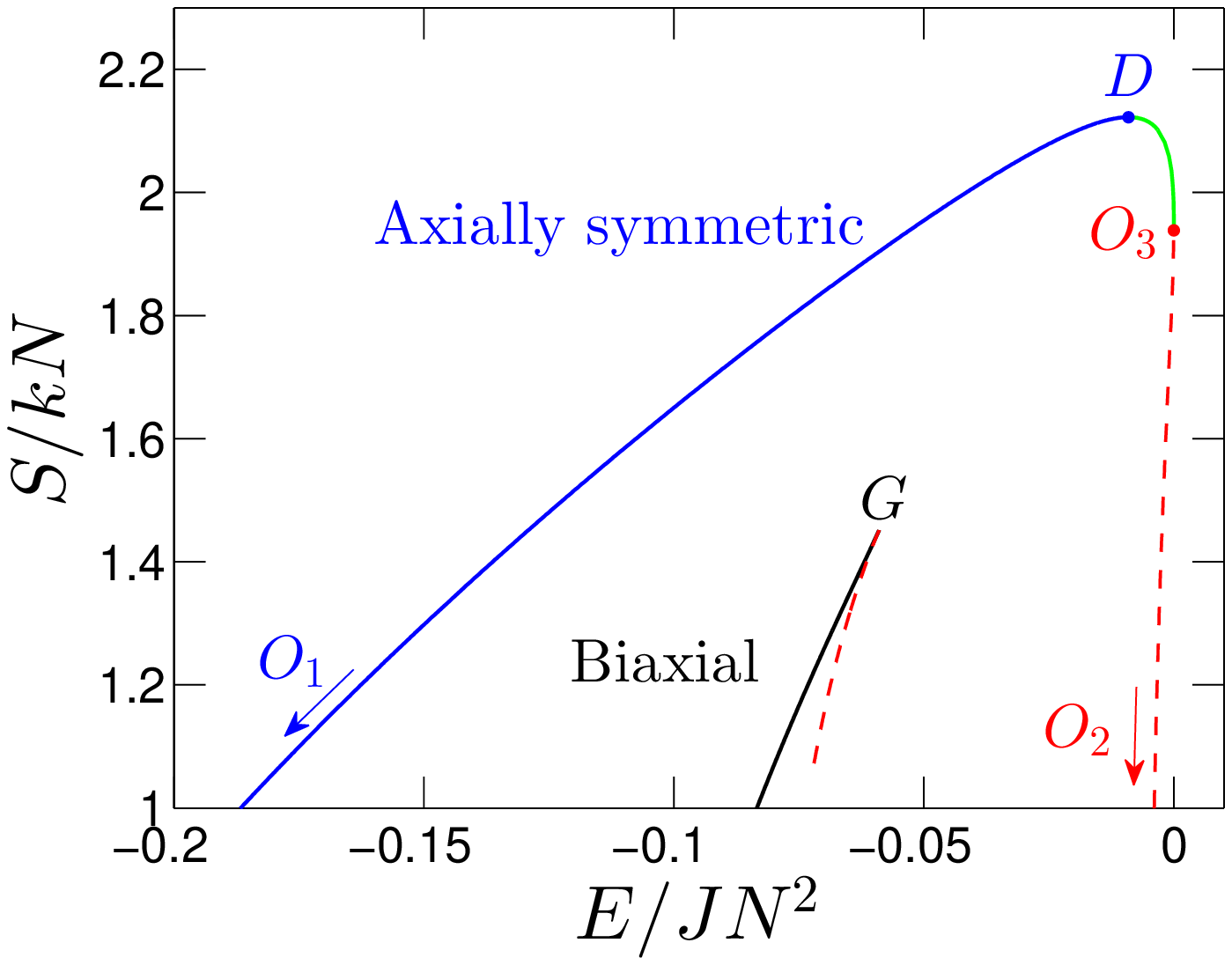} 
\caption{
The entropy vs.\ VRR energy for non-axisymmetric (biaxial) and axisymmetric series of equilibria with $L/(Nl) = 0.5$ 
(cf.\  Figure~\ref{fig:S_E_L_9e-1}). 
The series of non-axisymmetric equilibria shown by the solid black curve represents metastable equilibria in the microcanonical ensemble, which has lower entropy than the corresponding axisymmetric equilibria with the same energy along the 
$O_1D$ branch. Dashed curves show unstable series of equilibria. Clearly, there are no negative temperature non-axisymmetric equilibria since $\beta=\partial S/\partial E|_{L}>0$ (see also Figure~\ref{fig:caloric_F_tr}).} 
	\label{fig:S_E_L_5e-1_tr}
\end{center} 
\end{figure}

Figure~\ref{fig:f_QW} shows the two-dimensional probability density distribution of the orbit normals for the non-axisymmetric equilibria. When the angular-momentum vectors are concentrated near a single direction, as in the left panels, the distribution resembles a disk in physical space in which each object orbits in the same sense. Thus, in physical space, the metastable non-axisymmetric equilibria in the left panels are comprised of two misaligned but otherwise identical disks with axes along $(\cos \theta, \phi) = (\LA s\RA, 0)$ and $(\LA s\RA, \pi)$; half of the bodies are in each disk. The mutual inclination of the disks is approximately
\begin{equation}\label{eq:cosi}
\cos \iota \simeq 2\LA s\RA^2 - 1\,.
\end{equation}
We call these ``biaxial'' equilibria. 

Let us now perform a thought experiment in which we heat the system, and visualize how the distribution of angular momenta changes. At $T=0$, the metastable state is $O_5$ (analogous to $\Sigma_5$ in Figure~\ref{fig:F_q_w}), consisting of two razor-thin disks in physical space with angular momenta pointing along directions $(\theta,\phi)=(\cos^{-1}(L/(Nl)),\pm \pi)$\footnote{This configuration is metastable up to a global rotation of the coordinate system since the maximum entropy equilibrium state does not specify the direction of the eigenvectors of $\bm{Q}$ perpendicular to $\bm{\omega}$.
Integrating Hamilton's equations of motion for the VRR system shows that the two disks precess uniformly around the total angular-momentum vector \citep{Kocsis_2015}.}. As we heat the system, we increase the scatter of the angular-momentum vectors, or in physical space, the thickness of the disks. As the temperature continues to increase, the scatter of the angular-momentum vectors eventually becomes so large that the two angular-momentum direction distributions become connected, forming an inhomogeneous thick arc (or possibly a complete ring for small $\LA s\RA$) in the $x$--$z$ plane with two density maxima. Eventually, above some maximum temperature $T_I$, only axisymmetric states exist. A second non-axisymmetric sequence begins at $T=0$ with a razor-thin arc along the $x$--$z$ plane, which is the unstable $O_4$ ground state analogous to $\Sigma_4$ in Figure~\ref{fig:F_q_w}. As the temperature increases the arc thickens. At point $G$ in Figure \ref{fig:QW_T_tr}, the distribution function becomes bimodal and continues to thicken, until it finally connects with the $O_5$ sequence at $T_I$. 

Figure \ref{fig:QW_T_tr} shows that for any fixed total angular momentum, non-axisymmetric equilibria only exist below a maximum temperature $T_I$ corresponding to point $I$ in the figure. For $T<T_I$ both metastable and unstable configurations are allowed in the canonical ensemble. $T_I$ decreases with increasing $L$ or $\LA s\RA$. These equilibria are similar to\footnote{$O_1$ and $O_5$ are equivalent in the limit $L \rightarrow 0$, and similarly for $O_2$ and $O_4$.} the $O_1$ and $O_2$ branches of zero angular momentum axisymmetric equilibria for $T<T_A$ shown in Figure~\ref{fig:qsz_T_L0}. At this maximum temperature an instability\footnote{
The term ``gravitational zeroth order phase transition'' has been applied to analogous instabilities of the self-gravitating gas in \cite{deVega_Sanchez_2002a} where it is used to describe the collapse of classical Newtonian self-gravitating gas and has been used in \cite{Chavanis_2002} to describe the condensation of a fermionic Newtonian self-gravitating gas.
} 
arises in the canonical ensemble, which is also evident from the caloric curve shown in the left panel of Figure~\ref{fig:caloric_F_tr}, and the system's transition to the $W=0$ axisymmetric phase. Indeed, the left panel of Figure~\ref{fig:caloric_F_tr} shows that the stable axisymmetric equilibrium has lower free energy than the metastable non-axisymmetric equilibrium of the same temperature and angular momentum. 
Similarly, Figure~\ref{fig:S_E_L_5e-1_tr} shows that axisymmetric equilibria have higher entropy than  non-axisymmetric equilibria of the same energy and angular momentum.

\begin{figure*}[tbp]
\begin{center}
    	\includegraphics[scale = 0.45]{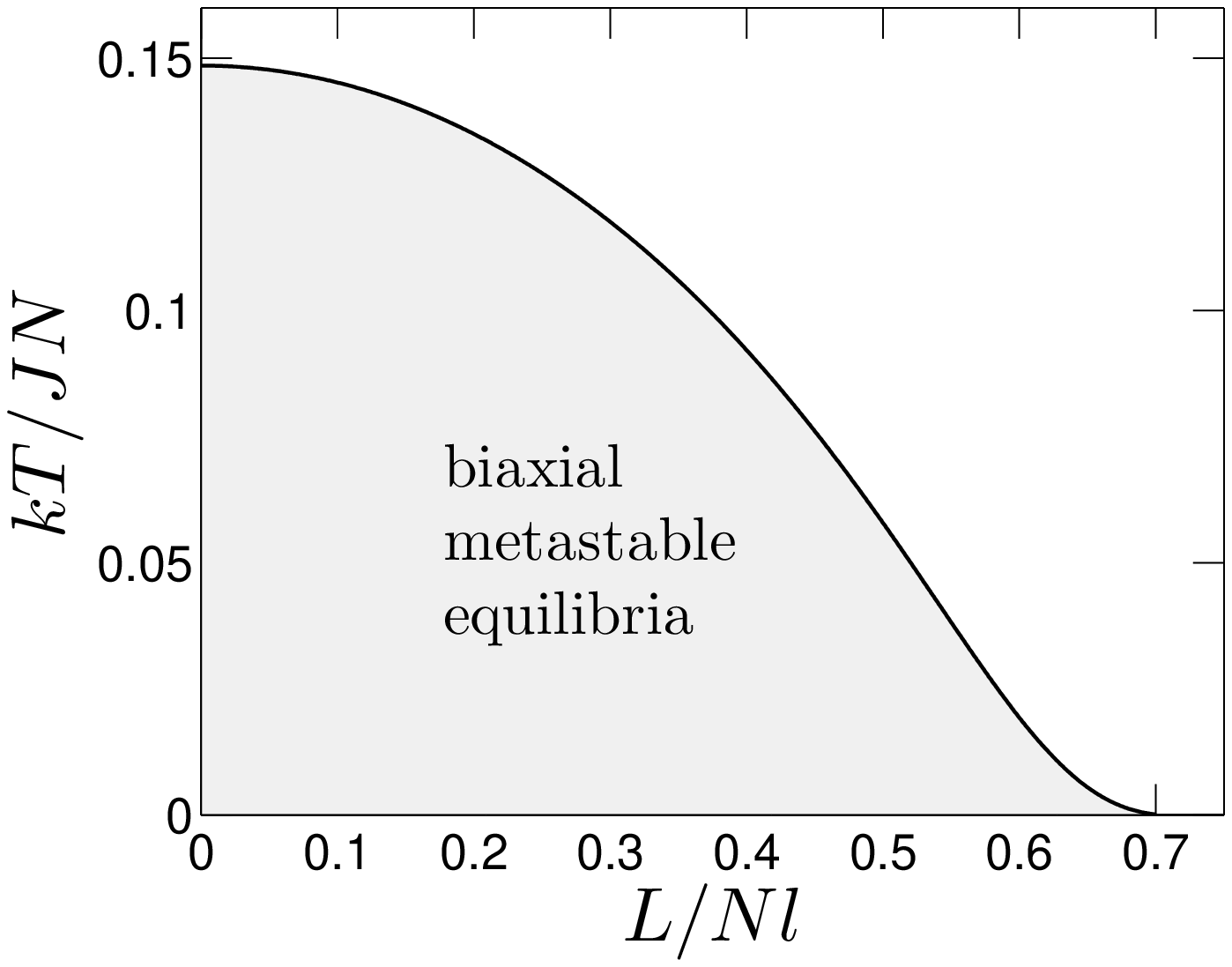} 
    	\includegraphics[scale = 0.45]{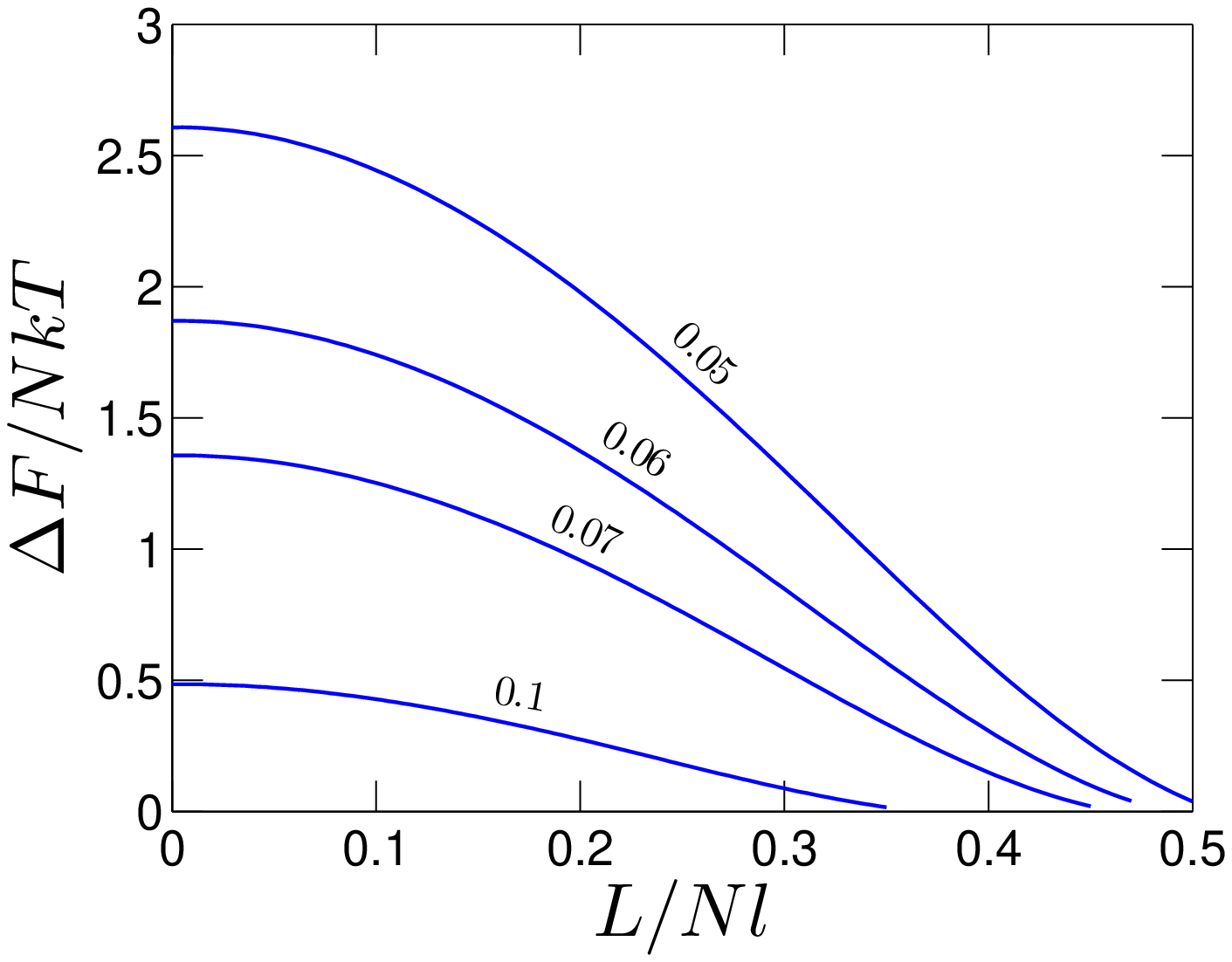} 
    \caption{\textit{Left:} 
    The range of temperature and mean angular momentum which allows biaxial metastable equilibria. The boundary marked by a solid curve corresponds to $I$ in Figures~\ref{fig:QW_T_tr} and \ref{fig:caloric_F_tr}. There are no such equilibria for $L/(Nl) \geq 1/\sqrt{2}$ at any temperature. Physically this bound is set by the condition that the two disks have relative inclination $\iota>90^{\circ}$ seen in Figure~\ref{fig:f_QW}). \textit{Right:} The free energy barrier $\Delta F$ between the metastable biaxial equilibria and the stable axisymmetric ones at the same temperature (i.e., $\Sigma_5\rightarrow \Sigma_4\rightarrow \Sigma_1$ in Figure~\ref{fig:F_q_w}). The numbers above each curve denote the corresponding value of $kT/(JN)$. Biaxial metastable states are long-lived compared to $t_{\rm vrr}$, their lifetime is between $e^{\Delta F/(NkT)} t_{\rm vrr}$ and $e^{\Delta F/(kT)} t_{\rm vrr}$. }
    \label{fig:tr_meta}
\end{center} 
\end{figure*}

The left panel of Figure~\ref{fig:tr_meta} displays the region in the VRR temperature and total angular momentum space where biaxial equilibria exist. The maximum biaxial temperature $T_{I}$ is a decreasing function of $\LA s\RA\equiv L/(Nl)$, and there are no biaxial states above $\LA s\RA=\LA s\RA_{\rm cr}\sim0.7$. The value of $\LA s\RA_{\rm cr}$ can be derived analytically by using the distribution function in angular-momentum direction space shown in Figure~\ref{fig:f_QW}, and exploiting the fact that $T_I=0$ at $\LA s\RA_{\rm cr}$. At $T=0$, we may assume that the bodies are arranged in $K$ razor-thin disks with normals $(\theta_i,\phi_i)$ and $s_i\equiv \cos\theta_i$: 
\begin{align}\label{eq:O5stability1}
f(\bm n) =& \sum_{i=1}^{K} N_i \delta(s - s_i) \delta(\phi-\phi_i)\\
E =&\ffrac16 J N^2 -\ffrac12 \sum_{i,j=1}^{K} J N_i N_j (\bm{n}_i\cdot\bm{n}_j)^2,\\
L =& \sum_{i=1}^{K} N_i s_i,\\
\bm{n}_i&=\Big(\sqrt{1-s_i^2}\cos\phi_i\,,\;\sqrt{1-s_i^2}\sin\phi_i\,,\;s_i\Big)\,.
\end{align}
In particular, the biaxial $O_5$ state has $K=2$ with 
\begin{align}
N_{1}=N_{2}=\ffrac12N, \quad s_{1}=s_{2}=\frac{L}{Nl},\quad \phi_{1}=0\,,\;\phi_{2}= \pi\,.
\label{eq:O5stability2}
\end{align}
To determine whether this state is stable or unstable to splitting each peak of the distribution function 
to two peaks centered at slightly smaller and larger $s_i$, we examine if the energy of those 
configurations is lower or higher than that of $O_5$ using Eqs.~(\ref{eq:O5stability1}--\ref{eq:O5stability2}).
At fixed $N$, $L$, and $T=0$, we find that these perturbations result in an energy decrease exactly if $L/(Nl) \geq 1/\surd{2}$ corresponding to $\LA s\RA_{\rm cr}=1/\surd{2}$. Hence the mutual inclination between the two axes in the zero-temperature biaxial state must be larger than $90^{\circ}$ for metastability. We conjecture that $O_5$ is metastable at higher inclinations and smaller $\LA s\RA$ against arbitrary perturbations of the distribution function\footnote{We find stability against a variety of perturbations including more general bifurcations into four modes with $N_i\rightarrow N_i+\delta N_i$, $\bm{n}_i = \bm{n}_i+\bm{\delta n}_i$ for $i\leq 4$ where $\sum_i \delta N_i = 0$ and for the additional modes $N_3=N_4=0$ with $\delta\bm{n}_i$ arbitrary.}. 

A rough estimate of the lifetime of the biaxial metastable states is obtained by assuming that the transition probability for each body in a unit VRR relaxation time $t_{\rm vrr}$ (see Section~\ref{s:Intro}) is of order $\exp[\Delta F/(NkT)]$, where $\Delta F$ is the free-energy barrier between the metastable biaxial state $\Sigma_5$ and the stable axisymmetric state $\Sigma_1$. To find $\Delta F$, note that the transition from $\Sigma_5$ to $\Sigma_1$ requiring the least free energy passes through $\Sigma_4$, the unstable non-axisymmetric state (see Figure~\ref{fig:F_q_w}). In the most conservative estimate \citep{Chavanis_2005}, the $\Sigma_5\rightarrow \Sigma_4 \rightarrow \Sigma_1$ transition occurs when all $N$ objects climb the barrier coincidentally during their thermal motion, which happens in a time 
\begin{equation}
	t_{\rm meta} \sim [e^{\Delta F/(NkT)}]^N t_{\rm vrr}\ \mbox{where}\ \ \Delta F\equiv F(\Sigma_4)-F(\Sigma_5).
\end{equation}
The right panel of Figure~\ref{fig:tr_meta} shows $\Delta F/(NkT)$. Based on this simple estimate,  the lifetime of the biaxial metastable states increases exponentially with $N$ (recall that at fixed values of the order parameters and coupling constant $F\sim N^2$ and $T\sim N$), and thus the metastable state is very long-lived for $N\gg 1$. In the opposite extreme, the transition to the $\Sigma_1$ state could occur gradually with different objects jumping over the barrier in succession.  This happens over time $\exp[(\Delta F/(NkT)] t_{\rm vrr}$, which is much shorter. Numerical simulations are needed to determine the correct scaling with $N$. 

\section{\texorpdfstring{\boldmath$\omega TN$}--ensemble}
\label{sec:g_canonical}

\begin{figure*}[tbp]
\begin{center}
    	\includegraphics[scale = 0.45]{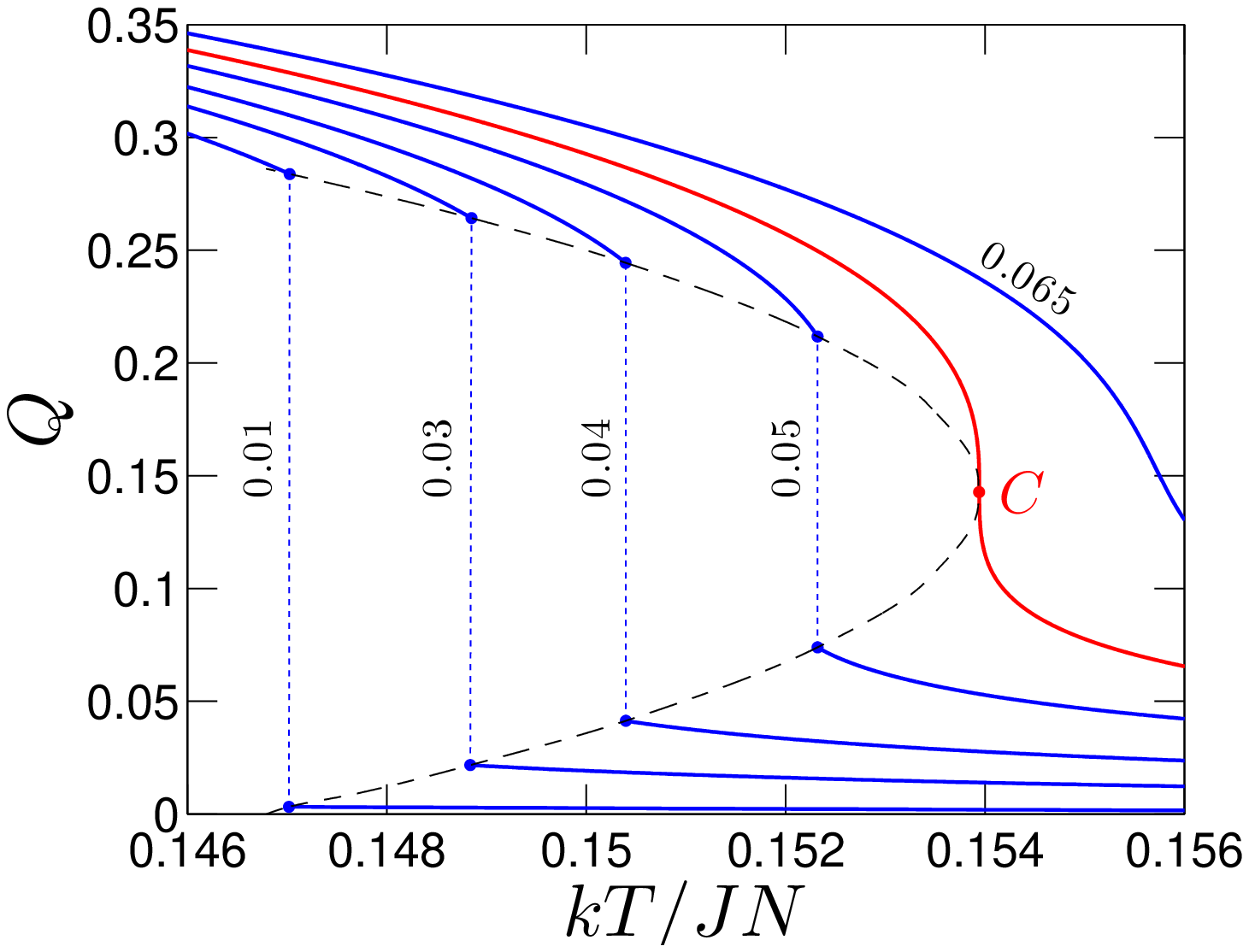} 
    	\includegraphics[scale = 0.45]{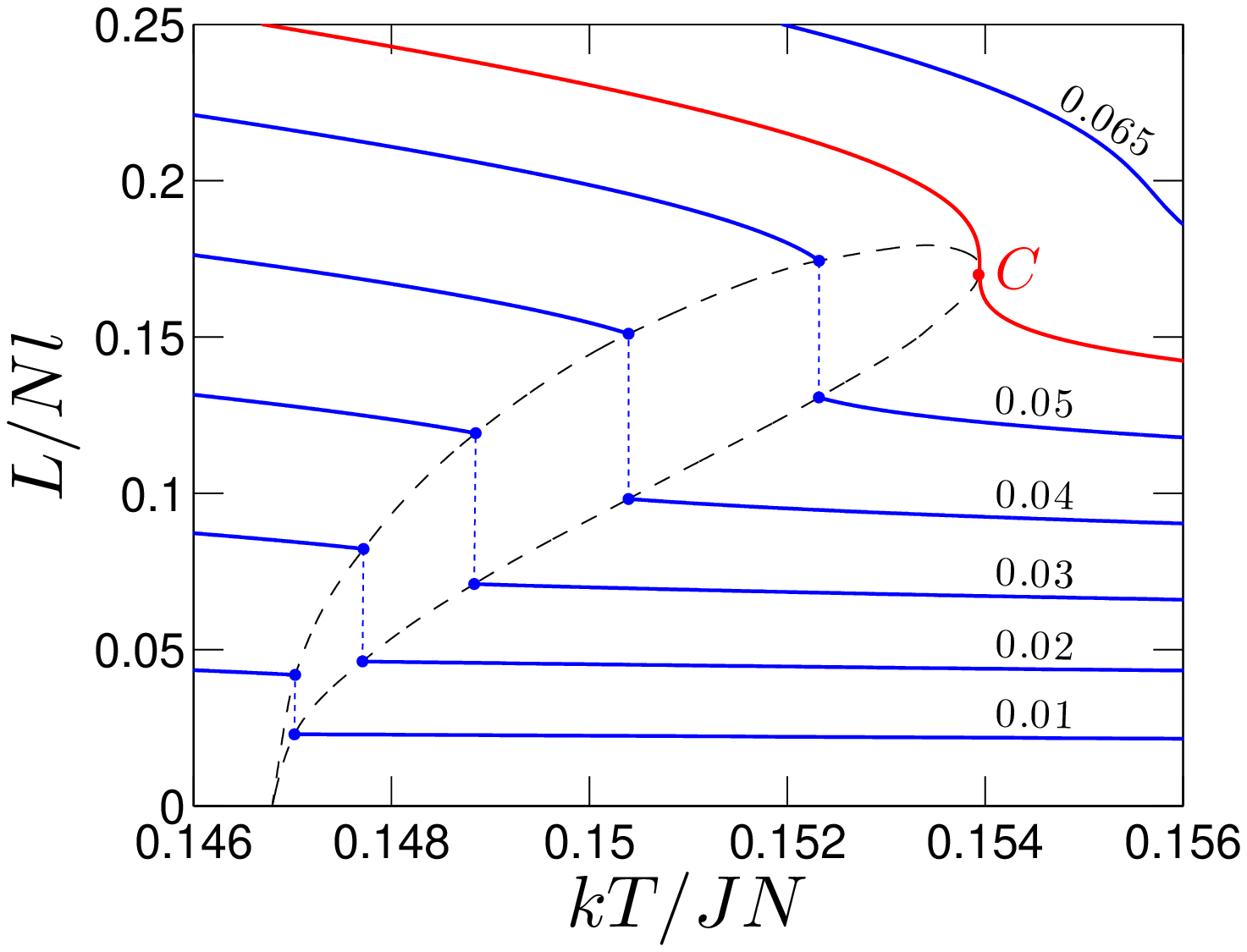} 
    \caption{The order parameter \textit{(left)} and total angular momentum \textit{(right)} for axisymmetric states in the $\omega TN$-ensemble, in which the system may exchange angular momentum and energy with the environment. 
    The contour labels denote the value of $\omega l/(JN)$ which is fixed along each contour (cf.\ Figure~\ref{fig:phase_trans_L} for the canonical ensemble). The system undergoes a first-order transition for temperatures corresponding to each dotted line for $\omega < \omega_C = 0.057285JN/l$. A system that is heated slowly from zero temperature jumps through an out-of-equilibrium process from the ordered nematic phase at high $Q$ and $L$ to the disordered phase at lower $Q$ and $L$ at the phase-transition temperature. 
    At the critical line $\omega_C$, marked in red, the phase transition becomes second-order, and for $\omega>\omega_C$ there is no phase transition. 
    }
    \label{fig:gensemble_phT}
\end{center} 
\end{figure*}

\begin{figure*}[tbp]
\begin{center}
    	\includegraphics[scale = 0.45]{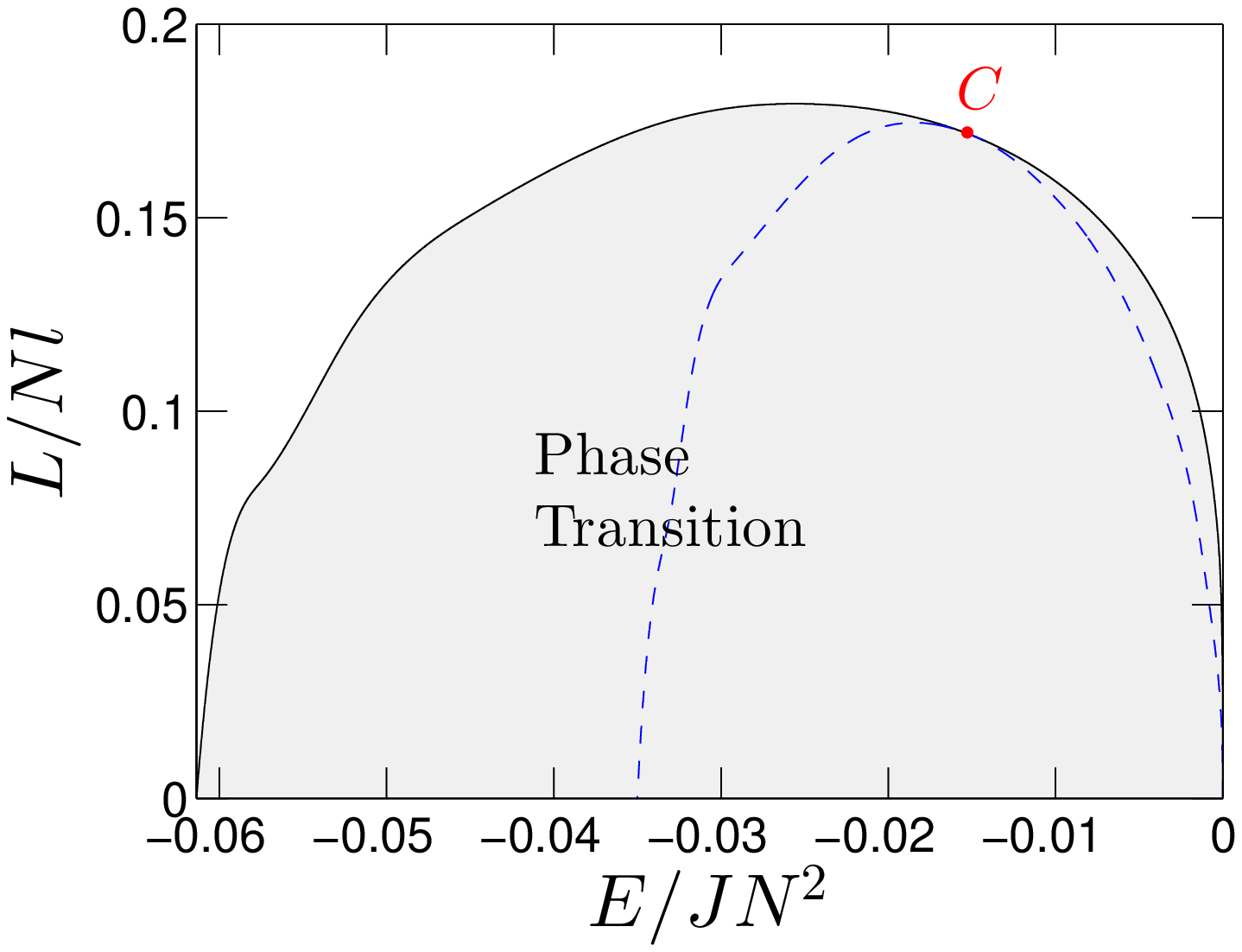} 
    	\includegraphics[scale = 0.45]{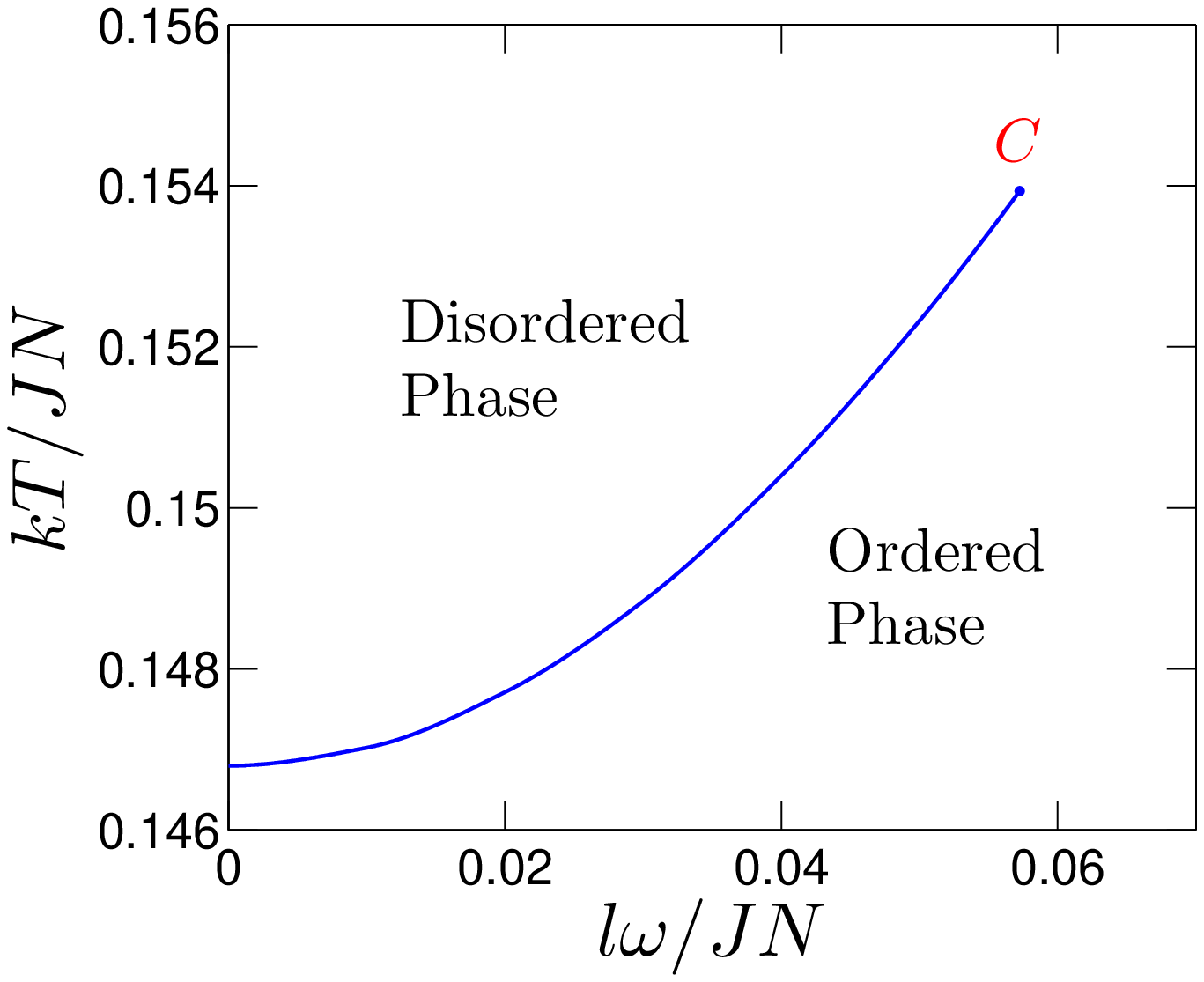} 
    \caption{\textit{Left:} The phase-transition region for the $\omega TN$-ensemble. \textit{Right:} The phase-transition temperature with respect to the rotation parameter $\omega$ in  the $\omega TN$-ensemble. At the critical point $C$, where the phase transition becomes second-order, we have $\omega_C = 0.057285 JN/l$, $T_C = 0.15394 JN/k$, $L_C = 0.17173 Nl$ and $Q_C = 0.14273$.}
    \label{fig:gensemble_Omega}
\end{center} 
\end{figure*}

In Section \ref{s:VRRequilibrium} we introduced the $\omega TN$-ensemble. The system is assumed to be in a heat bath, a much bigger system that surrounds our small one-component system, with which it can exchange VRR-energy and angular momentum but not bodies. Thus in this ensemble the temperature $T$, the number of bodies $N$, and the rotation parameter $\bm{\omega}$ (Eq.\ \ref{eq:omega}) conjugate to angular momentum are held fixed. 
The generalized thermodynamic potential of this ensemble is given in equation (\ref{eq:G_vrr_can_1}),
$
	G(\bm{\omega},T,N) = E - \bm{\omega} \cdot \bm{L} - TS.
$
We focus our attention to the axisymmetric case here. We derive a parametric solution to the self-consistency equations in Appendix~\ref{s:app:partition:rotating}.
Similar to the canonical ensemble, we find a phase transition, but the values of the order parameters at the transition points are modified as shown in Figure~\ref{fig:gensemble_phT} (cf.\ Figures~\ref{fig:phase_trans_L} and \ref{fig:L_E_Phase_Transition}). A first-order phase transition occurs for $\omega < \omega_C$; the transition becomes second-order for $\omega_C = 0.057285 JN/l$. This behavior resembles the canonical ensemble but the critical values are different with $T_C = 0.15394 JN/k$, $L_C = 0.17173 Nl$ and $Q_C = 0.14273$ as depicted in Figures \ref{fig:gensemble_phT} and \ref{fig:gensemble_Omega}.

\section{Separable multi-component systems}

\label{sec:multi}

Let us define a \textit{separable multi-component system} to be a system in which the energy of each individual component is dominated by its self-energy and the small interaction between components leads to thermodynamic equilibrium among different components.
More specifically, we require that $V^{(i)}_{\mu\nu}\approx J_{ii}Q^{(i)}_{\mu\nu}$ for all $i$ in Eq.~(\ref{eq:epsiloni_Q}), which is typically\footnote{This does not hold if $Q^{(i)}_{\mu\nu}\lesssim \sum_j N_{j} J_{ij} Q^{(j)}_{\mu\nu}/(N_i J_{ii})$ for some $i$.} satisfied if 
\begin{equation}\label{eq:J_sep}
	J_{ii} N_i \gg \sum_{j\neq i} J_{ij} N_{j} > 0,\quad \mbox{for all }i. 
\end{equation}

In such a system each individual component is subject to the one-component model developed above and thus the analysis (including all of the figures) applies to each one of them. In the special case where the mutual interaction of a particular component $i$ with all other components is exactly zero then this component behaves as an isolated system and it may have a different temperature $T_i$ from the rest of the system in equilibrium in a microcanonical ensemble. Otherwise, if the mutual interactions between components are small but non-zero, then different components settle at a common temperature $T_i=T$ and $\gamma_i=\gamma$.
However note that for each component, the equilibrium distribution function (Eq.~\ref{eq:f_qw}) depends explicitly on the dimensionless quantities
\begin{equation}\label{eq:tau}
\tau_i = \frac{kT}{J_{ii}N_i} \mbox{  and  } c_i = l_i\gamma = \frac{l_i \omega_i}{kT}
\end{equation}
as shown in the figures above. 
Thus, different components may exhibit different phases (uniaxial, biaxial, or disordered) in equilibrium, depending on their respective values of $J_{ii}N_i$ and $c_i$ at the same $T$ and $\gamma$.

In the special case where $J_{ii}N_i=J_{jj}N_j$ for all $i$ and $j$ and either $L=0$ or $l_i=l_j$ for all $i$ and $j$, all components' phase transition occurs at the same temperature. Then depending on the microscopic initial conditions, some components lie in the ordered and the rest in the disordered phase. In this sense, during the phase transition, different phases coexist.

In multi-component separable systems with radially non-overlapping components, those with larger values of $J_{ii}N_i \propto N_i m_i^2/a_i$ have smaller $\tau_i$ and therefore form thinner disks than components with lower $N_im_i^2/a_i$
(see Figures \ref{fig:qsz_T_L0} and \ref{fig:qomega_T_L}). In particular, if the radial number density follows $n\propto a^{-\gamma}$  and the stellar mass distribution is independent of radius then $N_i \propto a_i^{3-\gamma}$ and 
$\tau_i \propto N_i m_i^2/a_i \propto a_i^{\gamma-2}$ (see Eq.~\ref{eq:tau}). Thus if the radial number density is steeper than $a^{-2}$, as it is for the massive stars in the Galactic center \citep{Bartko_2009,Yelda_2014}, then the fractional thickness of the equilibrium disk increases outwards (i.e., the disk flares) and the distribution may be nearly isotropic (disordered) beyond a transition radius. Conversely, if the number density is shallower than $a^{-2}$ then the disk thickness increases inwards and the system is nearly isotropic inside some transition radius. Similarly, higher mass objects form thinner disks. We emphasize however that the multi-component system observed in the Galactic center is probably non-separable and higher order multipoles beyond the quadrupole are likely to play a significant role in the dynamics. We leave further study of multi-component systems to future work. 
 
\section{Conclusions}\label{s:con}

In self-gravitating systems subject to a dominant central potential, such as that of a massive central object, the bodies typically follow bounded planar orbits about the center. Due to rapid in-plane precession, each planar orbit can be represented by an axisymmetric surface density profile (over timescales $t$ given by $t_{\rm in-plane} \ll t \ll t_{\rm vrr}$, see Eq.\ \ref{eq:hierarchy}). On longer timescales the angular-momentum directions, which describe the orientation of the orbital planes, relax and attain thermal equilibrium long before other degrees of freedom do (see Section~\ref{s:Intro}). This process is called vector resonant relaxation (VRR).
We determined the mean-field thermodynamic VRR equilibrium states using the quadrupole approximation for the gravitational interactions between orbits. 
For a list of our assumptions see the introduction to Section \ref{s:mean_vrr}.
The equilibria exhibit a remarkable variety of behavior including three qualitatively different VRR-phases:

\begin{enumerate}[(i)]

 \item a \textit{uniaxial ``nematic'' ordered phase} which represents a disk in physical space containing both prograde and retrograde orbits, surrounded by a dilute halo (Figures~\ref{fig:f_s_rot} and \ref{fig:f_s_var_rot});
 
 \item a \textit{biaxial ``nematic'' ordered phase} consisting of two disks in physical space; the two disks have the same thickness and mass and their mutual inclination is between $90^{\circ}$ and $180^{\circ}$ (Figure~\ref{fig:f_QW});
 
  \item an \textit{axially symmetric disordered phase} representing a nearly isotropic distribution of orbits in both physical space and angular-momentum space (Figure~\ref{fig:f_s_rot}).
  
\end{enumerate}

The term ``nematic'' highlights the analogy with liquid crystals, where the molecules specified by their symmetry axes $\bm{n}_i$ are concentrated towards a symmetry axis $\hat{\bm{s}}$, in the sense that $\bm{n}_i$ is preferentially parallel or antiparallel to $\hat{\bm{s}}$. In both liquid crystals and the quadrupole VRR model the Hamiltonian is invariant under the inversion $\bm{n}_i\to -\bm{n}_i$, but the direction of the total angular momentum $\bm{L}$ in the VRR model breaks this symmetry. 

The ordered and disordered phases are stable at low and high temperatures, respectively\footnote{See definition of temperature below Eq.~(\ref{eq:deltaS_L1}.}. 
The biaxial ordered phase, consisting of two highly inclined disks, requires the temperature to be limited to the range shown in Figure~\ref{fig:tr_meta} and the mean angular momentum to be limited to  $L \leq Nl/\surd 2$ where $l$ is the angular momentum of one object (Eq.~\ref{eq:l}). Biaxial states are metastable in both the canonical and microcanonical ensembles (i.e., in the canonical [microcanonical] ensemble they are a local minimum of the VRR free energy [negative of the entropy] but this minimum is larger than that of axisymmetric ordered states with the same temperature [energy]). Nevertheless, for practical purposes the lifetime of the two-disk state may be very long, especially if the disks are thin.

The canonical ensemble exhibits a first-order gravitational phase transition between the axisymmetric ordered and disordered states 
for a limited range of total angular momentum $L < 0.137\, N l$ (where $l$ is the magnitude of the angular-momentum vector of a single object, see Figure~\ref{fig:phase_trans_L}). The phase-transition temperature $T_P$ depends on the total angular momentum ($0.146 \leq kT_P(L)/(JN) \leq 0.151$). The disordered state is only stable if either $T>T_P$ or $T<0$. Negative-temperature states are allowed due to the fact that the VRR energy has a maximum possible value, which corresponds to the isotropic distribution for $L/Nl<1/\surd 3$ (see Section~\ref{s:neg_T} and Figures~\ref{fig:S_E_L_9e-1} and \ref{fig:S_E_L_5e-1_tr}). 

At a critical total angular momentum $L_{\rm cr}=0.137 Nl$ the gravitational phase transition becomes second-order in the canonical ensemble and there is a smooth crossover between the ordered and disordered states for larger $L$. For high angular momentum, equilibria resemble an axisymmetric disk in physical space for any positive temperature.

Due to the non-additivity of the quadrupolar mean-field model and the existence of negative heat-capacity equilibrium states, the canonical and microcanonical ensembles are inequivalent. In particular, the phase-transition of the canonical ensemble is replaced by a stable sequence of equilibria without a phase transition in the microcanonical ensemble including a region of negative specific heat (between $A$ and $B$ in Figure~\ref{fig:Fbeta_L}). Furthermore, negative heat capacity biaxial equilibria (between $I$ and $G$ in Figure~\ref{fig:caloric_F_tr}) are metastable in the microcanonical ensemble but unstable in the canonical ensemble. Outside these regions, all VRR equilibrium states are identical in the microcanonical and canonical ensembles.

We furthermore introduced an ensemble, the $\omega TN$-ensemble, in which the system is embedded in a much bigger bath with which it can exchange not only VRR energy but also angular momentum. This ensemble also exhibits a first-order phase transition and a critical point where it becomes second-order, but at different values of the thermodynamic variables (see Section \ref{sec:g_canonical}). 

The one-component quadrupolar VRR model discussed in this paper is reminiscent of the Hamiltonian Mean Field (HMF) model \citep{Dauxois_2002a}, although the VRR model has two degrees of freedom per body while the HMF model has only one, and the HMF model has a kinetic energy term in the Hamiltonian where VRR does not.  Both the VRR and the HMF model exhibit ensemble inequivalence, first- and second-order phase transitions, and negative heat capacity, but the biaxial metastable equilibria and the negative-temperature equilibria do not appear in the HMF model. 

There are strong similarities between the equilibrium configurations of VRR and liquid crystals. The Newtonian gravitational interaction between rapidly precessing elliptical orbits, which trace out axisymmetric punctured disks, is similar to the Coulomb interaction between axisymmetric molecules. In particular, the quadrupole mean-field Hamiltonian of VRR for a one-component system with zero total angular momentum is equivalent to the Maier--Saupe model of liquid crystals \citep{Maier+Saupe_1958}, which exhibits a nematic-isotropic phase transition. In the nematic phase of VRR, the bodies are configured as a disk embedded in a dilute halo. 
The onset of the nematic-isotropic transition in VRR depends on the total angular momentum, just as the onset of the transition in liquid crystals depends on the external magnetic field, as can easily be verified by comparing our Figure \ref{fig:phase_trans_L} with the corresponding ones of \cite{Wojtowicz_1974,Gramsbergen_1986} for example.

However, there are important differences between the quadrupolar VRR system and liquid crystals. In contrast to the  diamagnetic term $-\ffrac12\Delta\chi (\bm{B}\cdot\bm{n}_i)^2$ that arises in the free energy due to an external magnetic field in liquid crystals, the term that arises in VRR due to the angular-momentum constraint is $-l\bm{\omega}\cdot \bm{n}_i$, which resembles  a paramagnetic term of a spin system in an external magnetic field, $-\chi \bm{B}\cdot \bm{n}_i$. 

We have restricted our attention in this paper to the mean-field approximation, in which each body is drawn independently from a distribution function and correlations are ignored. In the limit of zero temperature ($T\rightarrow 0^{+}$), we find two distinct distributions that are in stable or metastable thermodynamic equilibrium: 
 (i) a single razor-thin disk containing bodies with orbit normals that are both aligned and anti-aligned to the total angular momentum, with the relative numbers of each population depending on the total angular momentum (state $O_1$); and 
 (ii) two razor-thin disks of equal mass, with normals inclined to the total angular-momentum vector by $\ffrac{1}{2}\iota$ where $\iota$ depends on the total angular momentum and lies between $90^{\circ}$ and $180^{\circ}$ set by $L/(Nl)$ (state $O_5$). We call these states ``uniaxial" and ``biaxial", respectively. 
 
The zero-temperature equilibria $O_1$ and $O_5$ are discrete configurations of angular-momentum vector directions $\bm{n}$ (i.e., the distribution function has compact support on the unit sphere) that are time-invariant up to a rigid-body rotation at uniform angular speed. These are not the only such configurations. When $L=0$ other examples include 
 (i) $\bm{n}$ oriented along the vertices of a regular polyhedron, 
 (ii) $\bm{n}$ oriented along the vertices of a regular planar polygon in an arbitrary plane, 
 (iii) $\bm{n}$ oriented along the vertices of a regular planar polygon and either or both of the two directions perpendicular to the polygon's plane. 
In the simplest of these configurations the mass in each component is equal. These configurations do not appear in our maximum-entropy analysis, but in the limit where the number of components is large, we may recover the axisymmetric continuous zero-temperature states derived earlier in this paper: for example, configuration (ii) $\rightarrow O_2$. 

Our analysis is based on two important simplifications in addition to the mean-field approximation (see Section~\ref{s:mean_vrr} for a detailed inventory of our simplifying approximations). First, we approximated the gravitational torques between orbit annuli by their quadrupole component.  Multipoles beyond the quadrupole may be important for systems with radially overlapping or closely spaced orbits. Second, we focused on a one-component model, in which all bodies have the same scalar angular momentum and all of the pairwise interactions have the same coupling coefficients (we briefly discussed a simple extension of this model to multi-component systems in Section~\ref{sec:multi}). As we have shown in this paper, the statistical mechanics of this simplified model can be described completely yet exhibits a remarkable range of behavior. We expect that many of the features we have observed will also be present in more general models of stellar systems dominated by a central mass that do not depend on these approximations.

\acknowledgments{We are grateful to Julien Barr\'e for generously helping us to clarify several aspects of the stability analysis and to Benjamin Beri for pointing out that the quadrupole mean-field Hamiltonian is described by the Maier-Saupe model of liquid crystals, and to the referee for thoughtful comments that substantially improved the paper. We thank Zoltan R\'acz for useful discussions. This work was supported in part by the European Research Council under the European Union's Horizon 2020 Programme, ERC Starting Grant \#638435 (GalNUC), by the U.S.\ National Science Foundation through grant AST-1406166, and by NASA through grant NNX14AM24G.}

\bibliographystyle{apj}

\bibliography{ms}

\appendix

\section{Hamiltonian for vector resonant relaxation}\label{s:app:RR}

Here we describe the interaction Hamiltonian that governs VRR, that is, the interaction energy between Keplerian orbits after averaging over orbital phase and apsidal precession (equivalently, mean anomaly and longitude of pericenter). The most general form of this Hamiltonian may be written \citep{Kocsis_2015}
\begin{equation}\label{eq:Hdef}
H= -\ffrac{1}{2}\sum_{p\neq q}^N C_{pq}\sum_{\ell=0}^{\infty} P_{\ell}(0)^2 s_{pq\ell} \alpha_{pq}^{\ell}
\,P_{\ell}(\cos \theta_{pq}),
\end{equation}
where $P_{\ell}(x)$ are Legendre polynomials and $p$ and $q$ label the $N$ bodies, which orbit around the central point mass with angular-momentum unit vectors $\bm{n}_i$. In this formula  
$\cos\theta_{pq} = \bm{n}_p \cdot \bm{n}_q$, 
\begin{equation}
C_{pq} =  G\frac{m_p m_q}{a_{\Out}}\,,\quad
\alpha_{pq} = \frac{a_{\In}}{a_{\Out}}\,,
\end{equation}
and 
\begin{align}\label{e:s_ijl}
 s_{pq\ell} &=
\frac{1}{\pi^2}\int_0^{\pi} \D \phi \int_0^{\pi} \D \phi'
\frac{\min\left[\; (1 + e_{\In}\cos\phi),\; \alpha_{pq}^{-1}(1 + e_{\Out} \cos\phi')\;\right]^{\ell+1}
}{ \max\left[\; \alpha_{pq}(1 + e_{\In} \cos\phi),\; (1 + e_{\Out}  \cos\phi')\;\right]^{\ell}}.
\end{align}
Here $m_p$, $a_p$, $e_p$ denote the mass, semimajor axis, and eccentricity of the orbit of body $p$ around the central object, all of which are fixed during VRR, and 
``${\Out}$'' and ``${\In}$'' label the index $p$ or $q$ with the
larger and the smaller semimajor axis, respectively, i.e.,
$\alpha_{pq}\leq 1$. Note that only terms with even $\ell$ contribute to the sum, 
and the $\ell=0$ term only contributes an unimportant constant to the Hamiltonian. 

For circular orbits $s_{p q \ell} = 1$ for all $\ell$, and more generally, for
eccentric orbits with $a_{\Out}(1-e_{\Out}) > a_{\In}(1+e_{\In})$ (radially non-overlapping orbits) 
\begin{equation}
s_{pq\ell} = \frac{\chi_{ \Out }^{\ell}}{\chi_{ \In }^{\ell+1}}
 P_{\ell+1}(\chi_{ \In })P_{\ell-1}(\chi_{ \Out })
\label{e:xxxyyy}
\end{equation}
for $\ell > 0$. Here $\chi_p=a_p/b_p=(1-e_p^2)^{-1/2}$. For radially non-overlapping orbits
the sum in Eq.~(\ref{eq:Hdef}) converges exponentially as a function of $\ell$ and the $\ell=2$ term dominates the dynamics for arbitrary mutual inclinations. In contrast, for radially overlapping orbits, $a_{\Out}(1-e_{\Out}) \leq a_{\In}(1+e_{\In})$, the contribution of the $\ell^{\rm th}$ multipole decays as $\ell^{-2}$, and multipoles up to order $\ell_{\max}\sim (1-\bm{n}_p\cdot\bm{n}_q)^{-1/2}$ must be included when calculating the Hamiltonian (see Figure B1 in \citealt{Kocsis_2015}). However, if most orbits have relatively large inclinations and $N\gg1$, the net torque on an object is still dominated by the $\ell=2$ quadrupole interaction. In particular, for a spherical cluster the contribution of the $\ell^{\rm th}$ multipole to the net torque decays as fast as $\ell^{-3}\ln \ell$ (see Appendix D in \citealt{Kocsis_2015}). For clusters with overlapping orbits, the effects of multipoles beyond the quadrupole may be significant.

For simplicity, let us keep only the $\ell=2$ term and assume that a large number $N_i$ of bodies have similar $(m_i, a_i)$ but different 
$\bm{n}$ with several such components $i=1,2,\dots,K$. Then the interaction energy is a sum over the components and the bodies therein,
\begin{align}
H 
&= -\ffrac{1}{2}\sum_{i, j}^{K}\sum_{p_i,q_j}^{N_i, N_j} P_{2}(0)^2 C_{ij}s_{ij2}\alpha_{ij}^{2}\,\ffrac{3}{2} g(\bm{n}_{p_i},\bm{n}_{q_j})
= -\ffrac{1}{2}\sum_{i, j}^{K} J_{ij} \iint f_i(\bm{n}) f_j(\tilde{\bm{n}}) \,g(\bm{n},\tilde{\bm{n}})\,d\Omega d\tilde \Omega,
\end{align}
where $J_{ij} = \ffrac{3}{2} P_{2}(0)^2 C_{ij} s_{ij2}\alpha_{ij}^{2}=\ffrac{3}{8}C_{ij}s_{ij2}\alpha_{ij}^2$ and $g(\bm{n}_p,\bm{n}_q)=\ffrac{2}{3}P_{\ell}(\bm{n}_p\cdot\bm{n}_q)=(\bm{n}_p\cdot\bm{n}_q)^2-\frac{1}{3}$. In the second line the sums over $p$ and $q$ of components $i$ and $j$ respectively are written as integrals using the corresponding distribution functions $f_i$ and $f_j$ of the angular-momentum unit vectors. For circular orbits, $s_{ij2}=1$, and this is Eq.~(\ref{eq:Econt}) used in the main text. For eccentric, radially overlapping orbits $s_{ij2}$ is given analytically by Eqs.~(B12) and (B31) in \cite{Kocsis_2015}.

\section{Alignment of total angular momentum}\label{s:app:Rot}

In this appendix we show that, at equilibrium, the angular momentum and the Lagrange multiplier $\bm{\gamma}$ lie along one of the eigenvectors of the matrix $Q_{\mu\nu}$. We work in a coordinate system aligned with these eigenvectors so $Q_{\mu\nu}$ is diagonal (Eq.~\ref{eq:Qmatrix}) and we can write $Q_{\mu\nu}\equiv Q_\mu\delta_{\mu\nu}$ (there is no summation over $\mu$ here, that is, $Q_{\mu\nu}\equiv  \mbox{diag}[\mathbf{Q}]$, where $\mathbf{Q}$ is the vector whose elements are the eigenvalues of $Q_{\mu\nu}$). The self-consistency condition (\ref{eq:self_cons}) is then
\begin{equation}
Q_{\mu}\delta_{\mu\nu}=\frac{\int d\Omega\, (n_\mu n_\nu-\frac{1}{3}\delta_{\mu\nu})e^{JN \beta Q_\sigma n_\sigma^2+l\gamma_\sigma n_\sigma}}{\phantom{\Big|}\int d\Omega\,e^{JN \beta Q_\sigma n_\sigma^2+l\gamma_\sigma n_\sigma}}.
\end{equation} 
In the case $\mu=1$, $\nu=2$ this condition simplifies to 
\begin{align}
0=&\int d\Omega\,n_1 n_2 e^{JN \beta Q_\sigma n_\sigma^2+l\gamma_\sigma n_\sigma}\nonumber \\=&\int_0^\pi \sin\theta d\theta \int_0^{2\pi}d\phi\, \sin^2\theta\cos\phi\sin\phi \,
e^{JN \beta (Q_1\sin^2\theta\cos^2\phi+Q_2\sin^2\theta\sin^2\phi+Q_3\cos^2\theta)}
e^{l\gamma_1\sin\theta\cos\phi+l\gamma_2\sin\theta\sin\phi+l\gamma_3\cos\theta};
\end{align} 
in the last equation we have introduced a spherical polar coordinate system. Now divide the interval $\phi\in[0,2\pi)$ into four quadrants: in the first quadrant, $0\le\phi<\pi/2$, use $\psi=\phi$ as the integration variable; in the second quadrant, $\pi/2 \le\phi <\pi$, use $\psi=\pi-\phi$; in the third quadrant, $\pi\le\phi<3\pi/2$, use $\psi=\phi-\pi$; and in the fourth quadrant, $3\pi/2\le\phi<2\pi$, use $\psi=2\pi-\phi$. Adding the sub-integrals we find
\begin{align}
0=&4\int_0^\pi \sin\theta d\theta \int_0^{\pi/2}d\psi\, \sin^2\theta\cos\psi\sin\psi 
e^{JN \beta (Q_1\sin^2\theta\cos^2\psi+Q_2\sin^2\theta\sin^2\psi+Q_3\cos^2\theta) +l\gamma_3\cos\theta}
\nonumber\\&\times 
\sinh(\gamma_1\sin\theta\cos\psi)\sinh(\gamma_2\sin\theta\sin\psi).
\end{align}
Since the integration range is from 0 to $\pi/2$, $\sin\psi$ and $\cos\psi$ are positive, so the integral is positive-definite if $\gamma_1\gamma_2>0$, negative-definite if $\gamma_1\gamma_2$ is negative, and zero (as required) if and only if at least one of $\gamma_1$ and $\gamma_2$ is zero. Repeating this argument for $\mu=1$, $\nu=3$ and $\mu=2$, $\nu=3$ we conclude that at least two of the $\gamma_\mu$ must be zero. Therefore $\bm{\gamma}$ must be aligned with one of the coordinate axes and thus with one of the eigenvectors of $Q_{\mu\nu}$. 

Without loss of generality we may assume that $\bm{\gamma}$ points along $\bm{n}_3$. Then the angular-momentum vector is
\begin{equation}
L_{\mu}=l\frac{\int d\Omega\, n_\mu e^{JN \beta Q_\sigma n_\sigma^2+l\gamma_3n_3}}{\phantom{\Big|}\int d\Omega\,e^{JN \beta Q_\sigma n_\sigma^2+l\gamma_3n_3}}.
\end{equation} 
If $\mu=1$ or 2, the integrand in the numerator is odd under the transformation $\phi\to\phi+\pi$ so the integral must vanish. Thus only $L_3$ is non-zero so the angular momentum $\bm{L}$ is parallel or anti-parallel to $\bm{\gamma}$.

\section{Second-order variation}\label{s:app:S2}

Here we calculate the second variation of entropy and free energy under general perturbations of the mean-field distribution function and show the stability of the negative-temperature configurations and the disordered phase at infinite temperature, labeled $D$ in Figure \ref{fig:qsz_T_L0}. 

Let $f$ be the equilibrium distribution function, 
\begin{equation}\label{app:eq:feq}
	f = \frac{N}{Z}e^{-\beta\varepsilon+l\bm{\gamma}\cdot \mathbf{n}}\quad\mbox{where}\quad Z=\int e^{-\beta\varepsilon+l\bm{\gamma}\cdot \mathbf{n}}\, d\Omega; 
\end{equation}
and let $\delta f$ be a perturbation about this equilibrium. The entropy changes to
\begin{equation}
S+\delta S=-k\int (f+\delta f)\ln (f+\delta f)\,d\Omega.
\end{equation}
Expand this to second order in $\delta f$. Then the change in entropy is
\begin{equation}
\delta S=-k\int \left[\delta f(\ln f+1)+\frac{(\delta f)^2}{2f}\right]d\Omega + \mbox{O}(\delta f^3).
\end{equation}
Substituting Eq.\ (\ref{app:eq:feq}) and dropping terms higher than second order, 
\begin{align}
\delta S=&-k\int \left[\delta f\left(\ln \frac{N}{Z} -\beta\varepsilon+l\bm{\gamma}\cdot\bm{n}+1\right)+\frac{(\delta f)^2}{2f}\right]d\Omega.
\end{align}
Conservation of the number of bodies and the total angular momentum implies that
\begin{equation}\label{eq:NL_constraints}
\int \delta f\, d\Omega=0, \qquad \int n_\mu\,\delta f\,d\Omega=0.
\end{equation}
Thus the equation for the entropy variation simplifies to
\begin{equation}\label{eq:dels}
\delta S=k\int \left[\beta\varepsilon \delta f-\frac{(\delta f)^2}{2f}\right]d\Omega.
\end{equation}
The energy change can be written as
\begin{equation}\label{eq:dele}
E+\delta E=-\ffrac{1}{2}J\iint q_{\mu\nu}q_{\mu\nu}'(f+\delta f)(f'+\delta f')\,d\Omega d\Omega'
\end{equation}
which implies\footnote{We denote $M_{\mu\nu}^2=\sum_{\mu\nu} M_{\mu\nu} M_{\mu\nu}$.}
\begin{align} \label{eq:de}
\delta E&=-J\iint  (\delta f) f'q_{\mu\nu}q_{\mu\nu}'\,d\Omega d\Omega'-\ffrac{1}{2}J\left(\int   q_{\mu\nu}\, \delta f\,d\Omega\right)^2
=\int \varepsilon\, \delta f \,d\Omega- \ffrac{1}{2}J\left(\int q_{\mu\nu}\,\delta f\right)^2,
\end{align}
where the second line follows from the relation $\epsilon=-q_{\mu\nu}\int \,f'q_{\mu\nu}'d\Omega'$ (Eq.\ \ref{eq:epsiloni_Q}).
Thus if the energy is fixed 
\begin{equation}\label{eq:E_constraint}
\delta E = \int \varepsilon \delta f d\Omega-\ffrac{1}{2}J\left(\int q_{\mu\nu}\,\delta f\,d\Omega \right)^2=0.
\end{equation}
Substituting in Eq.\ (\ref{eq:dels}) to eliminate the terms depending on $\varepsilon$, we find 
\begin{equation}\label{eq:delsa}
\delta S= \delta^2 S + \mbox{O}(\delta f^3),
\end{equation}
where $\delta^2 S$ is the second-order variation of entropy
\begin{equation}\label{eq:dS2nd}
\delta^2 S = \frac{kJ\beta}{2}\left(\int \,q_{\mu\nu}\,\delta f\,d\Omega\right)^2-k\int\frac{(\delta f)^2}{2f}d\Omega.
\end{equation}
As it should, the entropy variation (\ref{eq:delsa}) contains no first-order terms, since the perturbation is performed about the equilibrium distribution, at which the entropy has vanishing first-order variation by construction, if the constraints are satisfied (see Section \ref{sec:StEq}).

In the canonical ensemble the equilibria are extrema of the free energy $F=E-TS$ at constant temperature $T$. Using Eqs.\ (\ref{eq:dels}) and (\ref{eq:E_constraint}) we find the second-order variation of free energy to be
\begin{equation}\label{eq:delf}
\delta^2 F =-\frac{J  }{2}\left(\int q_{\mu\nu}\,\delta f\,d\Omega\right)^2+\int\frac{(\delta f)^2}{2 \beta f}d\Omega.
\end{equation}

The condition for stability of an equilibrium with respect to any perturbation is given in the microcanonical ensemble by the condition
\begin{equation}\label{eq:d2S_ineq}
	\delta^2 S < 0,
\end{equation}
that is, the entropy must be a local maximum. In the canonical ensemble the stability condition is
\begin{equation}\label{eq:d2F_ineq}
	\beta \delta^2 F > 0,
\end{equation}
that is, the free energy must be a minimum if the temperature is positive, or a maximum if the temperature is negative. 
    
Now, since $\delta^2 S = -k\beta \delta^2 F$, as is evident from (\ref{eq:dS2nd}) and (\ref{eq:delf}), stability in the microcanonical ensemble is determined by the same inequality condition as in the canonical ensemble (Eqs.\ \ref{eq:d2S_ineq} and \ref{eq:d2S_ineq})). However, equilibria do not necessarily possess the same stability properties in the two ensembles, because the perturbations are subject to different constraints. In the canonical ensemble, the perturbations are subject only to constraints (\ref{eq:NL_constraints}), while in the microcanonical ensemble they are subject additionally to the fixed energy constraint (\ref{eq:E_constraint}). Thus, modes that render the canonical ensemble unstable may not exist in the microcanonical ensemble. If this happens, the system presents inequivalence of ensembles, as described in Section \ref{s:E_In}.
\\

\paragraph{Negative-temperature states}

For negative temperature, $\beta<0$, both of the terms in the entropy second variation (\ref{eq:dS2nd}) are negative-definite and so $\delta^2 S<0$ for perturbations that keep $E$, $N$, and $\bm{L}$ fixed. Therefore negative-temperature equilibria are 
local maxima of the entropy and hence they are (meta)stable in the microcanonical ensemble.

Similarly, in the canonical ensemble both terms in (\ref{eq:delf}) are positive-definite and so $\beta \delta^2 F>0$ for perturbations which keep $T$, $N$, and $\bm{L}$ fixed\footnote{The temperature $T$ is defined by the angular momentum vector distribution function of the heat bath; it is constant because the heat bath has a much larger angular momentum
than the system we are examining. Note that fixing $T$ does not impose any constraint on $\delta f$ in Eq.~(\ref{eq:delf}).}. 
Therefore negative-temperature equilibria are local minima of $\beta F$ and hence they are (meta)stable in the canonical ensemble.
\\

\paragraph{State $D$}
The state $D$ is the disordered phase at infinite temperature and therefore corresponds to axisymmetric solutions with $\beta \rightarrow 0^+$. Its stability is straightforwardly implied by Eq.\ (\ref{eq:dS2nd})
\begin{equation}
 \lim_{\beta\rightarrow 0} \delta^2 S =-k\int\frac{(\delta f)^2}{2f}d\Omega <0 .
\end{equation}

\section{Stability of off-diagonal elements}\label{s:app:Qmunu}

We show here that for positive temperature, there can always be found a coordinate system with one axis aligned with $\bm{\omega}$, such that the off-diagonal elements of $Q_{\mu\nu}$ do not give rise to instabilities and are zero in equilibrium. We follow the notation of Section \ref{s:VRRequilibrium}.

The dominant contributions to the integral (\ref{eq:Xi_fg_g_can_Q}) for the partition function of the $\omega TN$-ensemble come from the minima of $g(\Qmat,\beta,\gamma)$, which occur at a subset of the extrema where $\partial g/\partial Q_{\mu\nu}=0$. The extrema are located at $\Qmat^\eq$ given by the implicit equation 
\begin{equation}\label{eq:qeqdef}
Q_{\mu\nu}^\eq =\langle q_{\mu\nu}\rangle_{\Qmat^\eq};
\end{equation}
here the angle brackets denote the average
\begin{align}
\langle X&(\bm{n})\rangle_{\Qmat}
=\frac{\int d\Omega\,X(\bm{n}) \exp\Big[ \beta JN\sum_{\mu\nu}Q_{\mu\nu} q_{\mu\nu}+l\sum_\mu\gamma_\mu n_\mu\Big]}{\int d\Omega \exp\Big[\beta JN\sum_{\mu\nu}Q_{\mu\nu}q_{\mu\nu}+l\sum_\mu\gamma_\mu n_\mu\Big]}.
\end{align}

We now describe some properties of the matrix $\Qmat^\eq$. (i) Since $q_{\mu\nu}=q_{\nu\mu}$ we have  $Q_{\mu\nu}^\eq=Q_{\nu\mu}^\eq$, that is, $\Qmat^\eq$ is symmetric. (ii) Since $\sum_{\mu=1}^3q_{\mu\mu}=0$, we have $\mbox{Tr\,}\Qmat^\eq=0$, that is, $\Qmat^\eq$ is traceless. (iii) Since $q_{\mu\nu}$ is a tensor (Eq.\ \ref{eq:Qcont0}), Eq.\ (\ref{eq:qeqdef}) implies that $\Qmat^\eq$ also transforms as a tensor under rotations. Therefore without loss of generality we can choose the coordinate system in the $\Omega$-integral so that $\Qmat^\eq$ is diagonal\footnote{Thus systems with $Q^\eq_{xx}\not=Q^\eq_{yy}$ are an example of spontaneously broken symmetry.}. The whole system is free to change its principal axes perpendicular to $\bm{L}$. (iv) Using similar arguments to those in Appendix \ref{s:app:Rot} we may then show that $\bm{\gamma}$ lies along one of these axes, which we may choose to be the $z$-axis. 

Given these results, all of the elements of $\Qmat^\eq$ are either zero or linear combinations of the parameters $Q$ and $W$ defined in Eqs.\ (\ref{eq:Qmatrix})--(\ref{eq:w_def}). It is therefore useful to replace the nine coordinates $Q_{\mu\nu}$ by nine new coordinates $\bm{P}=(Q,W,X_0\ldots,X_3,Y_1,\ldots,Y_3)$ defined by
\begin{align}
\Qmat=\left(\begin{array}{ccc} \ffrac{1}{2}(W-Q) +X_0 & X_1+Y_1 & X_2+Y_2 \\ X_1-Y_1 & -\ffrac{1}{2}(W+Q)+X_0 & X_3+Y_3 \\ X_2-Y_2 & X_3-Y_3 &Q+X_0 \end{array}\right). 
\end{align}
The Jacobian $|\partial\bm{P}/\partial\Qmat|=\frac{1}{12}$. In the new coordinates
\begin{align}
	G(\bm{P},\beta,\gamma) &= -\frac{N}{\beta} \ln \int d\Omega 
\exp\Big[ \ffrac{1}{4}\beta JN(6qQ-3Q^2+2wW-W^2 - 6X_0^2 
+4X_1\sin^2\theta\sin2\phi-4X_1^2 +4X_2\sin2\theta\cos\phi 
\nonumber \\&\quad 
-4X_2^2+4X_3\sin2\theta\sin\phi-4X_3^2 - 4Y_1^2-4Y_2^2-4Y_3^2) 
+l\gamma\cos\theta \Big]\,.
    \label{eq:G_vrr_can_3}
\end{align}
Here $q=\cos^2\theta-\frac{1}{3}$ and $w=\sin^2\theta\cos2\phi$, as in  Eqs.\ (\ref{eq:q_def}) and (\ref{eq:w_def}). 
It is straightforward to confirm that the extrema of $G$ occur at $X_0=X_1=\cdots=Y_3=0$. 
An extremum is a minimum if the Hessian $\partial^2G/\partial P_\mu\partial P_\nu$ is positive definite (i.e., all of its eigenvalues are positive) and a maximum if it is negative definite; otherwise the extremum is a saddle point. It is straightforward to confirm that all off-diagonal elements of the Hessian are zero in the 7 rows and columns corresponding to $(X_0,X_1,\ldots,Y_3)$ and that
\begin{align}
\frac{\partial^2G}{\partial X_0^2}\bigg|_\eq &= 3JN^2; \nonumber \\
\frac{\partial^2G}{\partial X_j^2}\bigg|_\eq =
\frac{\partial^2G}{\partial Y_j^2}\bigg|_\eq &= 2JN^2, \qquad j=1,2,3; \nonumber \\
\frac{\partial^2G}{\partial X_i\partial X_j}\bigg|_\eq &=0, \qquad i\not=j, \nonumber \\
\frac{\partial^2G}{\partial X_i\partial Y_j}\bigg|_\eq &=0.
\end{align}
Thus $G$ is a minimum at the extrema as a function of these seven coordinates.

\section{Analytic results for the equilibria}
\label{s:app:partition}

Here we derive analytic expressions for the thermal equilibria.
We define the moment generating function
\begin{equation}\label{eq:Zdef}
Z_0(\kappa_1,\kappa_2,c) = \int  e^{\kappa_1 q + \kappa_2 w + c s}\,d\Omega\,.
\end{equation}
where $q=\cos^2\theta-\ffrac13$, $w=\sin^2\theta\cos2\phi$, and $s=\cos\theta$, the integral is over the unit sphere, and the parameters $\kappa_1$, $\kappa_2$, and $c$ are defined by Eq.~(\ref{eq:kappadef}). 
Once $Z_0(\kappa_1,\kappa_2,c)$ is known, all statistical quantities follow straightforwardly\footnote{We suppress the $N\ln N$ constants from $S$, $F$, and $G$.}: 
\begin{align}
 \LA q \RA &= \left.\frac{\partial}{\partial \kappa_1} \ln Z_0\right|_{\kappa_2,c}\,, \label{eq:q'}\\
 \LA w \RA &= \left.\frac{\partial}{\partial \kappa_2} \ln Z_0\right|_{\kappa_1,c}\,, \label{eq:w'}\\
   \frac{L}{l N} &= \left.\LA s \RA = \frac{\partial}{\partial c} \ln Z_0\right|_{\kappa_1,\kappa_2}\,, \label{eq:s'}
\end{align}
\begin{align}
 \frac{E}{J N^2} &= -\ffrac{3}{4} \LA q \RA ^2 -\ffrac{1}{4} \LA w \RA ^2\,, \label{eq:E'}\\
 \frac{S}{kN} &= -\kappa_1\LA q\RA -\kappa_2\LA w\RA -c\LA s\RA  
 + \ln Z_0 \,, \label{eq:S'}\\
 \frac{F}{NkT} &= \ffrac{1}{2} \kappa_1\LA q \RA + \ffrac{1}{2}\kappa_2 \LA w \RA + c\LA s\RA -\ln Z_0 \,,  \label{eq:F'}\\
 \frac{G}{NkT} &= \ffrac{1}{2} \kappa_1\LA q \RA + \ffrac{1}{2}\kappa_2 \LA w \RA -\ln Z_0 \,,  \label{eq:G'}\\
 \frac{kT}{JN} &= \frac{3 \LA q \RA }{2\kappa_1}  = \frac{\LA w \RA }{2\kappa_2}\label{eq:T'}\,,\\
\end{align}
and for the $\omega TN$-ensemble we have
\begin{equation}
 \frac{l \omega}{JN} = \frac{3 c\LA q \RA }{2\kappa_1}  = \frac{c\LA w \RA }{2\kappa_2}\label{eq:omega'}\,. 
\end{equation}
Eqs.~(\ref{eq:T'}) follow from Eq.~(\ref{eq:kappadef}) assuming $\LA q\RA \neq 0$ and $\LA w \RA\neq 0$. In the special case  $\LA q\RA = \LA w \RA = 0$, self-consistency requires that $c=0$. In this case $Z_0=4\pi$ is a trivial solution for arbitrary $T$, and $E=L=0$. 
This is the isotropic distribution $f(\bm{n})=N/(4\pi)$, which is an equilibrium for any temperature and is shown by the curve $O_3D$ in Figure \ref{fig:qsz_T_L0}.

\subsection{Axisymmetric equilibria with zero total angular momentum}\label{s:app:partition:L0}

Systems that are axisymmetric must have two of the diagonal elements of $Q_{\mu\nu}$ (Eq.\ \ref{eq:Qmatrix}) equal. This requires that either $\LA w\RA=0$ or $\LA w\RA=\pm 3\LA q\RA$. Since the total angular momentum $\bm{L}=0$ we are free to choose the $z$-axis to be aligned with any one of the three eigenvectors of $Q_{\mu\nu}$ so without loss of generality we can assume that $\LA w\RA=0$. Then $\kappa_2=0$ by Eq.\ (\ref{eq:kappadef}) and $c=0$ by Eq.\ (\ref{eq:s_cons}). Eq.~(\ref{eq:Zdef}) simplifies to
\begin{equation}
Z_0(\kappa_1,0,0)= 2 \pi^{3/2} \exp\left(-\ffrac{1}{3}\kappa_1\right)\frac{{\rm erf}(\sqrt{-\kappa_1})}{\sqrt{-\kappa_1}}\,.
\end{equation}
Note that this is real for either positive or negative $\kappa_1$. Eqs.~(\ref{eq:q'}) and (\ref{eq:T'}) yield the parametric solution
\begin{align}
\LA q\RA &=  -\frac{1}{3} - \frac{1}{2\kappa_1} - \frac{e^{\kappa_1}}{\sqrt{-\pi\kappa_1}{\rm erf}\sqrt{-\kappa_1}}\,,\label{eq:q_sw0}\\
\frac{k T}{JN} &= 
-\frac{3}{2\kappa_1}\left( \frac{1}{3} + \frac{1}{2\kappa_1} + \frac{e^{\kappa_1}}{\sqrt{-\pi\kappa_1}{\rm erf}\sqrt{-\kappa_1}}\right)\,.\label{eq:T_sw0}
\end{align}
As $\kappa_1$ increases from $-\infty$ to $\infty$, $\LA q\RA$ increases monotonically from $-\frac{1}{3}$ to $\frac{2}{3}$. Both limits $\kappa_1\to\pm\infty$ correspond to $T\to 0^+$ so the two limits are the zero-temperature states $O_2$ and $O_1$ in Figure \ref{fig:qsz_T_L0}. As $\kappa_1$ increases from $-\infty$ the temperature grows monotonically from $T=0$ to $kT_{A}/(JN)=0.148556$ at $\kappa_1= 2.178289$, then it decreases monotonically for higher $\kappa_1$, approaching zero at $\kappa_1\rightarrow \infty$. At $T_{A}$, $\LA q\RA_{A}=0.21573$, Point $B$ in Figure \ref{fig:qsz_T_L0} has $\kappa_1\rightarrow 0$, $\LA q\RA_B = 0$, and $T_{B} = \frac{2}{15}$.

\subsection{Axisymmetric, rotating equilibria}\label{s:app:partition:rotating}

Next consider systems with non-zero angular momentum. Since $Z_0$ is an even function of $\kappa_2$, $\kappa_2=0$ implies $\LA w\RA=0$ (see Eq.~\ref{eq:w'}).  These states are axisymmetric around the angular-momentum axis with  $0\leq \LA s\RA < 1$ and $\LA q\RA\neq 0$. We evaluate $Z_0$ (\ref{eq:Zdef}) at $\kappa_2=0$ 
\begin{align}
Z_0(\kappa_1,0,c) =& \frac{\pi^{3/2}}{\sqrt{-\kappa_1}}
\exp\left(-\frac{\kappa_1}{3}  - \frac{c^2}{4\kappa_1}\right) 
 \left[ 
  {\rm erf}\left(\frac{c-2\kappa_1}{2\sqrt{-\kappa_1}}\right)
  +{\rm erf}\left(-\frac{c+2\kappa_1}{2\sqrt{-\kappa_1}}\right) 
  \right] 
\label{eq:Zaxi}
\end{align}
which is real for all $-\infty<\kappa_1<\infty$. The quantities $\LA s\RA$, $\LA q\RA$, and $T$ may be obtained from Eqs.~(\ref{eq:s'}), (\ref{eq:q'}), and (\ref{eq:T'}), which yields the parametric solution
\begin{align}
\LA s \RA &= -\frac{c}{2\kappa_1} + \frac{2\pi e^{\frac23 \kappa_1}}{\kappa_1 Z_0}\sinh c \label{eq:s''}\\
\LA q\RA &= \frac{c^2}{4\kappa_1^2} - \frac{1}{2\kappa_1} -\frac{1}{3} + \frac{2\pi e^{\frac23 \kappa_1}}{\kappa_1 Z_0}\left(\cosh c - \frac{c}{2\kappa_1}\sinh c\right)\label{eq:q''}\\
\frac{kT}{JN} &=\frac{3}{2\kappa_1}\Big[\frac{c^2}{4\kappa_1^2} - \frac{1}{2\kappa_1} -\frac{1}{3} + \frac{2\pi e^{\frac23 \kappa_1}}{\kappa_1 Z_0}\left(\cosh c - \frac{c}{2\kappa_1}\sinh c\right)\Big]\label{eq:T''}
\end{align}
For any fixed $-\infty<\kappa_1<\infty$, Eq.~(\ref{eq:s''}) defines a monotonic function of $0<c<\infty$ mapping onto $L/(Nl)\equiv \LA s\RA\in[0,1)$.  For any fixed $0<\LA s\RA < 1$, the temperature assumes all values from $-\infty$ to $\infty$ since the expression for $T$ is singular at $\kappa_1\rightarrow 0$ (state $D$). In the canonical and microcanonical ensembles with fixed $\LA s\RA$, $\LA q\RA$ is a monotonically increasing function of $\kappa_1$, but the temperature has three local extrema as a function of $\kappa_1$ for $\LA s\RA < 0.13714$, one local maximum for $0.13714  < \LA s\RA < 1/\sqrt{3}$  and no extrema for $1/\sqrt{3}<\LA s \RA<1$. Note that $T\rightarrow 0$ at $\kappa_1\rightarrow \infty$ (state $O_1$), $\kappa_1\rightarrow -\infty$ (state $O_2$), and at a finite $\kappa_1$ for $\LA s\RA < 1/\sqrt{3}$ (state $O_3$). Note that $\LA q\RA = 0$ at $O_3$ for $\LA s\RA < 1/\sqrt{3}$ due to Eq.~(\ref{eq:q''}).

In the $\omega TN$-ensemble, $\omega = k cT(\kappa_1,c)/l$ and Eq.~(\ref{eq:T''}) defines a monotonic function of $c$ for fixed $\kappa_1>0$, which can be inverted numerically to find $c(\kappa_1,\omega)$ for the equilibrium states between $O_1$ and $D$ for fixed $\omega$. Substituting in Eqs.~(\ref{eq:q''}) and (\ref{eq:T''}), $\LA q\RA$ is a monotonic function of $\kappa_1$ for fixed $0<\omega<\infty$, while $T$ has two local extrema as a function of $\kappa_1$ for $0< l\omega/JN < 0.057285$ and no local extrema for $l\omega/JN>0.057285$. The heat capacity is negative between the two local maxima and positive otherwise.

Next we provide asymptotic expressions near $O_2$, $O_1$, and $D$. We utilize the asymptotics of the error function 
\begin{equation}
{\rm erf}(x) \approx 1 - \frac{e^{-x^2}}{\sqrt{\pi}}\left( \frac{1}{x} - \frac{1}{2x^3} + \frac{3}{4x^5} -\dots \right)
\quad\mbox{as $x\to\infty$.}\end{equation}
In particular as $\kappa_1\rightarrow-\infty$ and $c\rightarrow \infty$ with fixed $c/\kappa_1$ (state $O_2$ of Figure~\ref{fig:qomega_T_L})\footnote{The expressions below contain factors of $\sigma=1+(c/2\kappa_1)^2$. The factor $c/(2\kappa_1)$, which is proportional to $\omega$ defined in Eq.~(\ref{eq:omega}), is retained because at fixed angular momentum $\LA s \RA$, $c\propto \kappa_1$ as $\kappa_1\to -\infty$ according to Eq.~(\ref{eq:SO2}).}
\begin{align}
\LA s\RA &= -\frac{c}{2\kappa_1} + \mbox{O}\left(\frac{e^{\sigma\kappa_1}}{|\kappa_1|^{1/2}}\right)\,,\label{eq:sO2}\\
\LA q\RA &= -\frac{1}{3} - \frac{1}{2\kappa_1} + \frac{c^2}{4\kappa_1^2}+\mbox{O}\left(\frac{e^{\sigma\kappa_1}}{|\kappa_1|^{1/2}}\right) 
= \LA s\RA^2 - \frac{1}{3} - \frac{1}{2\kappa_1}
+\mbox{O}\left(\frac{e^{\sigma\kappa_1}}{|\kappa_1|^{1/2}}\right)\,,\label{eq:qO2}
\end{align}
which imply that
\begin{align}
\frac{kT}{JN} &= \frac{3}{2\kappa_1}\left(\LA s\RA^2 - \frac{1}{3} - \frac{1}{2\kappa_1}\right)%
+\mbox{O}\left(\frac{e^{\sigma\kappa_1}}{|\kappa|^{3/2}}\right),
\label{eq:TO2}\\
\frac{S}{kN} &=  -\frac{\ln(-\kappa_1)}{2} +\frac{1}{2}+\ln(2\pi^{3/2}) 
+\mbox{O}\left(\frac{e^{\sigma\kappa_1}}{|\kappa|^{1/2}}\right),
\label{eq:SO2}\\
\frac{E}{JN^2} &= -\frac{3}{4}\left(\LA s\RA^2 - \frac{1}{3} - \frac{1}{2\kappa_1}\right)^2 
+\mbox{O}\left(\frac{e^{\sigma\kappa_1}}{|\kappa|^{1/2}}\right),
\label{eq:EO2}\\
\frac{F}{NkT} &= \frac{\ln(-\kappa_1)}{2}+ \frac{1-3\LA s\RA^2}{6}\kappa_1 - \frac{1}{4}  -\ln(2\pi^{3/2}) 
+ \frac{1}{4\kappa_1}+\mbox{O}\left(\frac{e^{\sigma\kappa_1}}{|\kappa|^{1/2}}\right)
\label{eq:FO2}
\end{align}
and for $\kappa_1\rightarrow\infty$ and $0\leq \lim_{\kappa_1\rightarrow\infty} c <\infty$ (state $O_1$ of Figure~\ref{fig:qomega_T_L})
\begin{align}
\LA s\RA &= \tanh c - \frac{c(1-\tanh^2 c)+\tanh c}{2\kappa_1}+\mbox{O}(\kappa_1^{-2}),
\label{eq:sO1}\\
\LA q\RA &=\frac{2}{3}-\frac{1}{\kappa_1}+\frac{c\tanh c-1}{2\kappa_1^2} +\mbox{O}(\kappa_1^{-3})
=\frac{2}{3}-\frac{1}{\kappa_1}+\frac{\LA s\RA \tanh^{-1} \LA s\RA-1}{2\kappa_1^2} +\mbox{O}(\kappa_1^{-3}),\label{eq:qO1}\\
\frac{kT}{JN} &= \frac{1}{\kappa_1}-\frac{3}{2\kappa_1^2}+\frac{3\LA s\RA \tanh^{-1} \LA s\RA-3}{4\kappa_1^3}+\mbox{O}(\kappa_1^{-4}).
\label{eq:TO1}
\end{align}
As $\kappa_1\to0$ (state $D$ of Figures~\ref{fig:caloric_S_micro}, \ref{fig:S_E_L_9e-1}, and \ref{fig:S_E_L_5e-1_tr}) the temperature diverges,
\begin{align}
\LA s\RA &= \coth c - \frac{1}{c}
+\left[\frac{\coth^2 c}{c} + \frac{\coth c}{c^2} - \frac{c^2+1}{c^3}\right]2\kappa_1 
+ \mbox{O}(\kappa_1^2),\label{eq:sOD}\\
\LA q\RA &=\frac{2(3+c^2-3c\coth c)}{3c^2}
+\left(\frac{5+2c^2-4c\coth c -c^2\coth^2c}{c^4}\right)4\kappa_1 
+\mbox{O}(\kappa_1^2),\label{eq:qD}\\
\frac{kT}{JN} &= \frac{3+c^2-3c\coth c}{c^2\kappa_1}
+\left(\frac{5+2c^2-4c\coth c -c^2\coth^2c}{c^4}\right)6\kappa_1 
+\mbox{O}(\kappa_1).\label{eq:TD}
\end{align}

These results show that for axisymmetric states $\LA q\RA$ generally increases from $\LA s\RA^2-1/3$ to $2/3$ as $\kappa_1$ changes from $-\infty$ to $\infty$, the endpoints being the $O_2$ and $O_1$ states (see Figure~\ref{fig:qomega_T_L}). The temperature approaches zero at both of these endpoints. For very small negative $\kappa_1$, the temperature is positive for $\LA s\RA<1/\sqrt{3}$ and negative for $\LA s\RA > 1/\sqrt{3}$. For intermediate values of $\kappa_1$ (these are not shown by the asymptotics), we find that $\LA q\RA$ increases monotonically as a function of $\kappa_1$ for fixed $\LA s\RA$ and the temperature assumes all values between $-\infty$ and $\infty$ with up to three local maxima as a function of $\kappa_1$, i.e., one with $\LA q \RA < 0$ for $\LA s\RA<1/\sqrt{3}$ and two with $\LA q\RA >0$ for $\LA s\RA<\LA s\RA_{\rm cr}=0.13714$. In the limit $\LA s\RA\rightarrow 0$, two of the local maxima of $T(\kappa_1)$ approach points $B$ and the third approaches point $A$ in Figure~\ref{fig:qsz_T_L0}.

\subsection{Non-axisymmetric rotating equilibria}\label{s:app:partition:rotating:triax}

For arbitrary $\kappa_2$ one of the integrals over the two polar angles may be evaluated in $Z_0$ Eq.~(\ref{eq:Zdef}), which gives (see Eq.~\ref{eq:Z})
\begin{align}
Z_0(\kappa_1,\kappa_2,c) &= 2\pi e^{-\frac13 \kappa_1}\int_{-1}^{1}e^{\kappa_1 s^2 + cs } I_0\left[\kappa_2(1-s^2)\right]  \,ds\,.
\label{eq:Z_general}
\end{align}
The zero-temperature non-axisymmetric equilibria correspond to $\kappa_2\rightarrow\infty$. We use 
\begin{equation}
I_{\alpha}(z)= \frac{e^{|z|}}{\sqrt{2\pi |z|}}\left(1-\frac{4\alpha^2-1}{8z}\right)+\mbox{O}\left(\frac{e^{|z|}}{|z|^{5/2}}\right)
\end{equation}
for $z\equiv \kappa_2(1-s^2)\rightarrow \pm\infty$. We obtain
\begin{align}
Z_0 \approx& \frac{\sqrt{2\pi}}{\sqrt{|\kappa_2|}} e^{|\kappa_2|-\frac13 \kappa_1}
\int_{-1}^{1}\frac{e^{(\kappa_1-|\kappa_2|) s^2 + cs} }{\sqrt{1-s^2}}
\,ds
\label{eq:ZO45}
\end{align}

The asymptotics near state $O_5$ in Figure~\ref{fig:QW_T_tr} may be obtained from Eq.~(\ref{eq:ZO45}) in the limit $\Delta\kappa\equiv\kappa_2-\kappa_1\rightarrow \infty$ and $c\rightarrow \infty$ such that 
\begin{equation}
\lambda\equiv \frac{c}{2\Delta\kappa}
\end{equation}
is fixed. Laplace's method\footnote{
For any twice differentiable function $g(s)$ with a unique minimum $\lambda$ in $[a,b]$, the following integral (if it exists) approaches
\begin{align}\label{eq:Laplace}
	\int_a^b f(s)e^{-x g(s)}ds \overset{x\rightarrow+\infty}{\longrightarrow} \frac{\sqrt{2\pi}}{\sqrt{a g''(\lambda)}}e^{-x g(\lambda)}\left[f(\lambda)+\mbox{O}(x^{-1})\right]\,.
\end{align}
We apply Eq.~(\ref{eq:Laplace}) in Eq.~(\ref{eq:ZO45}) with $x=\Delta \kappa$, $f(s)= (1-s^2)^{-1/2}$, and $g(s) = (s-\lambda)^2$.
}
yields
\begin{align}
Z_0 &= \frac{\sqrt{2}\pi e^{-\frac13 \kappa_1+\kappa_2 + \frac{\lambda}{2} c}}{\sqrt{(1-\lambda^2)\kappa_2(\kappa_2-\kappa_1) }}
\left[1+\mbox{O}(\Delta\kappa^{-1},\kappa_2^{-1})\right]
\label{eq:ZO5}
\end{align}

Substituting in Eqs.~(\ref{eq:q'})--(\ref{eq:s'}) gives
\begin{align}
\LA s \RA &= \lambda 
+ \frac{\lambda}{2\,(1-\lambda^2)\Delta\kappa}
+\mbox{O}\left( 
\kappa_2^{-1}\Delta \kappa^{-1} 
\right)\,, 
\label{eq:sO5'}\\
\LA q\RA &= -\ffrac{1}{3} + \lambda^2 + \frac{1+\lambda^2}{2(1-\lambda^2)\Delta\kappa}
+\mbox{O}\left( 
\kappa_2^{-1}\Delta\kappa^{-1} 
\right)\,,
\label{eq:qO5'}\\
\LA w\RA &= 1- \lambda^2 - \frac{1}{2\kappa_2}- \frac{1+\lambda^2}{2(1-\lambda^2)\Delta\kappa}
+\mbox{O}\left( 
\kappa_2^{-1}\Delta\kappa^{-1} 
\right)\,. \label{eq:wO5'}
\end{align}

Substituting in the self-consistency equation~(\ref{eq:T'}) gives a relation between $\Delta\kappa$ and $\kappa_2$ 
\begin{equation}
\frac{1}{\Delta\kappa} = \frac{1-\lambda^2}{(1-2\lambda^2)2\kappa_2} 
+ \frac{3\lambda^2(1-\lambda^2)}{4(1-2\lambda^2)^3\kappa_2^2}+ \mbox{O}\left(\kappa_2^{-3}\right)\,.
\end{equation}
We may now eliminate $\Delta\kappa$. Next solve for $\lambda$ using Eq.~(\ref{eq:sO5'}) and substitute back into Eqs.~(\ref{eq:T'}) and (\ref{eq:qO5'})--(\ref{eq:wO5'}) to get the asymptotics near state $O_5$ parameterized by $\kappa_2$ for any given $\LA s\RA$:
\begin{align}
\lambda &= \langle s\rangle - \frac{\langle s\rangle}{4(1-2\langle s\rangle^2)\kappa_2} 
+ \mbox{O}(\kappa_2^{-2})\,,\\
\LA q\RA 
&=-\frac{1}{3} + \langle s\rangle^2+\frac{1-\langle s\rangle^2}{4\left(1-2\,\langle s\rangle^2\right)\kappa_2}
 + \mbox{O}(\kappa_2^{-2})\,,
\label{eq:qO5}\\
\LA w\RA 
&=1 -\langle s\rangle^2 -\frac{3-5\,\langle s\rangle^2}{4\left(1-2\,\langle s\rangle^2\right)\kappa_2} + \mbox{O}(\kappa_2^{-2})\,,
\\
c &= \frac{4\langle s\rangle \left(1-2\langle s\rangle^2\right)\kappa_2}{1-\langle s\rangle^2}
-\frac{\langle s\rangle(1+\langle s\rangle^2-4\,\langle s\rangle^4)}{\left(1-\langle s\rangle^2\right)^2\left(1-2\langle s\rangle^2\right)}
+ \mbox{O}(\kappa_2^{-1})
\label{eq:wO5}\\
\frac{kT}{JN} &= \frac{1-\langle s\rangle^2}{2\kappa_2} -\frac{3-5\,\langle s\rangle^2}{8\left(1-2\,\langle s\rangle^2\right)\kappa_2^2} 
+ \mbox{O}(\kappa_2^{-3})\,.
\label{eq:TO5}
\end{align}
The $O_5$ state corresponds to the limit $\kappa_2\rightarrow \infty$.

The asymptotics near $O_4$ may be obtained from Eq.~(\ref{eq:ZO45}) in the limit that $\Delta\kappa$ approaches a finite value while $\kappa_2\rightarrow\infty$. In practice, we find numerically that $0\leq \Delta\kappa<1$ for $0\leq\LA s\RA < 0.52$, $\lim_{O_4}\Delta\kappa=2\LA s\RA^2$ for $0\leq\LA s\RA \ll 0.5$, and for any fixed $0<\LA s\RA<2^{-1/2}$, $c$ approaches a finite value. We derive an analytic approximation for the asymptotic behavior for $\LA s\RA < 0.52$ by expanding $Z_0$ Eq.~(\ref{eq:ZO45}) in $\Delta\kappa$ around 0 and using the identity (\ref{eq:Bessel}),
\begin{align}
Z_0&\approx\frac{\sqrt{2\pi^3}e^{\kappa_2-\frac13\kappa_1}}{\sqrt{\kappa_2}}
\left[I_0 + \frac{I_0+I_2}{2}\Delta\kappa 
+ \left(\frac{3I_0}{16}+\frac{I_2}{4}+ \frac{I_4}{16} \right)\Delta\kappa^2  + \mbox{O}\left(\Delta\kappa^{3},\kappa_2^{-1} \right)\right]\label{eq:ZO4}
\end{align}
where $I_n\equiv I_n(c)$ is the modified Bessel function\footnote{These equations become inaccurate for $s\gtrsim 0.52$. For $0.65<s< 2^{-1/2}$, we find numerically that $\Delta\kappa>10$, which implies that accurate analytic expressions exist in this regime (not shown), which may be derived with the Laplace method as in Eq.~(\ref{eq:Laplace}).}. Similarly, from Eqs.~(\ref{eq:Bessel}), (\ref{eq:ZO45}), and (\ref{eq:q'})--(\ref{eq:s'}), we get
\begin{align}
\LA s\RA &= \frac{I_1}{I_0}-\left(\frac{I_1}{4I_0} - \frac{I_1I_2}{2I_0^2} +\frac{I_3}{4I_0}\right)\Delta\kappa+\mbox{O}(\Delta\kappa^2)\,,
\label{eq:sO4'}\\
\LA q\RA &= \frac{1}{6} + \frac{I_2}{2I_0} -\left(\frac{1}{8}-\frac{I_2^2}{4I_0^2} + \frac{I_4}{8I_0}\right)\Delta\kappa+\mbox{O}(\Delta\kappa^2)\,,
\label{eq:qO4'}\\
\LA w\RA &= \frac{1}{2} - \frac{I_2}{2I_0} - \frac{3}{8\kappa_2} +\left(\frac{1}{8}-\frac{I_2^2}{4I_0^2} + \frac{I_4}{8I_0}\right)\Delta\kappa
+\mbox{O}(\Delta\kappa^2,\kappa_2^{-2})\,.
\label{eq:wO4'}
\end{align}
We may substitute in the self-consistency equation (\ref{eq:T'}) to get a relation between $\Delta\kappa$ and $\kappa_2$
\begin{align}
\Delta\kappa =& \frac{4 I_0 I_2}{I_0^2 - 2I_2^2 + I_0I_4} + 
\frac{I_0^2(3I_0^2+16I_0I_2-6I_2^2+3I_0I_4)}{4(I_0^2-2I_2^2+I_0I_4)^2\kappa_2}
+\mbox{O}\left( \kappa_2^{-2} \right)\,,
\end{align}
Now eliminate $\Delta\kappa$ from Eqs.~(\ref{eq:sO4'})--(\ref{eq:wO4'}) 
\begin{align}
&\LA s\RA = \frac{I_1}{I_0} - \frac{I_2}{I_0}\left(\frac{I_0I_1 - 2I_1I_2 + I_0I_3}{I_0^2-2I_2^2+I_0I_4}\right)
-\frac{(I_0I_1-2I_1I_2+I_0I_3)\left(3I_0^2+16I_0I_2-6I_2^2+3I_0I_4\right)}{16\left(I_0^2-2I_2^2+I_0I_4\right)^2\kappa_2}
+\mbox{O}\left(\kappa_2^{-2}\right)\,,
\label{eq:sO4}
\end{align}
\begin{align}
\LA q\RA &= \frac{1}{6} - \left[\frac{3}{32} + \frac{I_0I_2}{2\left(I_0^2-2I_2^2+I_0I_4\right)}\right]\frac{1}{\kappa_2}
+\mbox{O}\left(\kappa_2^{-2}\right)\,,
\label{eq:qO4}\\
\LA w\RA &= \frac{1}{2} - \left[\frac{9}{32} - \frac{I_0I_2}{2(I_0^2-I_2^2+I_0I_4)}\right]\frac{1}{\kappa_2} + \mbox{O}\left(\kappa_2^{-2}\right)\,,
\label{eq:wO4}\\
\frac{kT}{JN} &= \frac{1}{4\kappa_2} - \left[\frac{9}{32} - \frac{I_0I_2}{2(I_0^2-I_2^2+I_0I_4)}\right]\frac{1}{2\kappa_2^2} + \mbox{O}\left(\kappa_2^{-3}\right)\,.
\label{eq:TO4}
\end{align}

Therefore at zero temperature, the $O_2$ and $O_4$ equilibria depend on $\LA s\RA$, while $O_1$ and $O_5$ are independent of $\LA s\RA$. The latter series of equilibria depend on $\LA s\RA$ for $T\neq 0$ at first and second beyond leading order in $T$ for $O_5$ and $O_1$, respectively.
In the $T\rightarrow 0$ limit, Eqs.~(\ref{eq:qO5})--(\ref{eq:TO5}) show that the $O_5$ order parameters are in the range $-\frac13\leq \LA q\RA \leq \frac16$ and $1\geq \LA w\RA \geq \frac12$ depending on $\LA s\RA$ as long as it satisfies $0\leq \LA s\RA\leq 2^{-1/2}$. Outside of this range, $\Delta \kappa<0$ for all $\kappa_2$, hence the $\Delta\kappa\rightarrow \infty$ assumption cannot be satisfied, and there is no $O_5$ state. 
The asymptotics near $O_4$ (Eqs.~\ref{eq:sO4})--(\ref{eq:TO4}) show that $\LA q\RA=\frac16$ and $\LA w\RA=\frac12$, independent of $\LA s\RA$. 

In the $\LA s\RA\rightarrow 0$ limit, the non-axisymmetric equilibria near $O_4$ and $O_5$ (Eqs.~\ref{eq:sO4}--\ref{eq:TO4} and \ref{eq:qO5}--\ref{eq:TO5}) reduce to the axisymmetric asymptotics near $O_2$ and $O_1$ (Eqs.~\ref{eq:sO2}--\ref{eq:TO2} and \ref{eq:sO1}--\ref{eq:TO1}), respectively, in a rotated coordinate system (see Eq.~\ref{eq:Qmatrix}) 
\begin{align}
\LA q\RA_{\{O_5\}}&\rightarrow
-\ffrac{1}{2}\LA q\RA_{\{O_1\}}\,,\quad 
\LA w\RA_{\{O_5\}} \rightarrow \ffrac{3}{2}\LA q\RA_{\{O_1\}}\,,
\label{eq:qO1O5}\\
\LA q\RA_{\{O_4\}}&\rightarrow
-\ffrac{1}{2}\LA q\RA_{\{O_2\}}\,,\quad 
\LA w\RA_{\{O_4\}}\rightarrow -\ffrac{3}{2}\LA q\RA_{\{O_2\}}\,.
\label{eq:qO2O4}
\end{align}

The energy of non-axisymmetric equilibria is bounded between 
\begin{align}
E_{O_5} \leq  E \leq E_G  \mathrm{~~if~} \frac{L}{Nl} \leq \frac{1}{\sqrt{2}}\,.
\end{align}
where 
\begin{equation}
E_{O_5} = -\frac{1}{3} + \frac{L^2}{N^2l^2} \left(1-\frac{L^2}{N^2l^2}\right)
\end{equation}
and $E_G \leq E_{O_3}$ for all $L$ where $E_{O_3}$ is the upper energy bound in Eq.~(\ref{eq:Ebound}) of the main text (see Figure~\ref{fig:caloric_F_tr}).

\end{document}